\documentclass[prd, onecolumn, superscriptaddress, nofootinbib, notitlepage, floatfix, preprintnumbers]{revtex4-1}
\usepackage{graphicx} 
\usepackage{amsmath}
\usepackage{amsfonts}
\usepackage{amssymb}
\usepackage{natbib}
\usepackage{diagbox}
\usepackage{physics}

\usepackage{color}

\begin{document}

\title{Non-local Nucleon Matrix Elements in the Rest Frame}

\newcommand*{\Jlab}{Thomas Jefferson National Accelerator Facility, Newport News, VA 23606,  USA}\affiliation{\Jlab}    
\newcommand*{\WM}{William \& Mary, Williamsburg, VA  23187, USA }\affiliation{\WM}
\newcommand*{\ODU}{Department of Physics, Old Dominion University, Norfolk, VA 23606, USA}\affiliation{\ODU}

\author{Joe Karpie} \affiliation{\Jlab}  
\author{Christopher Monahan} \affiliation{\WM}
\author{Anatoly Radyushkin} \affiliation{\Jlab} \affiliation{\ODU}  

\begin{abstract}
Extracting parton structure from lattice quantum chromodynamics (QCD) calculations requires studying the coordinate scale $z_3$ dependence of the matrix elements of bilocal operators. The most significant contribution comes from the 
$z_3$ dependence induced by ultraviolet (UV) renormalization of  the Wilson line. We demonstrate that the next-to-leading order perturbative calculations of the renormalization factor can describe, to a few percent accuracy, the 
{  logarithm 
}
of the lattice QCD rest frame matrix elements with separations up to distances of 0.6~fm  on multiple lattice spacings. The residual discrepancies can be modeled by a leading effect from the structure of the nucleon.
\end{abstract}

\preprint{JLAB-THY-24-4126}
\maketitle

\section{Introduction}

Studying the internal structure of the nucleon requires calculations of the complex interactions of quarks and gluons. Very different types of effects generated by Quantum ChromoDynamics (QCD) can generate the scale dependence seen in observables calculated in lattice QCD. To extract quark distributions from lattice calculations one needs to consider correlations of quarks at different locations through the operator ${\cal O}^\mu(z) = \bar{\psi}(0)\gamma^\mu W( 0 , z ) \psi(z)$ 
of two quark fields connected by the Wilson line $ W( 0 , z )$, and separated by interval  $z$. The standard parton distribution function definition sets $z^2=0$ by separating the quarks in space and time. 
Due to the  Euclidean  nature of numerical lattice QCD, $z$ cannot be light-like. The usual choice~\cite{Ji:2013dva} on the lattice is to  take $z=(0,0,0,z_3)$, which introduces $z^2$ as a new scale in the problem, alongside the lattice spacing $a$ and the IR scale of QCD $\Lambda_{\rm QCD}$.
To  obtain  information about light-cone functions, it is necessary to handle the $z^2$-dependence of lattice data. It is crucial to disentangle the origins of the short-distance behavior generated by the ultraviolet (UV) regulator from the interesting long distance behavior that governs the parton structure of the hadron.

The strongest artifact of setting $z^2\neq 0$ is produced by the renormalization of the  Wilson line, which leads 
to linear $z/a$ divergences. Other UV effects come in the form of logarithms in $z^2/a^2$ and terms that vanish when the regulator is removed. Another subdominant effect is generated by the finite size of the hadron. That is, quark correlations must eventually decay strongly when the quarks are separated beyond the hadron's diameter. Finally, in matrix elements of moving states, the collinear divergence of the underlying parton distribution generates a logarithmic $z^2$ behavior. This behavior can be removed by looking at the rest frame matrix elements since these effects vanish when $p\cdot z=0$.

Our goal in this paper is  to analyze nonperturbative lattice data for the rest-frame matrix  elements 
 and investigate the extent to which the scale-dependence of the matrix elements follows the perturbative predictions. As part of this, we quantify the  deviations from perturbative expectations. 

 The paper is organized  as follows. 
 In Sec.~\ref{sec:matelem}, we discuss 
 basic effects leading to $z_3$ dependence of 
 matrix elements of bilocal operators. 
 First, in Sec.~\ref{sec:matelem}A we illustrate the structure of 
 $\langle p | {\cal O}(z_3) |p \rangle $  matrix element on the example of simplest tree diagrams in scalar theory, which we calculate in a Lorentz-invariant way. The results are derived  for arbitrary $p$ and $z$, and the rest-frame amplitude is obtained by simply taking $(pz)=0$ in the final expression. We also 
 use this example to show definitions of the basic objects of the pseudo-PDF approach. In Sec.~\ref{sec:matelem}B, we discuss the renormalization constant calculated with a Polyakov regularization. In Sec.~\ref{sec:matelem}C, we discuss
nonperturbative effects related to the finite size of hadrons. 
 In Sec.~\ref{sec:polyakov} we 
 describe  the early observations made in Ref.~\cite{Radyushkin:2019mye} that the 
 perturbative 
 expression based on Polyakov regularization for the link renormalization  factor  accurately reproduces lattice data for the 
 rest-frame matrix elements 
  up to rather large values of $z_3$.  In Sec.~\ref{sec:lattresults},
 we describe the lattice data of Ref. \cite{Karpie:2021pap} in which three different 
 lattice spacings have been  used. In Sec.~\ref{sec:polyakov_fits}, we fit these data with an analytic  expression 
 for the link renormalization factor $Z_{\rm PR}$
 obtained using Polyakov regularization.
 After that, we fit the data by a product of the perturbative 
formula and a Gaussian factor that models effects  due to the finite-size of the proton. We observe that inclusion of the Gaussian factor results in strong improvement of fits quality.
 In Sec.~\ref{sec:lpt},  we discuss  numerical results for 
 the renormalization  factor $Z_{\rm LPT}$ based on lattice perturbation theory and repeat the fitting procedure using $Z_{\rm LPT}$ instead of $Z_{\rm PR}$. 
 Our conclusions are summarized in Sec.~\ref{sec:conclusion}.

\section{Structure of matrix elements\label{sec:matelem}}

\subsection{Scalar toy model}  
 
\begin{figure}[h]
 	\centerline{\includegraphics[width=2.5in]{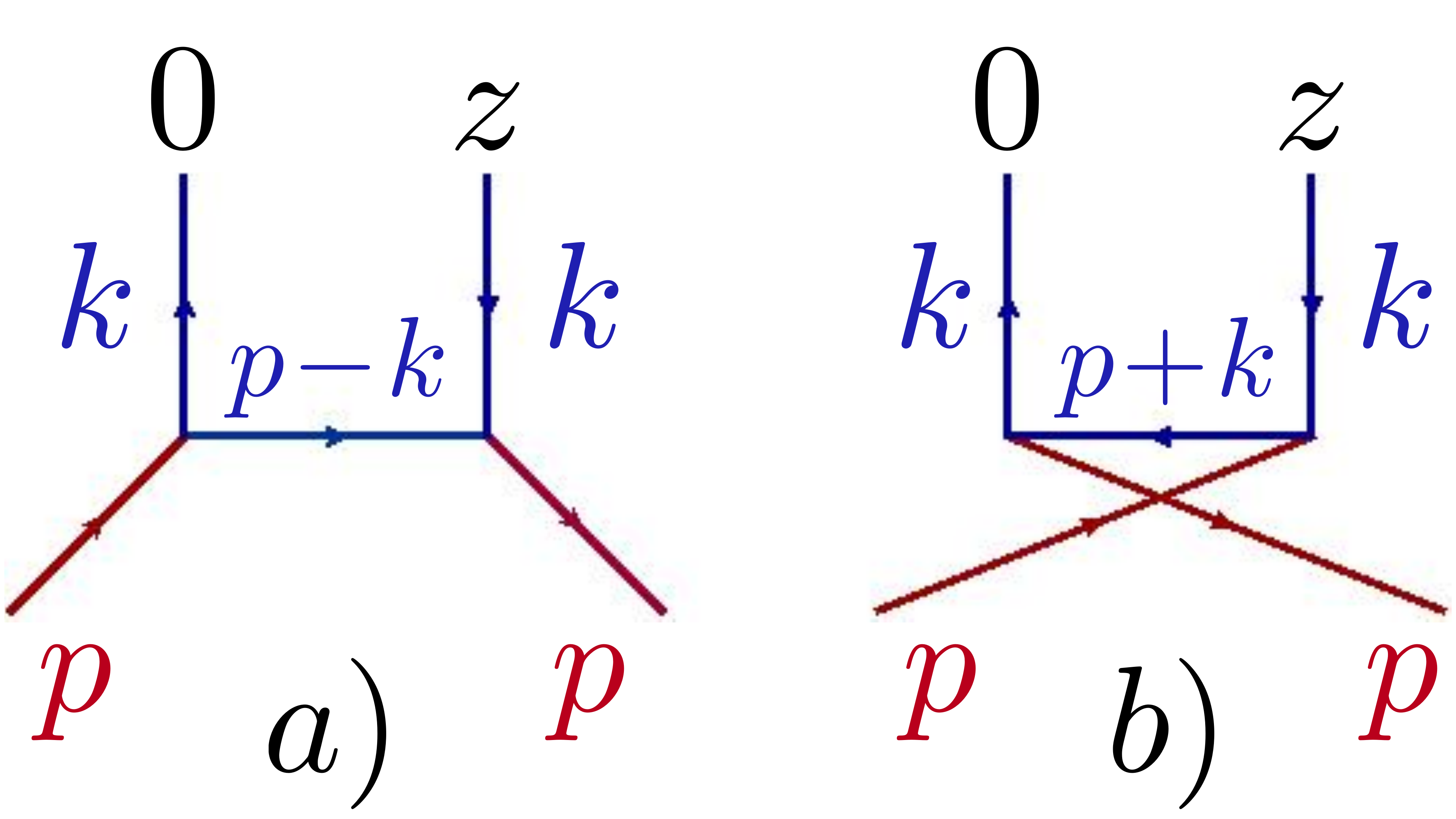}}
 	\caption{ Matrix element  $\langle p | {\cal O}(z) |p \rangle$ in the lowest order of a scalar model.
 		\label{dia}}
 \end{figure}

Before restricting our analysis to the rest frame, it is instructive to illustrate the  structure of the 
 $\langle p | {\cal O}(z) |p \rangle$ matrix element 
 with the example of the  simplest tree diagrams in  
 scalar theory. Here the bilinear operator is  ${\cal O}(z) =\varphi (0) \varphi (z)$, 
taking elementary scalar particles with momentum $p$ as external states. More details appear in, for example,  \mbox{Refs.~\cite{Radyushkin:2016hsy,Radyushkin:2019mye,DelDebbio:2020cbz}.}
 The ``$s$-channel'' diagram shown in Fig. \ref{dia}{\it a)} is given by the 
 momentum integral
  \begin{align}
T_s(z,p) \sim  \int \frac{\dd^4 k} {k^2 (p-k)^2 k^2{ }} e^{-i(kz)} \  . 
\end{align}

Using the $\alpha$-representation for propagators 
  \begin{align}
\frac{1} {k_i^2 +i \varepsilon}= \frac1{i} \int_0^\infty \dd\alpha_i 
e^{i \alpha_i(k_i^2 +i \varepsilon)}
\end{align}
we obtain
  \begin{align}
T_s(z,p) \sim  \int_0^\infty \frac{\dd\alpha_1 \dd\alpha_2 \dd\alpha_3}{ ( \alpha_1 +\alpha_2 +\alpha_3 )^2}  \exp \left [ -i(pz)
\frac{ \alpha_2 }{  \alpha_1 +\alpha_2 +\alpha_3} -i \frac{z^2}{4(  \alpha_1 +\alpha_2 +\alpha_3)} + i p^2 \frac{\alpha_2 (  \alpha_1+\alpha_3)}{ \alpha_1 +\alpha_2 +\alpha_3 } \right ] \ .
\label{Ts}
\end{align}
This representation explicitly shows that the amplitude $T(z,p) $ depends on the coordinate $z$ through two Lorentz 
invariants $(pz)\equiv -\nu $ (Ioffe time)  and the interval $z^2$, so that  we have  $T(z,p)= {\cal M} (\nu, -z^2)$,
where ${\cal M} (\nu, -z^2)$ is the Ioffe-time pseudodistribution (pseudo-ITD).

Introducing variables  $\lambda \equiv \alpha_1 +\alpha_2 +\alpha_3 $  and $x \equiv { \alpha_2 }/\lambda$, 
we can  write $T_s(z,p)$  in a ``pseudo-PDF'' representation 
 \begin{align}
T_s(z,p) = \int_0^1 \dd x \,  e^{-ix (pz)} {\cal P}_s(x, z^2) \ .  
\label{pPDF}
\end{align}
Since $\alpha_2 \leq \lambda$, the variable $x$ has the standard ``parton'' support region $0 \leq x \leq 1$ for  any $z^2$. 
 On the light cone $z^2=0$, the pseudo-PDF ${\cal P}_s(x, z^2)$ converts into usual ``quark'' PDF $f(x)$. 

Similarly, for  the ``$u$-channel'' diagram shown in Fig. \ref{dia}{\it b)}, we obtain
  \begin{align}
T_u(z,p) \sim  \int_0^\infty \frac{\dd\alpha_1 \dd\alpha_2 \dd\alpha_3}{ ( \alpha_1 +\alpha_2 +\alpha_3 )^2}  \exp \left [ i(pz)
\frac{ \alpha_2 }{  \alpha_1 +\alpha_2 +\alpha_3} -i \frac{z^2}{4(  \alpha_1 +\alpha_2 +\alpha_3)} + i p^2 \frac{\alpha_2 (  \alpha_1+\alpha_3)}{ \alpha_1 +\alpha_2 +\alpha_3 } \right ] \ . 
\label{Tu}
\end{align}
It  has opposite sign in front of  $(pz)$,   hence,  the $x$-variable in its  pseudo-PDF representation runs within the $-1\leq x \leq 0$  limits, \emph{i.e.}~the $u$-diagram corresponds to  the 
``antiquark''  distribution.

It should be emphasized that our derivation of the ``parton'' representation 
\eqref{pPDF}  is fully Lorentz-invariant. We have  not used  the  light-front decomposition  for momenta, 
did not pick out the  
plus-components
of $k$ and $p$, nor did we use Sudakov variables, \emph{etc}. 
The  variable $x$ is Lorentz-invariant, and, according to 
  Eq.~\eqref{pPDF},    
the parton carries the fraction $x$ of the whole  external  momentum $p$
(of course, if we  take $z$ on the light cone, with only $z^-$  being nonzero, then the   parton carries $xp^+$
momentum).
Furthermore, the support property $-1\leq x \leq 1$ that is often expected to hold 
just for  light-cone  intervals $z^2$=0, as we see, holds for all  $z^2$. In particular, we can take $z=(0,0,0,z_3)$.
If  $p$ is also taken in the $3^{\rm  rd}$ direction, then $\nu=0 $   corresponds to the rest frame ${\bf p}=0$.
In our example, to obtain ${\cal M} (0,z^2)$, one may    simply take $(pz)=0$  in the  final results, Eqs.~\eqref{Ts} and \eqref{Tu}.

This calculation scheme may be also  used for higher-order diagrams.  It is possible to show (see Refs.~\cite{Radyushkin:1983wh,Radyushkin:2016hsy,Radyushkin:2019mye}) that  any Feynman diagram contributing to $T(z,p)$  has the ``virtuality distribution function'' 
(VDF) representation \cite{Radyushkin:2014vla}  
 \begin{align}
 T(z,p)
=   
\int_{0}^{\infty} \dd \sigma \int_{-1}^1 \dd x\,  %
 \Phi (x,\sigma; p^2)  e^{-i x (pz) -i \sigma {(z^2-i \epsilon )}/{4}} \,   . 
 \
 \label{newVDFx}
\end{align} 
The parameter $\sigma$ is  Fourier-conjugate to $z^2$, so it may be loosely interpreted as virtuality,
hence the ``VDF'' name  for the function $ \Phi (x,\sigma; p^2)$. 

\subsection{Ultraviolet renormalization}

In the realistic  case of QCD,
the basic operator ${\cal O}^\mu(z) = \bar{\psi}(0)\gamma^\mu W( 0 , z ) \psi(z)$ contains an extra factor 
$ W( 0 , z )$,  the Wilson line. 
Perturbative corrections to this operator are ultraviolet-divergent 
in the continuum limit, and correspond to Wilson-line 
renormalization. On the lattice, one  deals
with linear $\sim 1/a$ and logarithmic $\sim \ln a$ singularities with respect to the lattice spacing $a$.

In the continuum case, the ultraviolet 
renormalization factor  may be  calculated in  perturbation theory,
using, \emph{e.g.}~the Polyakov regularization (PR) \cite{Polyakov:1980ca}, in which the coordinate-space gluon propagator $1/z^2$ is substituted (in Euclidean space) 
by $1/(z^2+a_{\rm PR}^2)$,  where $a_{\rm PR}$ is the UV regularization parameter. 
The expression for the sum of one-loop  link self-energy and vertex  corrections, as calculated in Ref.~\cite{Radyushkin:2017cyf},  is given by 
    \begin{align}
  \Gamma (\zeta) =  -A \left[ 2 \zeta  \tan^{-1}(\zeta ) -\log \left(1+\zeta^2\right)+1  -\left(\frac{1}{\zeta^2}+1\right) \log\left(1+\zeta ^2\right)  \right]\equiv - A s(\zeta ).
   \label{fla} 
       \end{align} 
The calculation gives $\Gamma $ as   a function of  $\zeta\equiv  z_3/a_{\rm PR}$,   
with an overall factor $A$,   which is $A=C_F\alpha_s/(2 \pi )  $ for quarks in QCD.
The large $\zeta$ expansion
\begin{equation}
    2 \zeta  \tan^{-1}(\zeta )\sim \pi\zeta -2 +\frac{2}{3\zeta^2}  +O\big(\frac{1}{\zeta^4}\big) \label{eq:arctan_exp}
    \end{equation}
produces a linear $\sim 1/a_{\rm PR}$ singularity for small $a_{\rm PR}$, while the expansion 
\begin{equation}
    \log \left(1+\zeta ^2\right) =\log\zeta^2 +  \log\left(1+\frac{1}{\zeta ^2}\right) \sim \log\zeta^2 +\frac1{\zeta^2} + O\left(\frac{1}{\zeta^4}\right ) \label{eq:log_exp}
\end{equation}
gives a logarithmic singularity.
 In Feynman gauge, the $2 \zeta \tan
   ^{-1}(\zeta) -\log \left(1+\zeta^2\right)$ contribution comes from the Wilson line self-energy correction, while 
  \mbox{ $1  -\left({1}/{\zeta^2}+1\right) \log
   \left(1+\zeta^2\right)$}   corresponds to vertex corrections.
 Note that the ``1'' and $\frac{1}{\zeta^2} \log
   \left(1+\zeta^2\right)$  terms in $\Gamma(\zeta)$ are not singular for $a_{\rm PR}=0$, and they  have   been omitted in the result for
   $\Gamma(\zeta)$  given in Ref.~\cite{Chen:2016fxx}.
   
   {
  The ultraviolet divergencies are also present in the quark self-energy corrections. Regularizing them using the  Polyakov regularization
   gives   logarithms of $\ln (a^2 \mu_{\rm IR}^2)$ form,  where $\mu_{\rm IR}^2$ is  an infrared  cut-off like $m^2$, the quark mass squared,
 quark  virtuality $-p^2$, etc.  These   contributions do not depend on $z^2$,  and do  not give   corrections to
 normalized matrix elements like $M(z_3)/M(0)$.  
 
 However, in Ref.~ \cite{Radyushkin:2017lvu}, it was proposed to  write the
   logarithm $\ln (a^2 \mu_{\rm IR}^2)$
 as the   sum $\ln (a^2 /z_3^2) + \ln(z_3^2 \mu_{\rm IR}^2)$, thus  separating   its UV and IR parts.  
 The   first term $\ln (a^2 /z_3^2)$ has  the  structure similar to the  logarithmic contributions in $   \Gamma (\zeta)$,
 and   may  be   added to it  to produce a function containing all UV terms appearing at the one-loop level. 
 The  second term $\ln(z_3^2 \mu_{\rm IR}^2)$ may  be  combined with the logarithmic 
    contribution from the diagram describing the  gluon 
 exchange between incoming and outgoing quarks. As  a result,  the two $\ln(z^2 \mu_{\rm IR}^2)$ terms  produce the 
 contribution to  the evolution kernel that  has the plus-prescription structure. Due to this property, the  evolution  
 contribution gives a vanishing  correction to the rest-frame matrix element. Adding the $\sim \ln (a^2 /z^2)$ to $ \Gamma (\zeta)$
 gives the  modified function $   \widetilde{\Gamma} (\zeta)$
     \begin{align}
  \widetilde{\Gamma}  (\zeta) = \Gamma  (\zeta)  - A \ln (\zeta)
   \label{flamod} 
       \end{align} 
 that  will be used in what  follows.    
  For large $\zeta$, the logarithmic contribution to $ \widetilde{\Gamma}  (\zeta)$ equals $\frac32 A \ln \zeta^2$,
  in agreement with Refs.\cite{LatticePartonLPC:2021gpi,Gao:2024gui}.   }

   Using lattice perturbation theory (LPT), it
 is possible to carry out similar calculations, with the lattice spacing $a$ serving as    the UV cut-off.
 It was  observed~\cite{Chen:2016fxx} that the result obtained using the continuum Polyakov regularization practically coincides 
 with that  of LPT, provided that one takes $a_{\rm PR}=a/\pi$, which is the inverse of the UV momentum cutoff in LPT.

In practical lattice QCD, the UV singularities complicate taking the continuum $a\to 0$ limit.
Fortunately, these divergences 
 are  multiplicative \cite{Polyakov:1980ca,Dotsenko:1979wb,Brandt:1981kf,Aoyama:1981ev,Craigie:1980qs,Dorn:1986dt,Bagan:1993zv}  (see also more 
 recent papers  
 \cite{Ishikawa:2017faj,Ji:2017oey,Green:2017xeu}). After summation over all loops, the divergences exponentiate and 
form  a  factor $Z(z_3/a)$, where $a$ is a UV cut-off, \emph{e.g.}, the Polyakov parameter $a_{\rm PR}$ or  
lattice spacing $a$ as calculated in Secs.~\ref{sec:polyakov} and~\ref{sec:lpt}, respectively.  Thus, one can renormalize hadron matrix elements $\langle p | {\cal O}^\mu(z) |p \rangle$
(where $p$ is the hadron momentum, usually taken as $p=(E, 0, 0, P_3)$) by dividing them  by $Z(z_3/a)$. 
After  exponentiation, 
   the linear term in $\zeta = z_3/a $  of the one loop diagrams gives a quickly  
   decreasing factor $\exp\left [-\zeta \alpha_s C_F/2 \right ] $
that brings a strong $z_3$ dependence into the matrix elements. This dependence is  an artifact of 
using a non-lightcone separation, and should be eliminated in the process of 
extrapolating  lattice results to the $z_3 \to  0$ 
limit.  


Thus, division by $Z(z_3/a)$ serves two purposes: \\ {\it i)} it eliminates the UV dependence of lattice results on the lattice spacing
and \newline  {\it ii)} it  cancels the strong $z_3$ dependence  induced by the UV singular effects. 

A natural question is:  how one can obtain $Z(z_3/a)$? Since the  spacing $a$ used in lattice PDF extractions 
is usually about or less than $0.1$ fm,  one may appeal to asymptotic
freedom of QCD and use the perturbative expressions \eqref{fla}, \eqref{flamod}, at least when   
$z_3$ is also small, say, a few lattice spacings.

\subsection{Nonperturbative effects for large $z$}

However, for large $z_3$, one may expect long distance effects which are generated by the low virtuality dependence of the VDF. An obvious source of such effects is the finite size of hadrons,
which results from the Fermi motion of quarks/gluons inside a hadron. In momentum representation, such effects lead to a $k_T$ dependence of momentum-dependent parton distributions (TMDs) $F(x,k_T)$.  It is common to take such a dependence to be a Gaussian $\sim e^{-k_T^2/\Lambda^2}$ and  the same for all $x$ values, \emph{i.e.}~to use 
$F(x,k_T) =f(x) e^{-k_T^2/\Lambda^2} $  
 (see  Ref.~\cite{Boussarie:2023izj}, Section  5.2.1 for further  references). 

 In the coordinate representation, such a  model  would correspond 
to an overall  Gaussian $e^{-z_T^2 \Lambda^2/4}$ factor.   
This model is   a particular case of {\it factorization} of $x$ and $k_T$  dependence  for   a TMD, 
$F(x,k_T) =f(x) K(k_T^2)$. 
For matrix elements $\langle p | {\cal O}(z_3) |p \rangle \equiv M(z_3, P_3)$,  such an  assumption corresponds to the $M(z_3, P_3)= {\cal M} (\nu) R(z_3^2) $ relation, where $\nu \equiv P_3 z_3$ 
is the Ioffe time \cite{Ioffe:1969kf,Braun:1994jq} and $R(z_3^2) $ is some function reflecting the finite size of the hadron.

To eliminate the unwanted  $z_3^2$ dependence in  $M(z_3, P_3) = \langle p | \mathcal{O}^\mu(z_3) | p \rangle$, from perturbative effects and from finite size effects
in the  pseudo-PDF approach  \cite{Radyushkin:2017cyf,Orginos:2017kos}, it was  proposed 
to measure  the  ratio 
\begin{align}
{\cal M}^\mu (z_3,p;a) =\frac{\langle p | {\cal O}^\mu(z_3) |p \rangle }{\langle p_0 | {\cal O}^\mu(z_3) |p_0 \rangle} 
 \label{redm}
\end{align}
of the basic matrix element $\langle p | {\cal O}^\mu(z_3) |p \rangle$ with boosted momentum $p=\{E, 0,0, p_3 \}$ to 
  the rest-frame matrix element $\langle p_0 | {\cal O}^\mu(z) |p_0 \rangle$,
in which  $p_0= (M,0,0,0)$, with $M$ the hadron mass.  This ratio
(that was considered, as a matter of fact,  as early as in 2010 \cite{Musch:2010ka})
has a finite $a\to 0$ limit, because the numerator and denominator have the same UV $Z(z_3/a)$ factor. 
Moreover, if the hadron finite-size effects result in an overall $z_3^2$-dependent factor, as suggested by popular TMD models,
then such  a factor  also cancels in the ratio. In such a situation, the ratio would depend on 
the Ioffe time only, as largely demonstrated by the pioneering work of Ref~\cite{Musch:2010ka}.  This expectation is also supported by the results of the first pseudo-PDF analysis \cite{Orginos:2017kos}  and further studies (see, e.g., 
Refs.~\cite{Joo:2019jct,Joo:2019bzr,Joo:2020spy,Bhat:2020ktg,Gao:2020ito,Karpie:2021pap,Egerer:2021ymv,HadStruc:2021qdf,HadStruc:2022nay,Bhat:2022zrw,Bhattacharya:2024qpp,HadStruc:2024rix})
based on the ratio scheme.  
In fact,  some deviation from a purely $\nu$-dependent function is observed for small $z_3$,
but it  was shown to be in  agreement with expected perturbative QCD evolution.

One may argue that there is no first-principle reason for factorization of $x$- and $k_T$-dependence,
or, what is the same, for  factorization of $\nu$- and $z_3^2$ dependence  in the  $M(z_3, P_3)$ matrix element. 
Still, we may  appeal to the physical interpretation of the non-perturbative $z_3^2$-dependence as reflecting 
the finite hadron size. Thus, we may    expect that, for all $\nu$,  the  function ${\cal M}(\nu, z_3^2)$ 
decreases with the increase of
$ z_3^2$.  As a result, the ratio  ${\cal M}(\nu, z_3^2)/{\cal M}(0, z_3^2)$ is expected to have a slower decrease  
as $z_3^2$  grows  than ${\cal M}(\nu, z_3^2)$.  

On the practical side, one should realize that  the nonperturbative $z_3^2$-dependence 
of  ${\cal M}(\nu, z_3^2)$ cannot   be predicted:
it  can be only extracted from lattice data by fitting. 
Thus, dealing with the ratio ${\cal M}(\nu, z_3^2)/{\cal M}(0, z_3^2)$  we simply  fit its nonperturbative 
\mbox{$z_3^2$-dependence}   rather than that of ${\cal M}(\nu, z_3^2)$. 
The advantage of using  the ratio is that,  due to  expected  partial (and maybe almost  complete)  cancellation 
of such dependence between  ${\cal M}(\nu, z_3^2)$ and ${\cal M}(0, z_3^2)$, the ratio has 
 much smaller 
higher-twist  corrections.

\section{Continuum Polyakov Regulator: Early Findings }\label{sec:polyakov}

In Ref.~\cite{Radyushkin:2019mye}, it was noted  that the rest-frame matrix element obtained in the 
initial quenched calculation \cite{Orginos:2017kos} based on the pseudo-PDF method
is   rather well described by the   
exponentiated version
 \begin{align} 
Z_{\rm pert}  (z_3/a)  = \exp \left  \{  \Gamma (z_3 \pi/a) \right \} \  
       \label{Zuv}
 \end{align}
 of the one-loop  formula in Eq.~\eqref{fla}, in which the Polyakov regularization parameter $a_{\rm PR}$ is  substituted  by the lattice spacing $a$,
using the correspondence $a_{\rm PR}=a/\pi$ proposed in Ref.~ \cite{Chen:2016fxx}.

\begin{figure}[h]
 	\centerline{\includegraphics[width=2.5in]{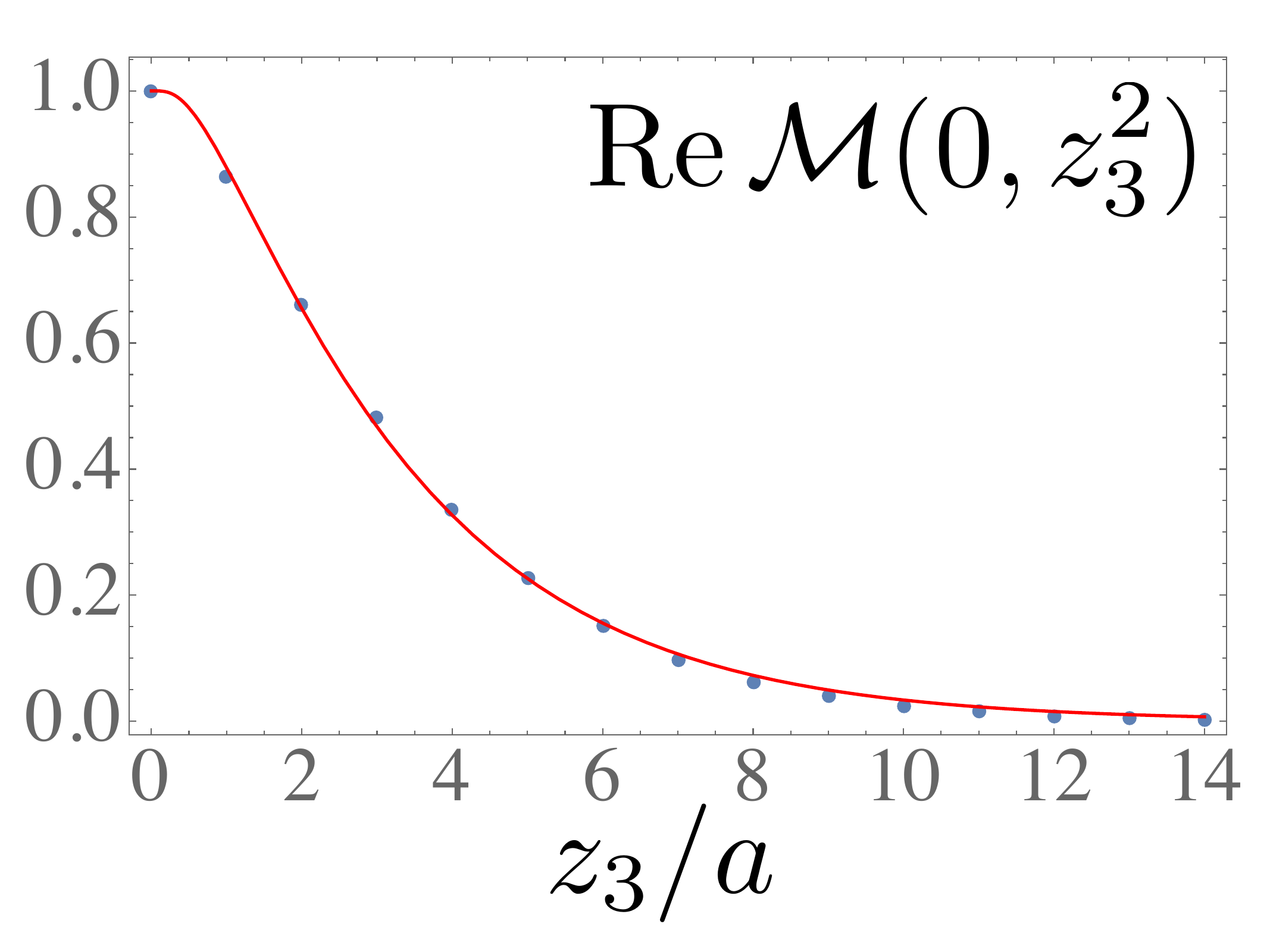}}
 	\caption{Real part of the rest-frame amplitude $M(z_3, p_3=0) ={\cal M}(0,z_3^2)$ in quenched calculation \cite{Orginos:2017kos}.
 		\label{Mrest}}
 \end{figure}

 The   $p_3=0$ data obtained in \cite{Orginos:2017kos} are shown in   Fig.~\ref{Mrest}, together with the 
 curve for $Z_{\rm pert}  (z_3/a) $. 
The value of  $\alpha_s$ obtained from the fit  is 0.19.  
One can see that the data for ${\cal M} (0, z_3^2)$ in this particular lattice
simulation  are rather  accurately reproduced by the perturbative formula.

\begin{figure}[h]
 	\centerline{\includegraphics[width=2.9in]{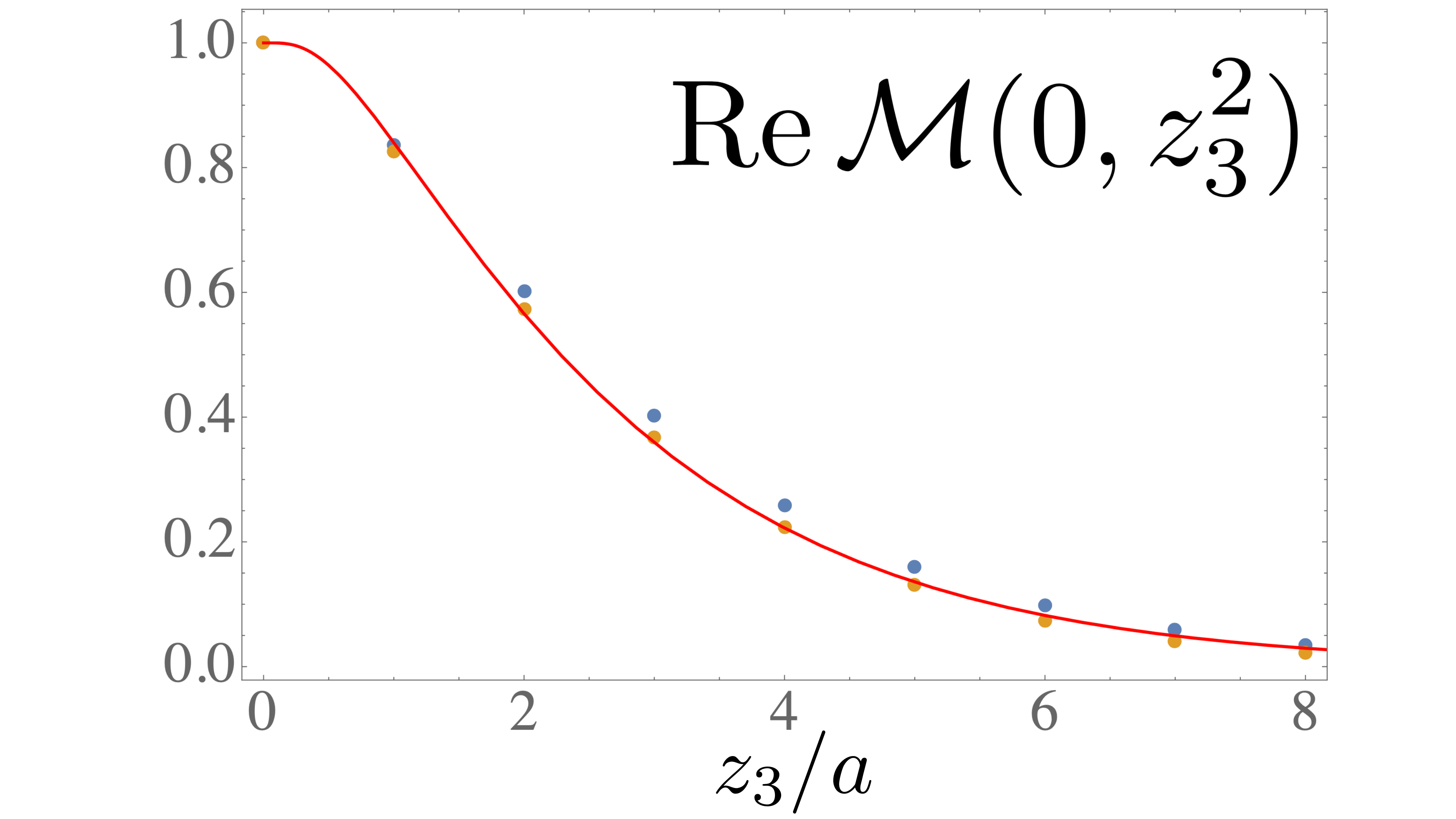}}
 	\caption{Real part of the rest-frame amplitude ${\cal M}(0,z_3^2)$
	for lattice spacings 0.094 fm (higher points) and 0.127 fm in  calculation with dynamical fermions \cite{Joo:2019jct} .
 		\label{Mrestdyn}}
 \end{figure}

In Ref.~\cite{Radyushkin:2019mye}, it was also shown that a
good  agreement between the  rest frame data  holds for the results 
of a dynamical-fermion calculations  described in Ref.~\cite{Joo:2019jct}.
In that paper, two  lattice spacings,   0.094~fm
and 0.127~fm were used.
In Fig.~\ref{Mrestdyn} we show the lattice results of  Ref.~\cite{Joo:2019jct}, together 
with the  fit of $a=0.127$~fm  data by  the perturbative formula of Eq.~\eqref{Zuv}, 
in which $\alpha_s=0.26$.  One  can see that, just as in the quenched calculation,
the    lattice   points  are well described by the perturbative formula, Eq.~\eqref{Zuv}.

Note that the points for the  two  different lattice spacings  are plotted as functions 
of the ratio $z_3/a$ rather than as functions of the physical distance $z_3$.
This choice is suggested by the perturbative calculation  that 
gives  the $Z$-factor as  a function of $z_3/a$. 
Indeed, one can see that the two sets of points in Fig. \ref{Mrestdyn}
are very close to each other. The points corresponding to the 
0.094~fm lattice spacing 
are just slightly above the curve in Fig. \ref{Mrestdyn} describing the 0.127~fm points. 
In fact, the 0.094~fm points  are also 
well described 
by the perturbative formula, Eq.~\eqref{Zuv}, if one uses a  smaller value  $\alpha_s=0.24$.

The reduced value of $\alpha_s$ extracted from a fit to the data at smaller lattice spacing suggests 
that the characteristic scale of the coupling, $\alpha_s(\mu)$, here should be something like $\alpha_s (1/a)$. Calculations at different values of the lattice spacing enable one to confirm this hypothesis. In particular, 
the lattice calculations in Ref.~\cite{Karpie:2021pap} have been performed using three lattice spacings, of approximately 0.075~fm, 0.065~fm, and 0.048~fm.
Our  main goal in  the rest of this paper is to  analyze the rest-frame matrix elements  obtained in Ref.~\cite{Karpie:2021pap}.
We first  fit the rest-frame matrix elements  using the perturbative  expression, Eq.~\eqref{Zuv}
{ (with $ \Gamma (\zeta) $  substituted  by $\widetilde  \Gamma (\zeta )$ of Eq.~\eqref{flamod}),  
}
and by a lattice perturbation theory result. We show that better fits are  obtained if we add  
a Gaussian factor,  that is, if we  fit the rest-frame  matrix elements by
\begin{equation}
M(z_3)  = Z_{\rm pert} (z_3/a )   \exp\left[ -{z_3^2 \Lambda^2}/{4}\right] \, . 
\label{gauss}
\end{equation}
In this formula $\Lambda$  is   a  parameter  proportional to the average transverse momentum or the inverse of the hadronic radius.  This physical interpretation suggest that this scale  also sets the scale for    power corrections in moving frames.

\section{Numerical Lattice QCD Results\label{sec:lattresults}}

The matrix element 
$\langle p | {\cal O}^\mu(z_3) |p \rangle $
for (Euclidean) time component, $\mu =t$,   has been calculated in  Ref.~\cite{Karpie:2021pap} on a series of ensembles with $N_f=2$ degenerate flavors of clover-improved Wilson fermions, with $m_\pi\sim 445$ MeV and varying in lattice spacing~\cite{Karpie:2021pap}. Tab.~\ref{tab:latt_deets} highlights the pertinent details of the ensembles. 
\begin{table}[ht!]
\centering
\def\arraystretch{2.0}
\begin{tabular}{ p{35pt} c c  c c c }
\hline\hline
Ens.~ID~ & ~$a$(fm)~ & ~$M_\pi$(MeV) & $L^3 \times T$ & $N_{\rm cfg}$\\\hline
A4p5 & 0.0749(8) & 446(1) &  $32^3 \times 64$ & 1904 \\
E5 & 0.0652(6) & 440(5) & $32^3 \times 64$ & 999 \\
N5 & 0.0483(4) & 443(4) & $48^3 \times 96$ & 477 \\\hline\hline
\end{tabular}
    \caption{Ensemble of configurations generated by CLS~\cite{Fritzsch:2012wq} and HadStruc~\cite{Karpie:2021pap}. The rest frame matrix element was calculated for the nucleon on eight randomly chosen sources per configuration.}
    \label{tab:latt_deets}
\end{table}

In lattice QCD, the gauge potential is replaced with gauge links as the dynamical variable. In perturbation theory with a lattice regulator, the individual gauge links can be approximated by the exponential of a gauge field at the center of the link. When comparing Polyakov-regulated expressions and 
Lattice-regulated expressions, a simple heuristic argument suggests this distinction may be quite important for data with small $z/a$. Consider the case of $z=a$. The Polyakov-regulated expression still allows the gluon to interact with the Wilson line at any point in the interval $[0,z]$, while the lattice-regulated expression is restricted to an interaction at $z/2$.  In a sense, a Wilson line composed of a single gauge link is less representative of the true continuum Wilson line than one of the same physical length and 100 individual gauge links,  all of which are changing and interacting dynamically. In  Ref.~\cite{Karpie:2021pap}, the finite $a/z$ effects of the matrix elements in a moving frame were studied on the same ensembles and it was found that the results extracted from separations $z=a$ were largely incompatible, despite different models used to control this discretization effect. In the following analysis, we determine results with $z_{\rm min}=a$, $2a$, and $3a$ to study this systematic error.

\begin{figure}
    \includegraphics[width=2.9in]{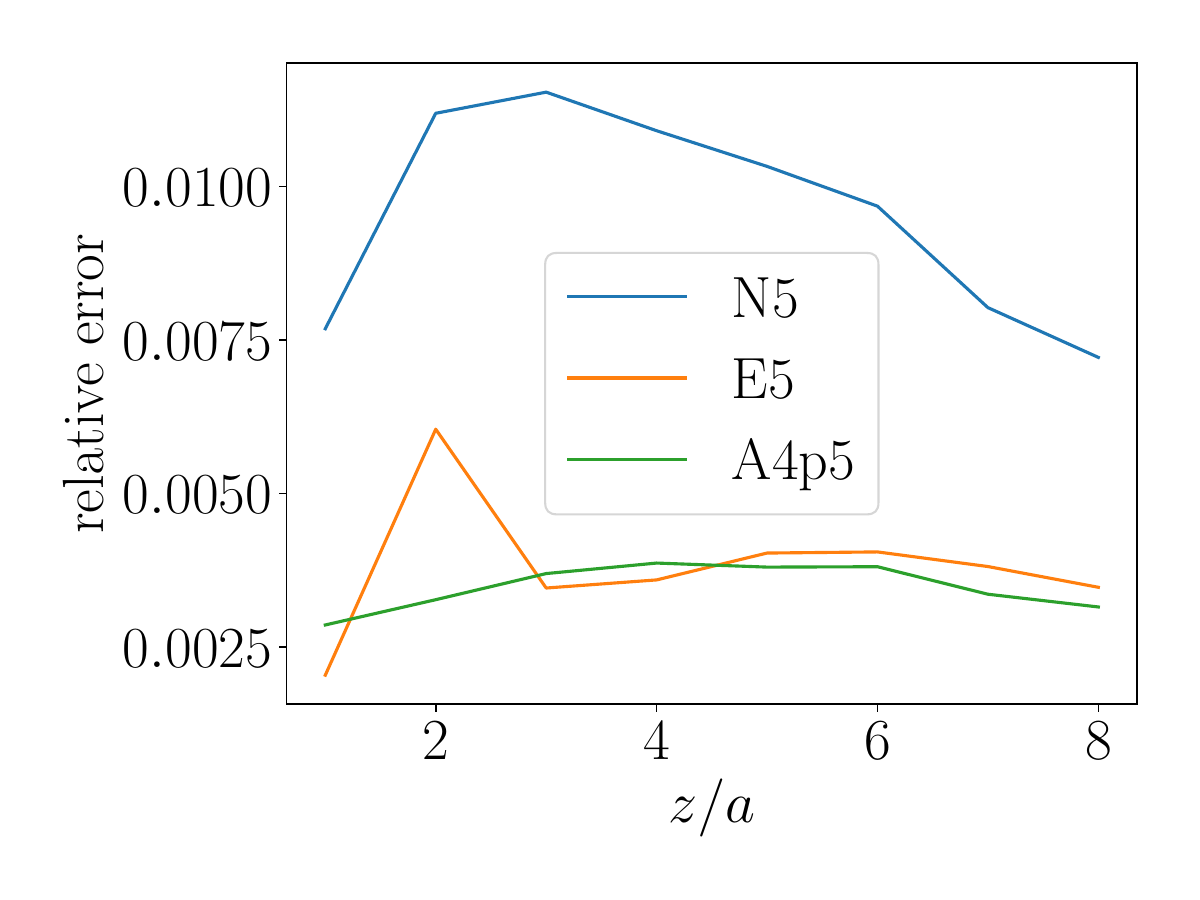}
    \caption{The relative error of the rest frame data on the three ensembles.}\label{fig:stat_error}
\end{figure}

The data having percent, and lower, precision is due to cancellation of correlated fluctuations in the ratio $M(z)/M(0)$. Fig.~\ref{fig:stat_error} shows the relative error of the data used. With such precision, it is quite possible that this analysis will suffer from inaccurate theory. Given the scales used in these lattice calculations, $\alpha^2\sim O(0.01)$, which could generate statistically meaningful deviations. This may generate poor $\chi^2$ when such terms are neglected, but without more accurate perturbation theory 
{ having large $\chi^2$  may  not necessarily reject the  hypothesis that  perturbation theory 
provides a  good representation of   the data in the region  of  our study, which extends to $z=0.6$ fm. 
}

 In what follows we perform the analyses neglecting correlations in the data. While this is typically unjustifiable, since $\chi^2$ is expected to be a poor metric in the first place, we do not wish to confuse matters by inclusion of the covariance. Fits of a model which poorly describes highly correlated data can have significant deviations for expected results. Until more accurate models are used, neglecting data correlations may be necessary to draw conclusions.

We note that when comparing perturbative expressions to data with a finite regulator, it is  unreasonable to utilize the expansions in Eqs.~\eqref{eq:arctan_exp} and~\eqref{eq:log_exp} when the statistical precision of the data is comparable to the subdominant terms. In this study, the largest separation is $\zeta=8$ which can leave percent level corrections to the linear and logarithmic divergences, and the shorter distances used can be much more.

\section{Fits to perturbative expression  with Polyakov Regulator}\label{sec:polyakov_fits}

This section contains analysis of the matrix element  $M^\mu (z,a) =\langle p_0 | {\cal O}(z) |p_0 \rangle$, with $\mu$ 
corresponding to (Euclidean) time component, $\mu =t$, on  the three ensembles.

{ As discussed in Sect III, the early  findings indicate  that the normalized rest-frame  matrix element 
$ R^t(z,a) \equiv { M^t(z_3,a)}/{M^t(0,a)}$ is well described by  the exponential of the perturbative function $ \Gamma (z_3 \pi/a)$. 
 In the present paper, we  will  analyze  fitting the data for   the logarithm  
\begin{equation}
    l (z_3,a) \equiv -\log    R^t(z_3,a)
    \label{lza}
\end{equation} 
 by the one-loop  perturbative   expression $\widetilde  \Gamma (z_3 \pi/a)$ given by Eqs. (\ref{fla}), (\ref{flamod}). 
 }
 
\begin{figure}[ht!]
    \centering
    \includegraphics[width=0.30\textwidth]{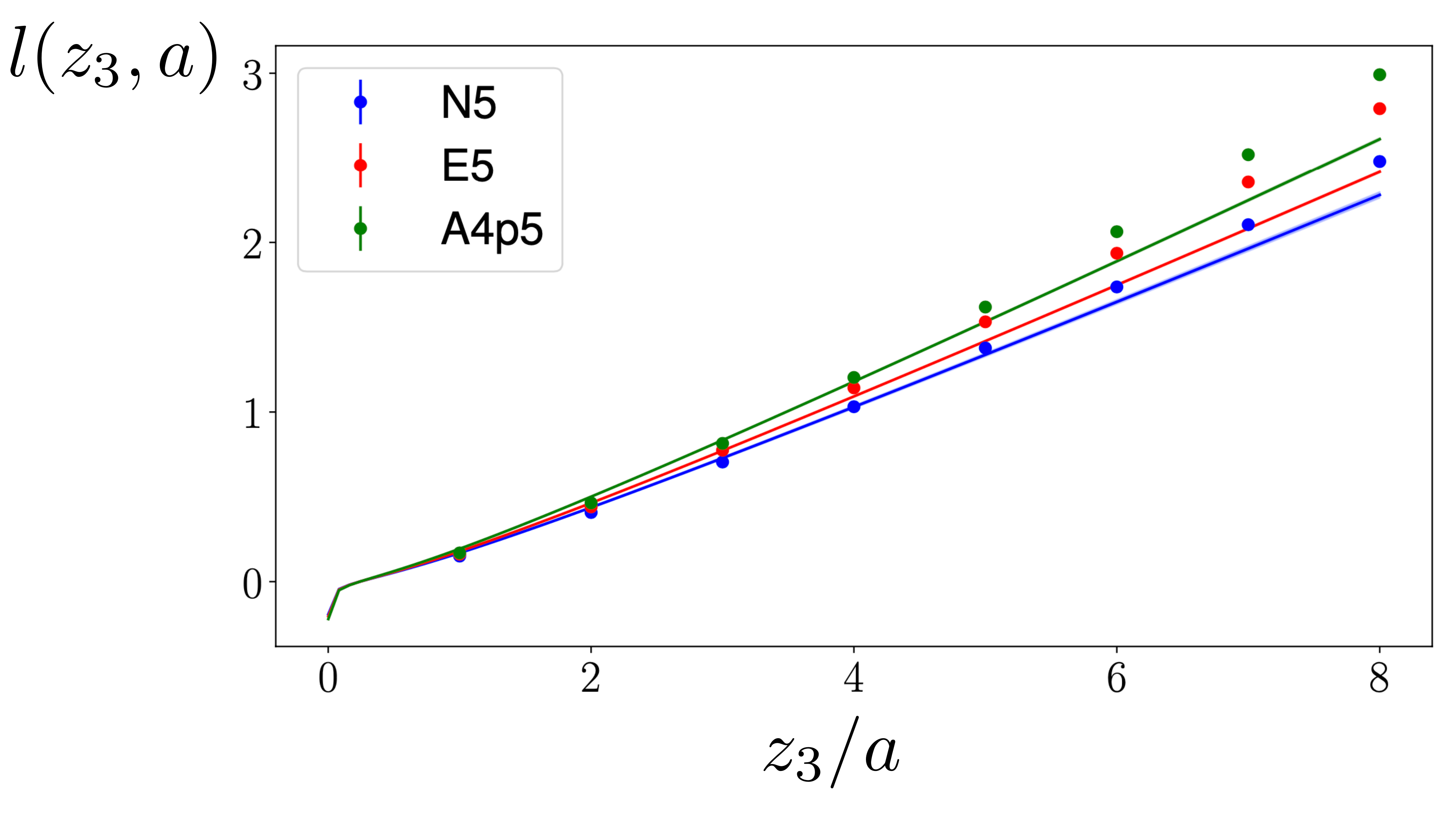}
   \includegraphics[width=0.30\textwidth]{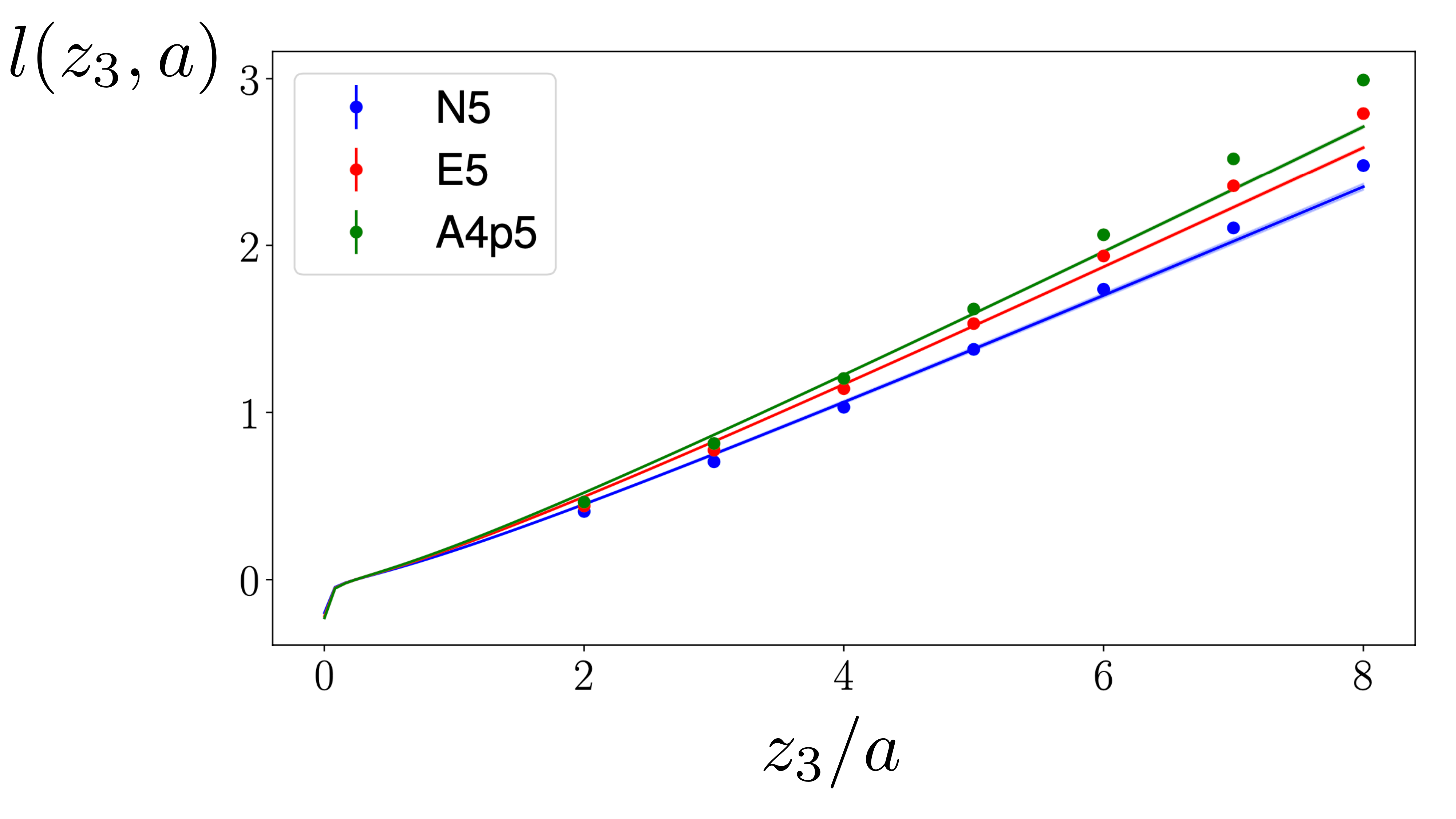}
    \includegraphics[width=0.30\textwidth]{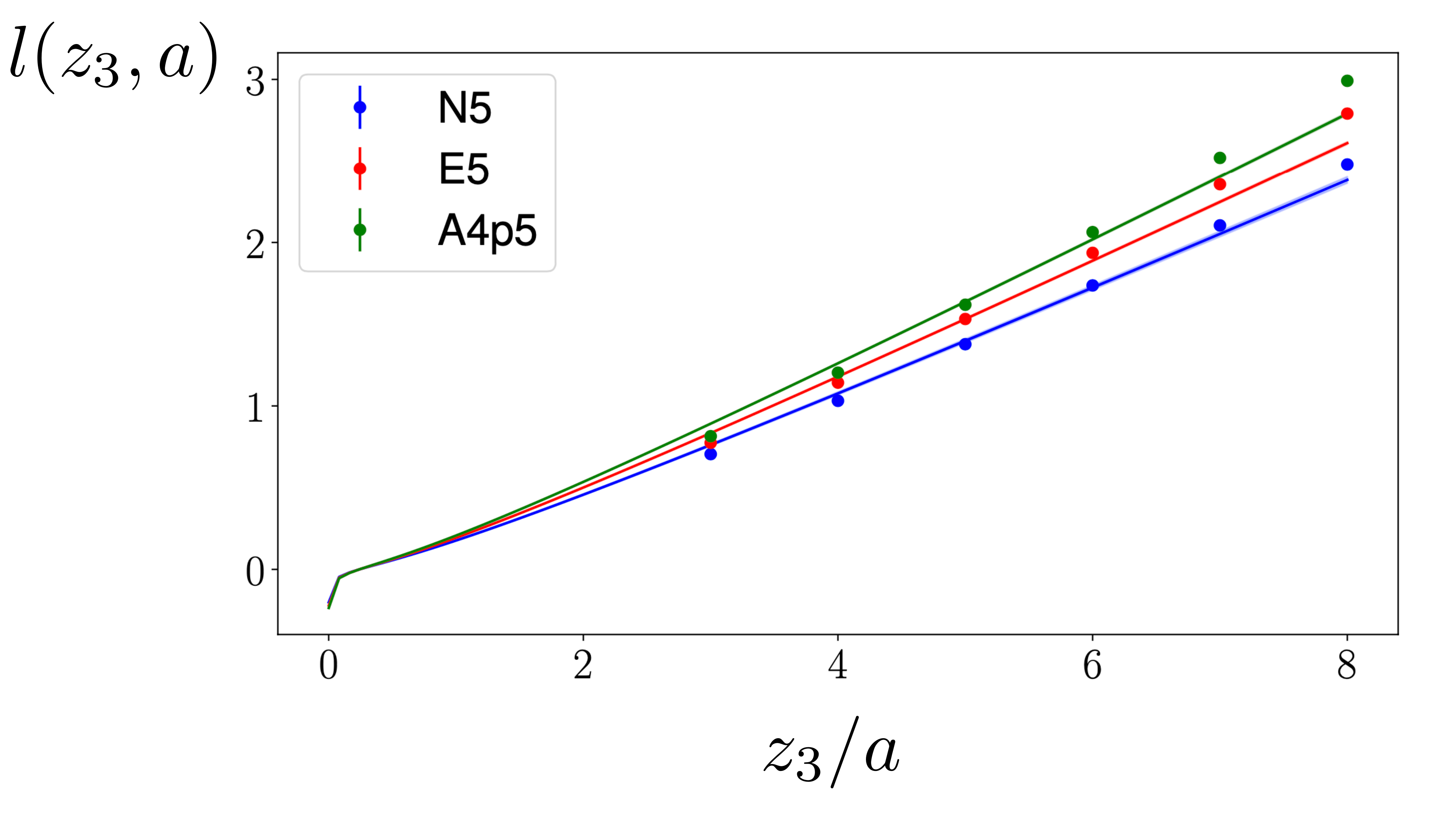}
    \caption{Results of fits on all three ensembles by  the Polyakov regulated perturbative  expression (\ref{lza})   with $z_{\rm min}=a$ (left), $z_{\rm min}=2a$ (center), and  $z_{\rm min}=3a$ (right)}
    \label{fig:pt_fit_noht}
\end{figure}

\begin{figure}
    \centering
    \includegraphics[width=0.30\textwidth]{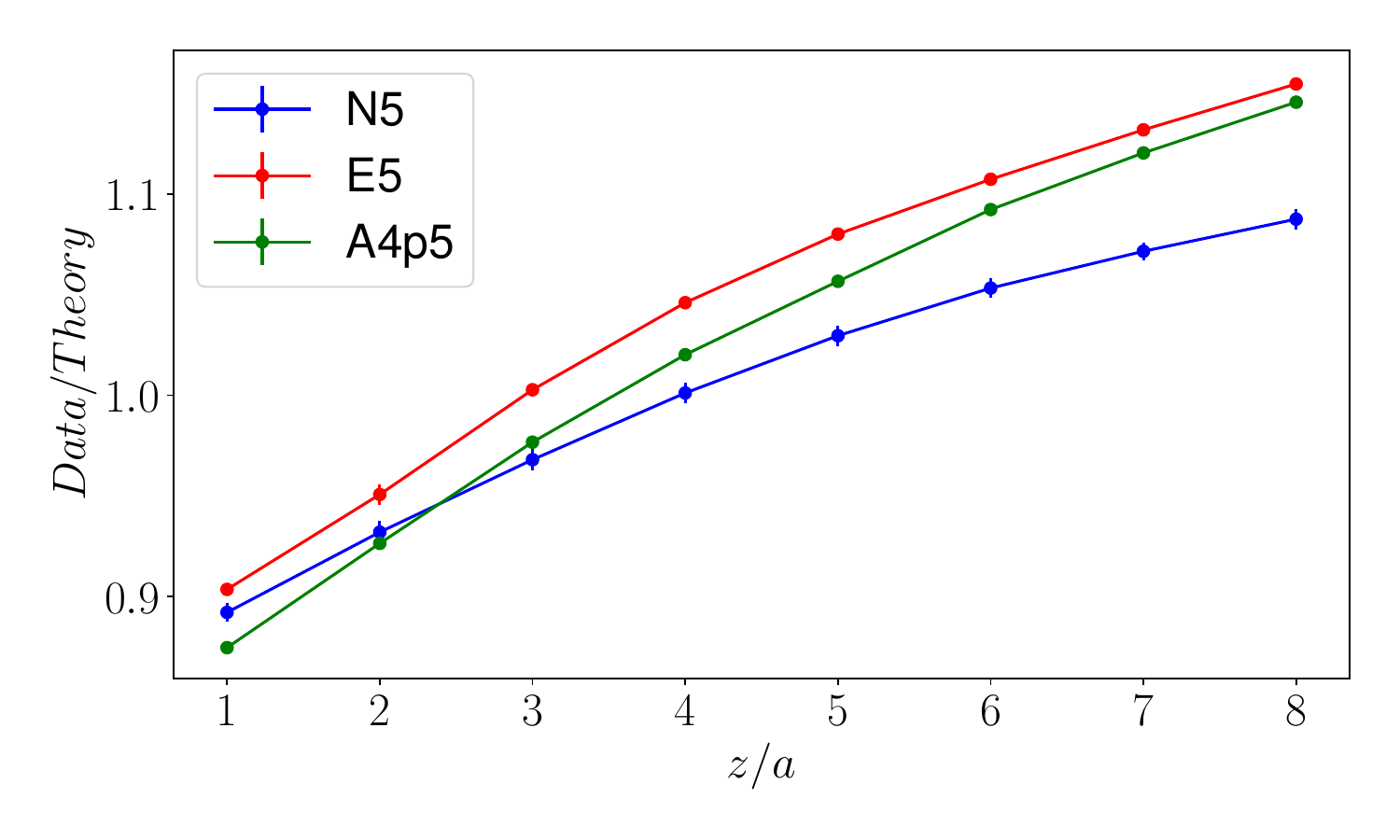}
    \includegraphics[width=0.30\textwidth]{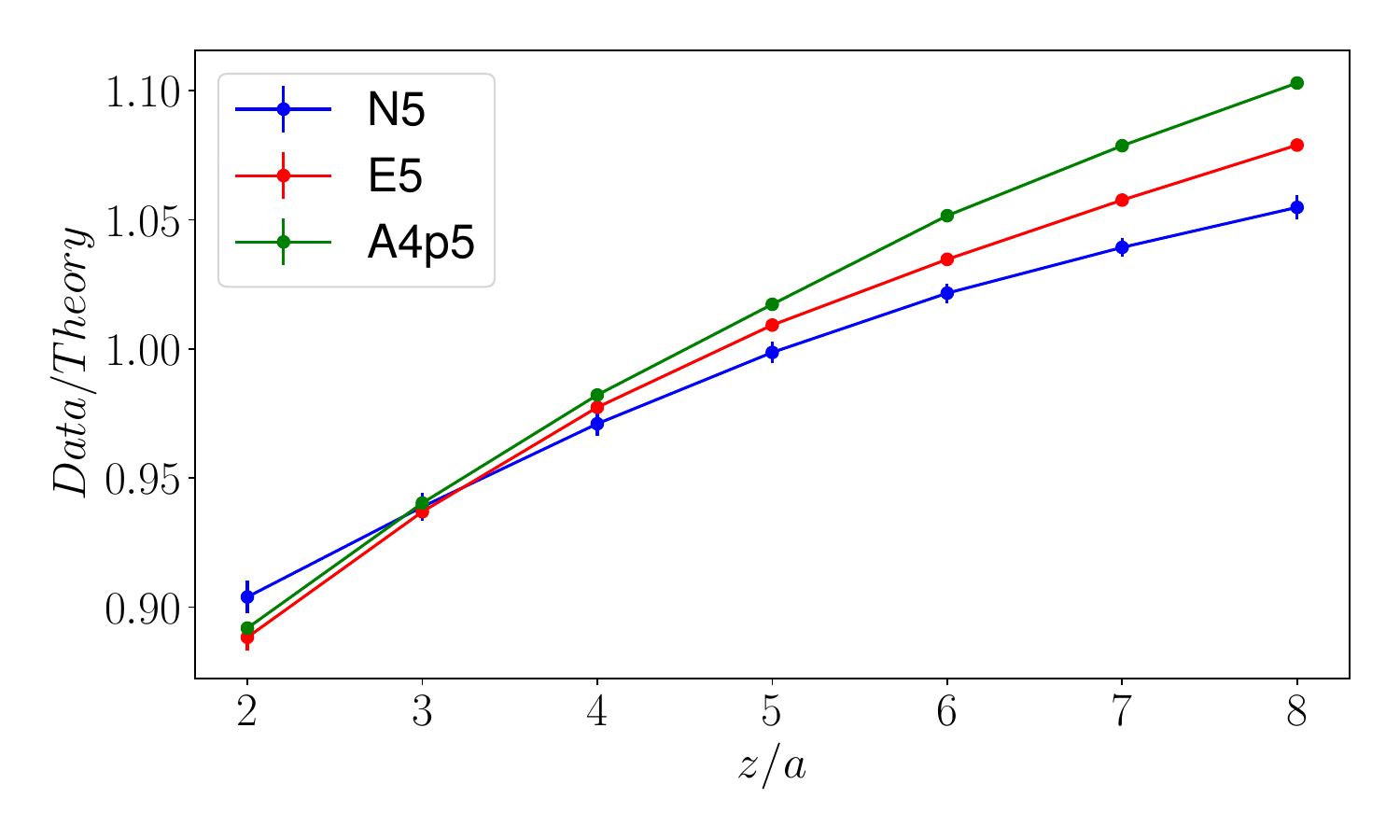}
    \includegraphics[width=0.30\textwidth]{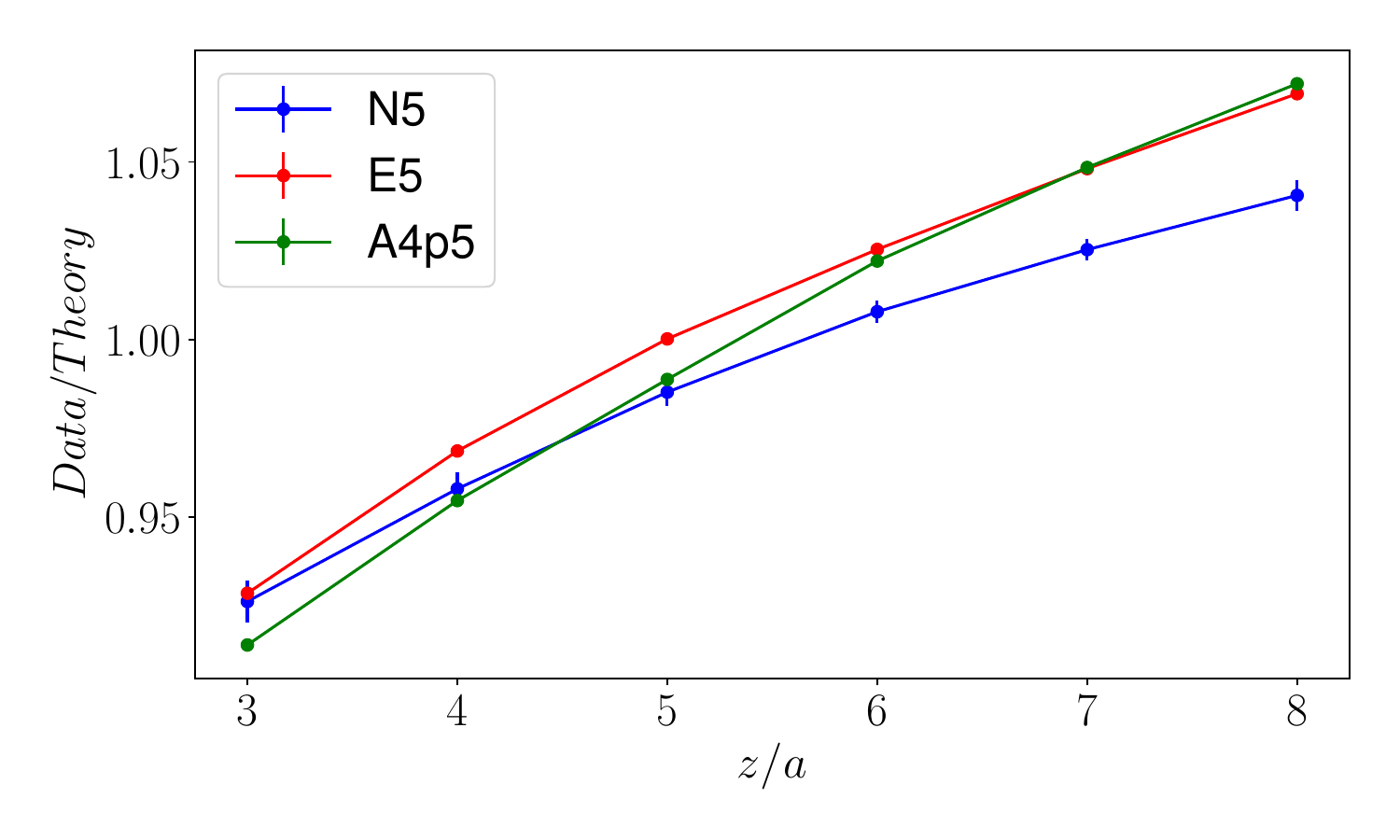}
    \caption{The ratio of the data to the  Polyakov regulated perturbative  model  (\ref{lza}) with $z_{\rm min}=a$ (left), $z_{\rm min}=2a$ (center), and  $z_{\rm min}=3a$ (right). }
    \label{fig:dot_noht}
\end{figure}

The results of the fit are shown in Fig.~\ref{fig:pt_fit_noht}. All the fit results are summarized in  Tab.~\ref{tab:fit_res_noht}. Clearly the $\chi^2/$dof are quite large for each ensemble, but this is due to the high precision of the data compared to the accuracy of the theoretical model. As can be seen in the comparison of data and theory presented in Fig.~\ref{fig:dot_noht}, there appears to be an effect of  just a few percent
{ in $ l (z,a) $ 
}
that is not captured by this model. 

\begin{figure}[ht!]
    \centering
    \includegraphics[width=0.30\textwidth]{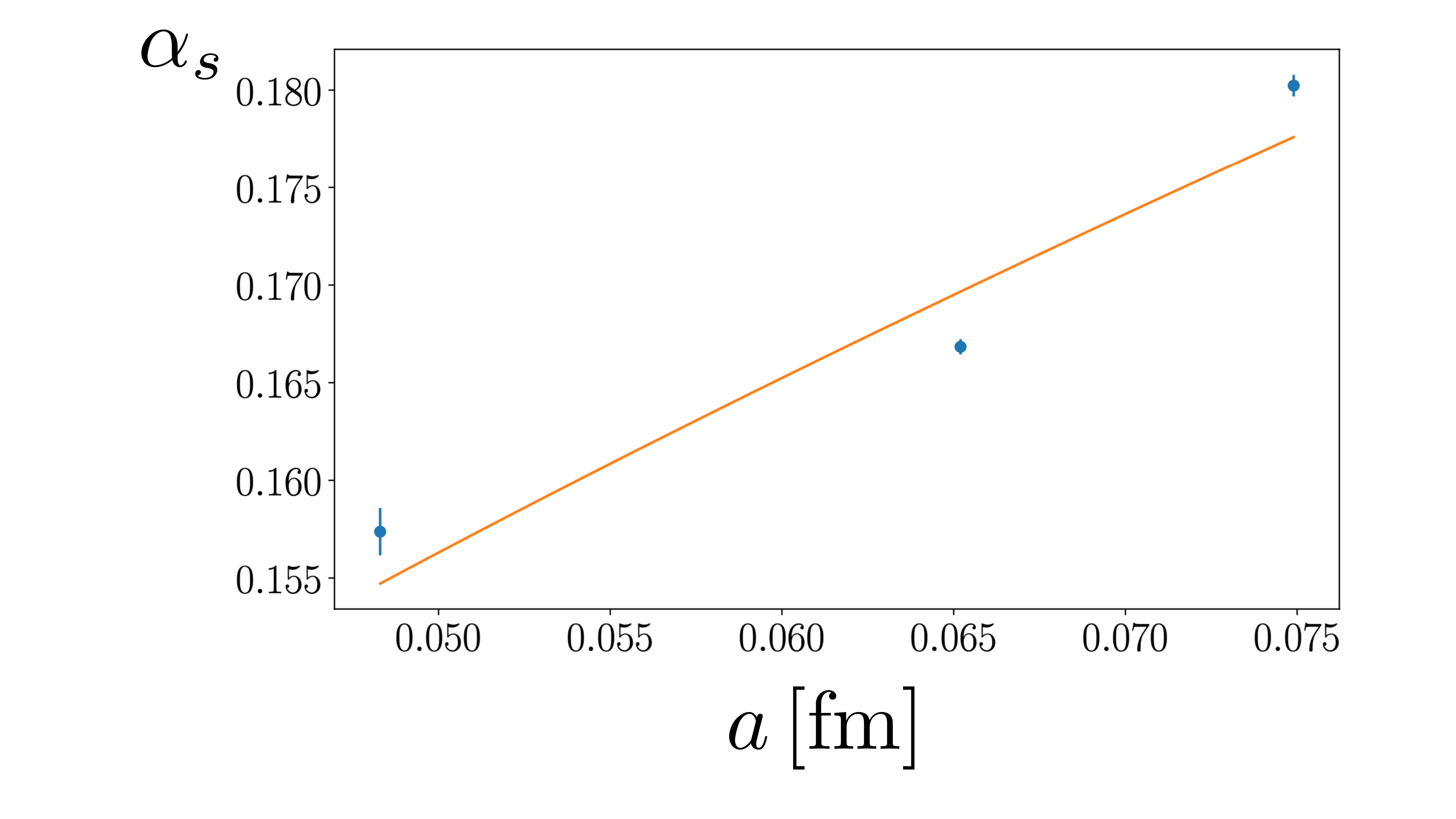}
    \includegraphics[width=0.30\textwidth]{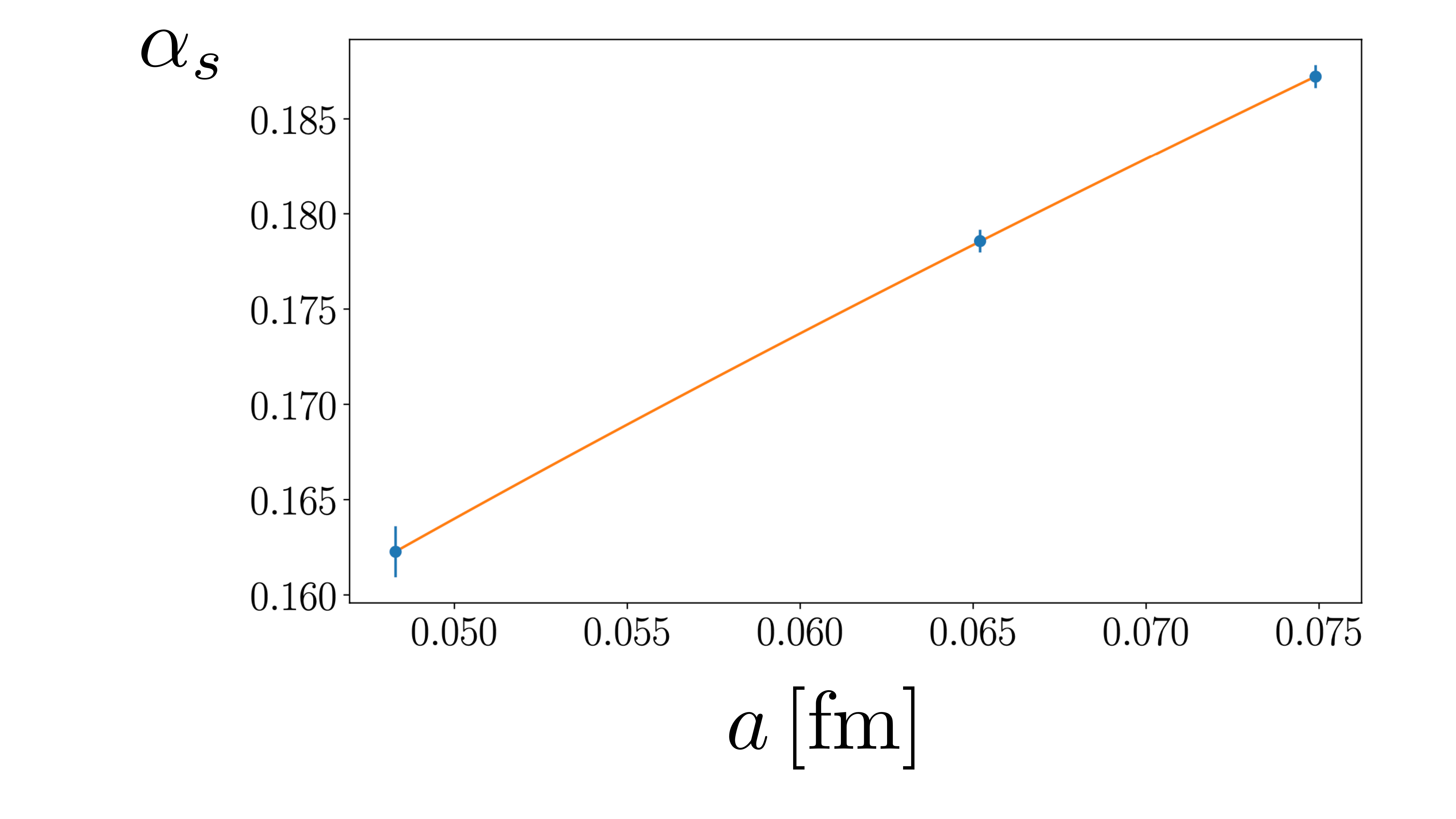}
    \includegraphics[width=0.30\textwidth]{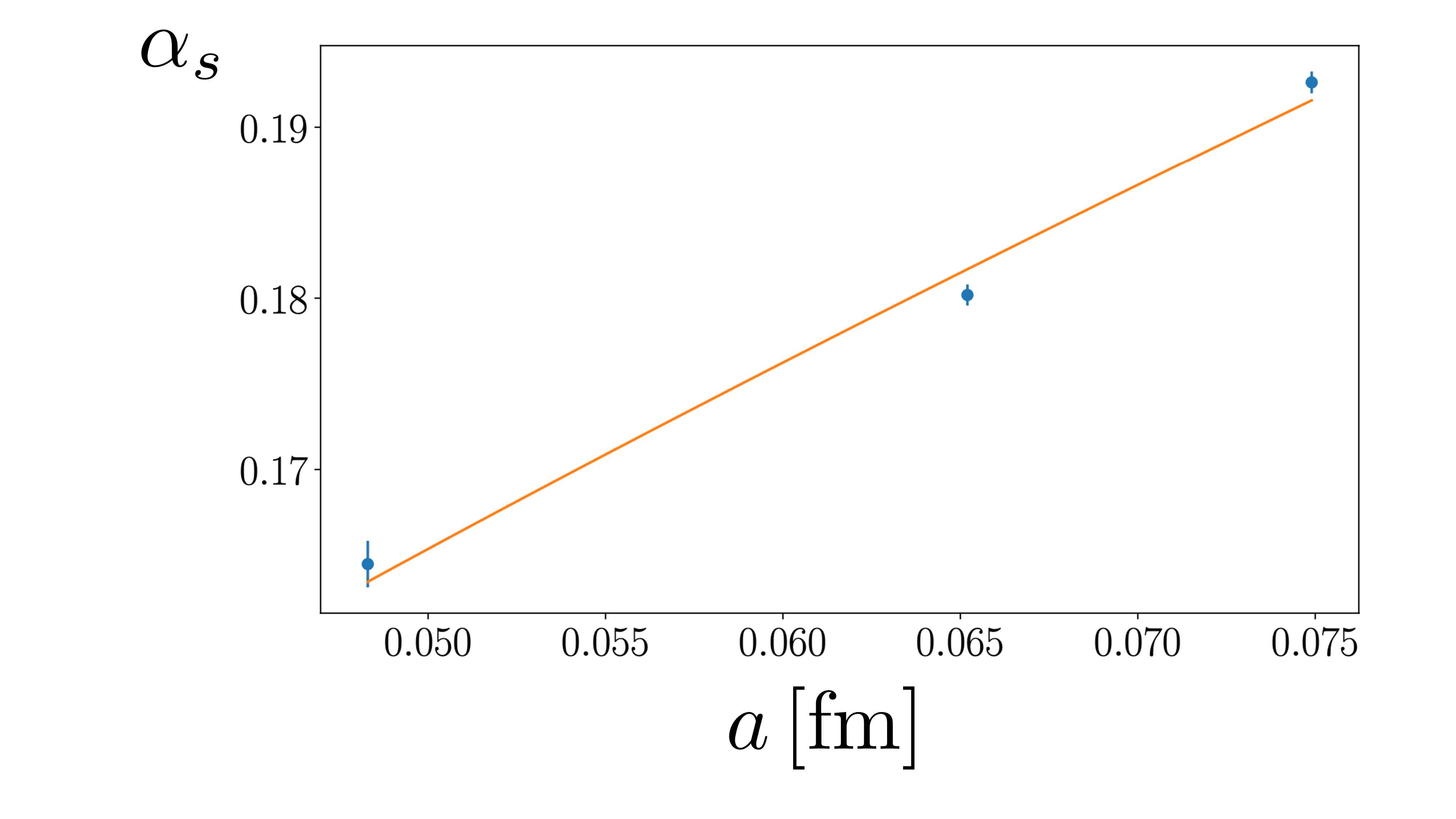}
    \caption{The values of $\alpha_s$ from  Polyakov regulator fits for $l(z_3,a)$  as a function of lattice spacing
    with $z_{\rm min}=a$ (left), $z_{\rm min}=2a$ (center), and  $z_{\rm min}=3a$ (right). The curve  represents a fit to the LO perturbative formula (\ref{eq:running_alpha}), 
     where $\Lambda_{\rm QCD}$ was the single fit parameter.  }
    \label{fig:alpha_fit_noht}
\end{figure}

The fit  results for $\alpha_s$ are shown in Fig.~\ref{fig:alpha_fit_noht}. 
One can  see that  $\alpha_s$ trends smaller  for smaller $a$. 
So, we assume  that $\alpha_s $ is  taken at the  coordinate scale 
given by lattice spacing $a$ or PR parameter $a_{\rm PR}$.
Since these scales differ by a factor of $\pi$, the relevant values of ``$\Lambda_{\rm QCD}$''  also differ by $\pi$. That is, we can take $a_{\rm PR}=a/\pi$ as the scale in the argument of $\alpha_s$ in continuum perturbative running. Using the leading-order running of the coupling, 
\begin{equation}
    \alpha_s({\pi}/{a}) = -\frac{2\pi}{(11-2 N_f/3)\log(a \Lambda_{\rm QCD}/\pi)}\label{eq:running_alpha}
\end{equation}
one can infer the value of $\Lambda_{\rm QCD}$ necessary to generate the coupling at that scale. 

\begin{figure}[ht!]
    \centering
    \includegraphics[width=0.30\textwidth]{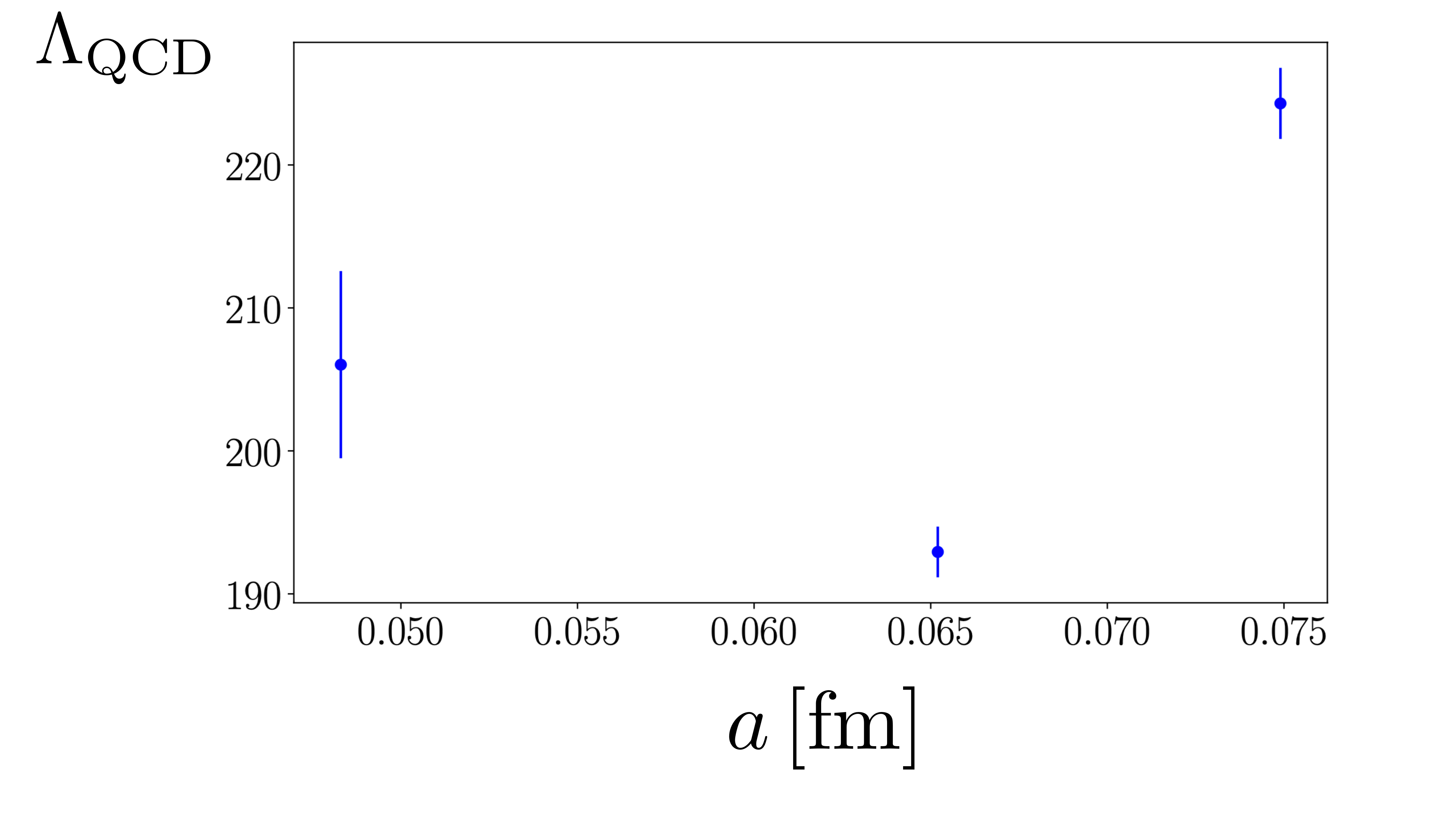}
    \includegraphics[width=0.30\textwidth]{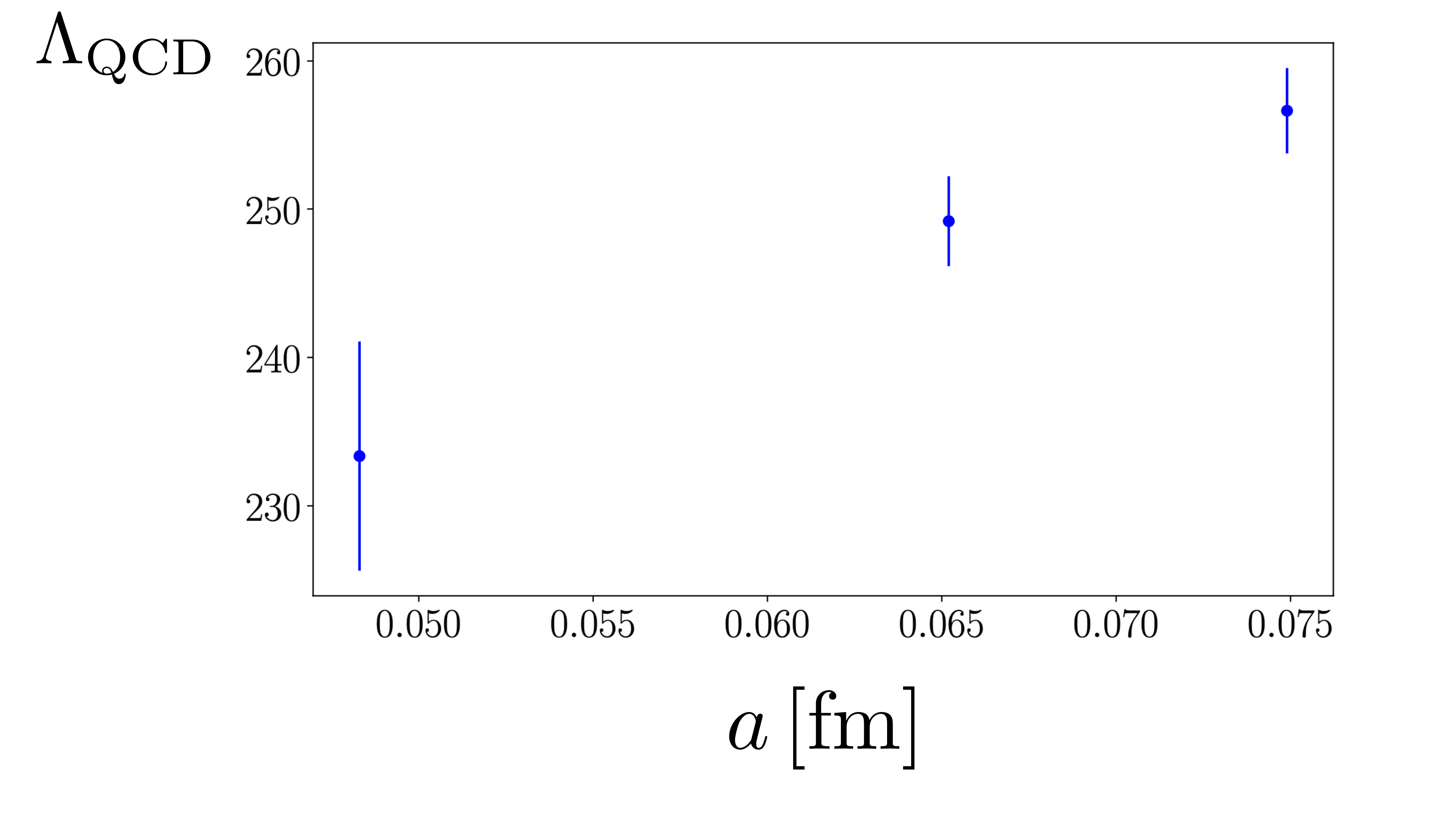}
    \includegraphics[width=0.30\textwidth]{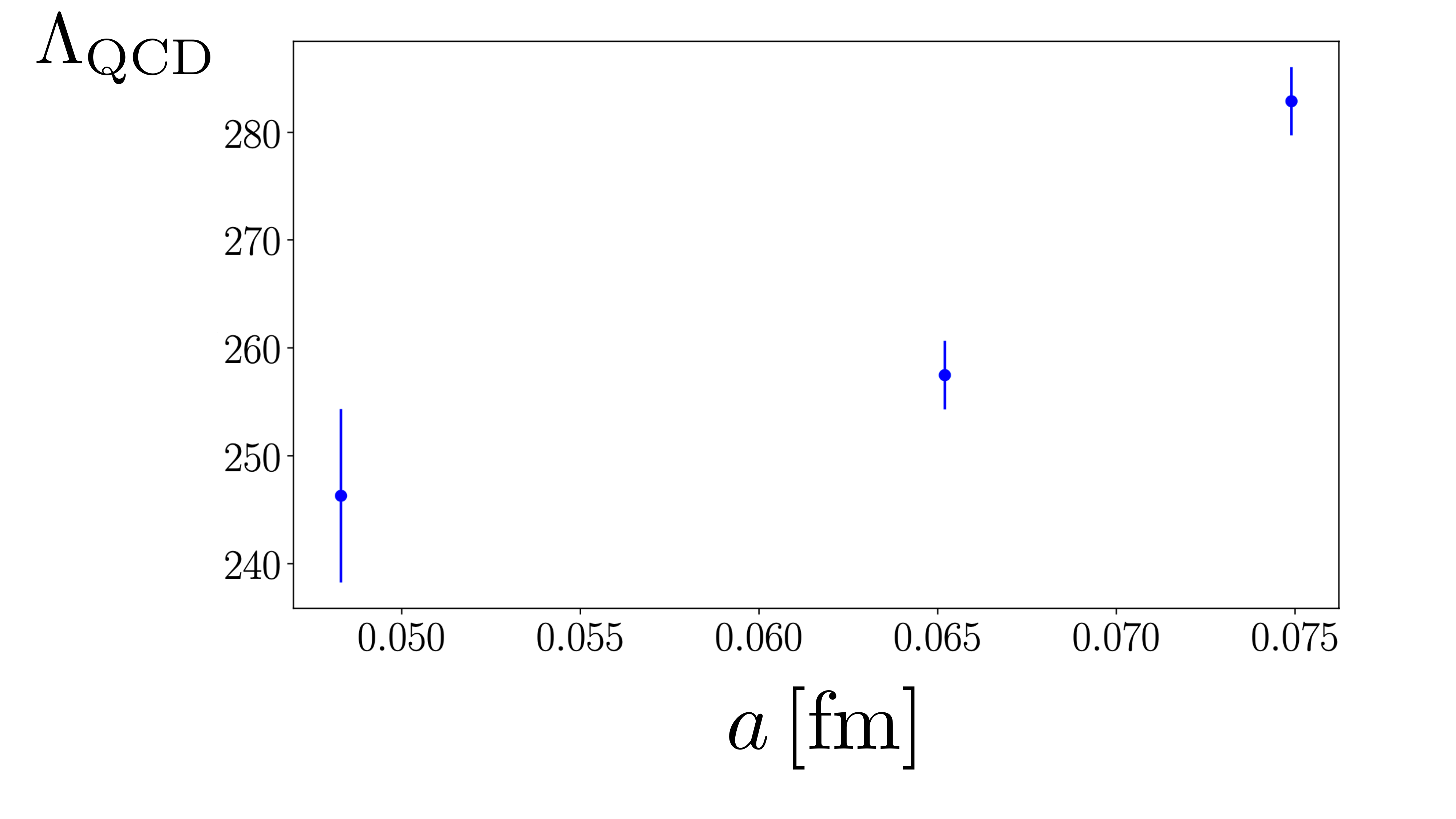}
    \caption{The value of $\Lambda_{\rm QCD}$ extracted from $\alpha_s$ using LO perturbation theory in Fig.~\ref{fig:alpha_fit_noht} with $z_{\rm min}=a$ (left), $z_{\rm min}=2a$ (center), and  $z_{\rm min}=3a$ (right))}
    \label{fig:lam_qcd_noht}
\end{figure}

The results are given in Fig.~\ref{fig:lam_qcd_noht}. One can see that the resulting 
values $\sim 250$~MeV of $\Lambda_{\rm QCD}$ are of the expected  order despite this atypical method of determining $\Lambda_{\rm QCD}$.

There exist a number of possible effects responsible for the discrepancy  between the model and lattice data. As mentioned previously, finite size effects could generated a Gaussian dependence on $z_3^2$
that becomes  more pronounced for large $z^2$ but  was  not taken into account in the numerics.
For small $z_3^2$, 
this discrepancy can be described by $a/z_3$ error contaminating $z_3=a$ data. In principle, a  model containing finite $a/z_3$ effects could be investigated, but  in this study, these effects are exposed by cutting the shortest separations. Another unexplored possibility are effects from a nonzero quark mass in the perturbative formula. The pion mass of these ensembles is nearly three times larger than physical and a finite quark mass would modify the perturbative formula, which was derived with vanishing quark mass. Finally, given the value of $\alpha_s$ from the fits,  an ${\cal O}(1\%)$ discrepancy could also be attributed to higher orders in perturbation theory. 
 With subpercent precision in the data and percent accuracy in the model, the $\chi^2\sim O(10)$ 
       {  values could simply  be    expected.}
  As such, the large $\chi^2$ of this analysis {\it alone}  does not conclusively state that ``perturbation theory cannot describe the matrix element 
          {\it completely}  at $z_3=0.6$ fm'', but only that one loop is insufficient. 
          
        {    In what follows,  we will investigate if  the discrepancy  between the 
        one-loop perturbative model and lattice data  may be explained by additional  Gaussian dependence of the data on $z^2$.}

    \begin{table}[ht!]
    \centering
    \def\arraystretch{2.0}
    \begin{tabular}{p{35pt}cccccc }
    \hline\hline
    & ($z_{\rm min}=a$) & ($z_{\rm min}=2a$)& ($z_{\rm min}=3a$) & ($z_{\rm min}=a$) &($z_{\rm min}=2a$) & ($z_{\rm min}=3a$) \\
      Ens & $\alpha_S$  &  $\alpha_S$ &  $\alpha_S$ & $\chi^2$/dof & $\chi^2$/dof &  $\chi^2$/dof \\\hline
    N5  & 0.1573(12) & 0.1622(13)& 0.1644(14) & 75(7) & 34(5) & 21(4) \\
    E5  & 0.1668(4) & 0.1786(6) & 0.1802(6) & 897(35) & 261(14) & 221(11) \\
    A4p5 & 0.1802(6) & 0.1872(6) & 0.1926(6) & 929(21) & 536(15) & 296(11) \\\hline \hline
    \end{tabular}
    \caption{Results of fits on all three ensembles by the perturbative expression in Eq.~\eqref{lza}.}
    \label{tab:fit_res_noht}
\end{table}

\subsection{Fits including Gaussian model for finite size effects}

\begin{figure}[ht!]
    \centering
    \includegraphics[width=0.31\textwidth]{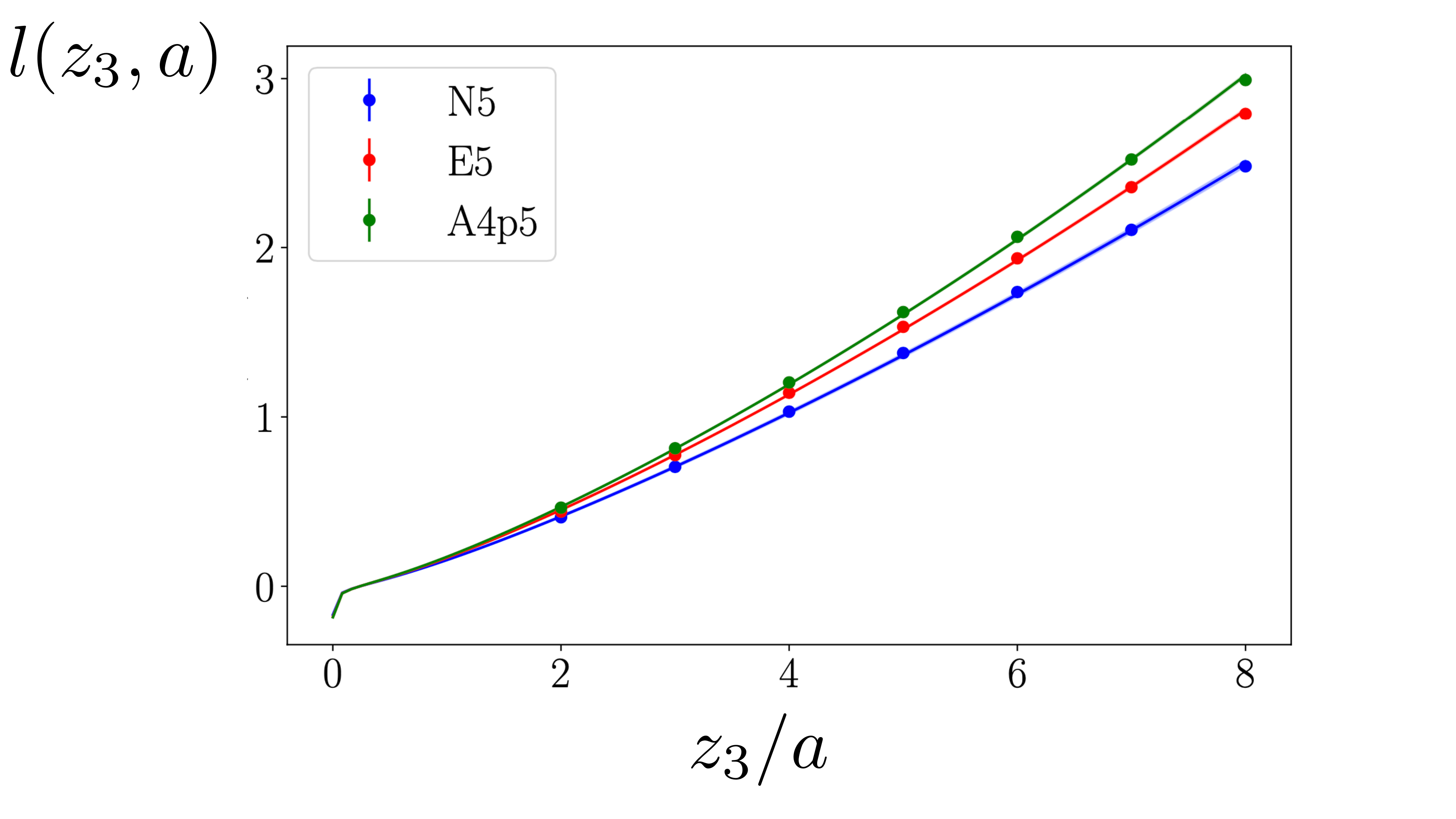}
    \includegraphics[width=0.30\textwidth]{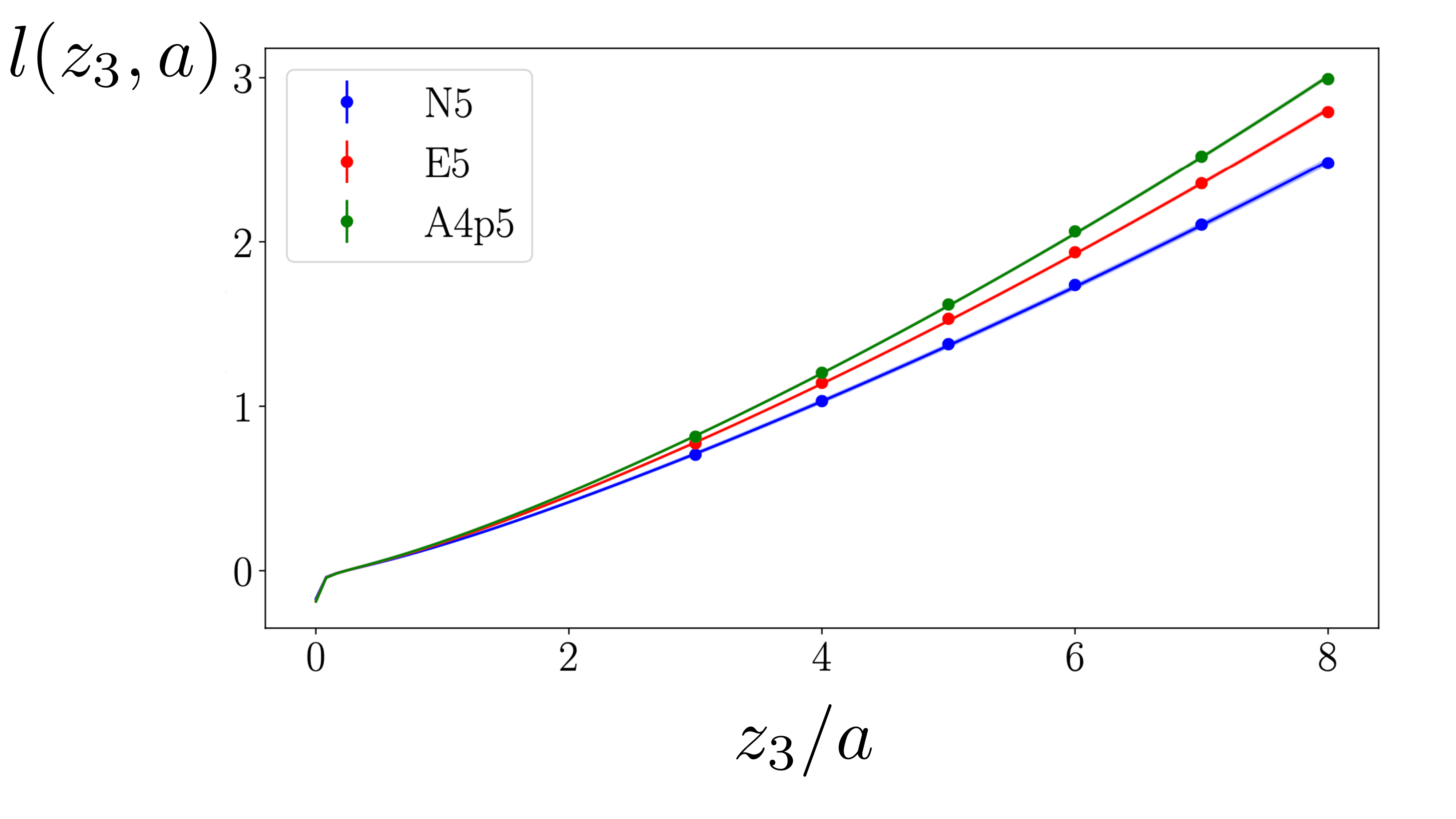}
    \caption{Results of fits for $l(z_3,a)$  on all three ensembles in  a  model containing additional Gaussian factor with $z_{\rm min}=2a$ (left), and  $z_{\rm min}=3a$ (right).}
       \label{fig:pt_ht_fit}
\end{figure}

\begin{figure}[h!]
    \centering
    \includegraphics[width=0.30\textwidth]{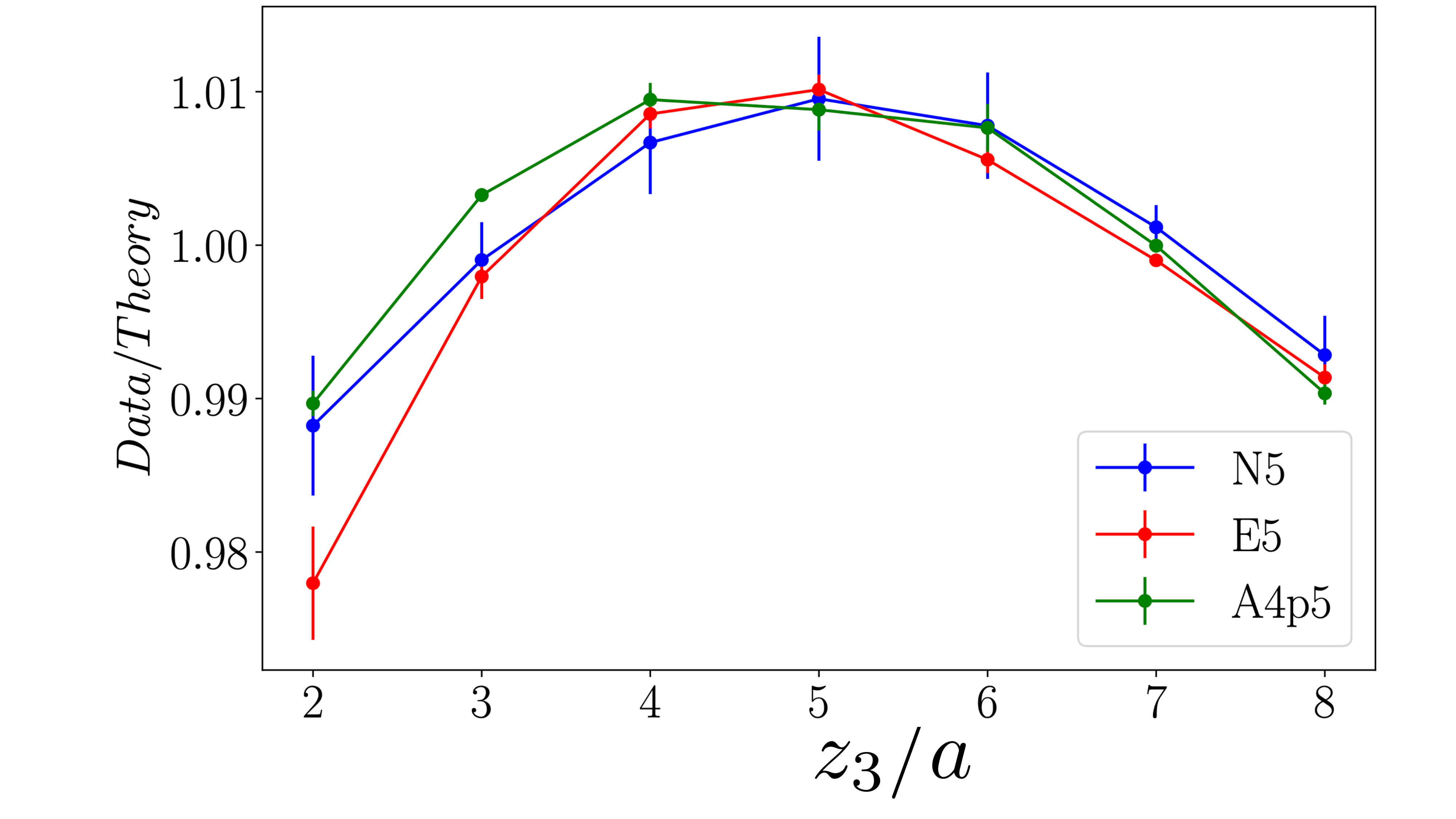}
    \includegraphics[width=0.31\textwidth]{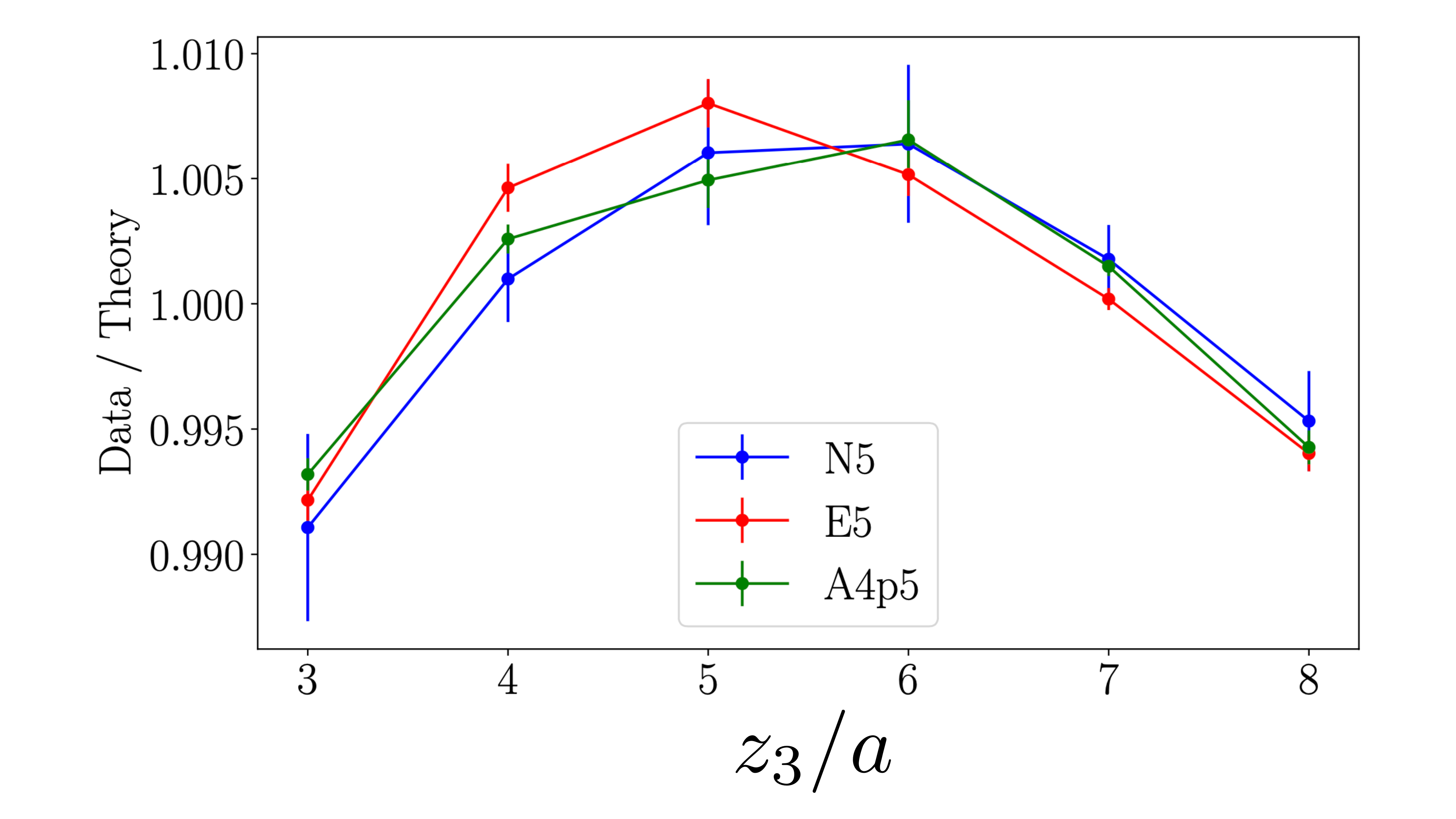}
    \caption{The ratio of the data for $l(z_3,a)$ to the results of  the  model containing additional Gaussian term, with $z_{\rm min}=2a$ (left),  and  $z_{\rm min}=3a$ (right).   }
    \label{fig:dot_ht}
\end{figure}

This section contains analysis of the three ensembles to the functional form
\begin{eqnarray}
  l (z_3,a) =  \widetilde  \Gamma (z_3 \pi/a)  -\frac{z_3^2\Lambda^2}{4}
     \label{lzaL}
\end{eqnarray} 
 where the Gaussian term  with $\Lambda^2>0$ is  assumed to reflect  the finite size of the nucleon 
and produces damping of  the matrix element  for large $z_3$. The results for
$l(z_3,a)$ (see Eq.~\eqref{lza}) are shown in Fig.~\ref{fig:pt_ht_fit}. 
All the fit results are summarized in Tab.~\ref{tab:fit_res_ht}. The fit with these finite size effects with  $z_{\rm min}=2a$ and $3a$ has a significant decrease in $\chi^2$. The comparison of data to the model is shown in Fig.~\ref{fig:dot_ht}. It is interesting to note that after inclusion of the scale dependent correction, the $z_{\rm min}=a$ results did not improve dramatically. It is possible that higher orders in perturbation theory are required to describe the discretization at this shortest interval. 

\begin{figure}[ht!]
    \centering
   \includegraphics[width=0.30\textwidth]{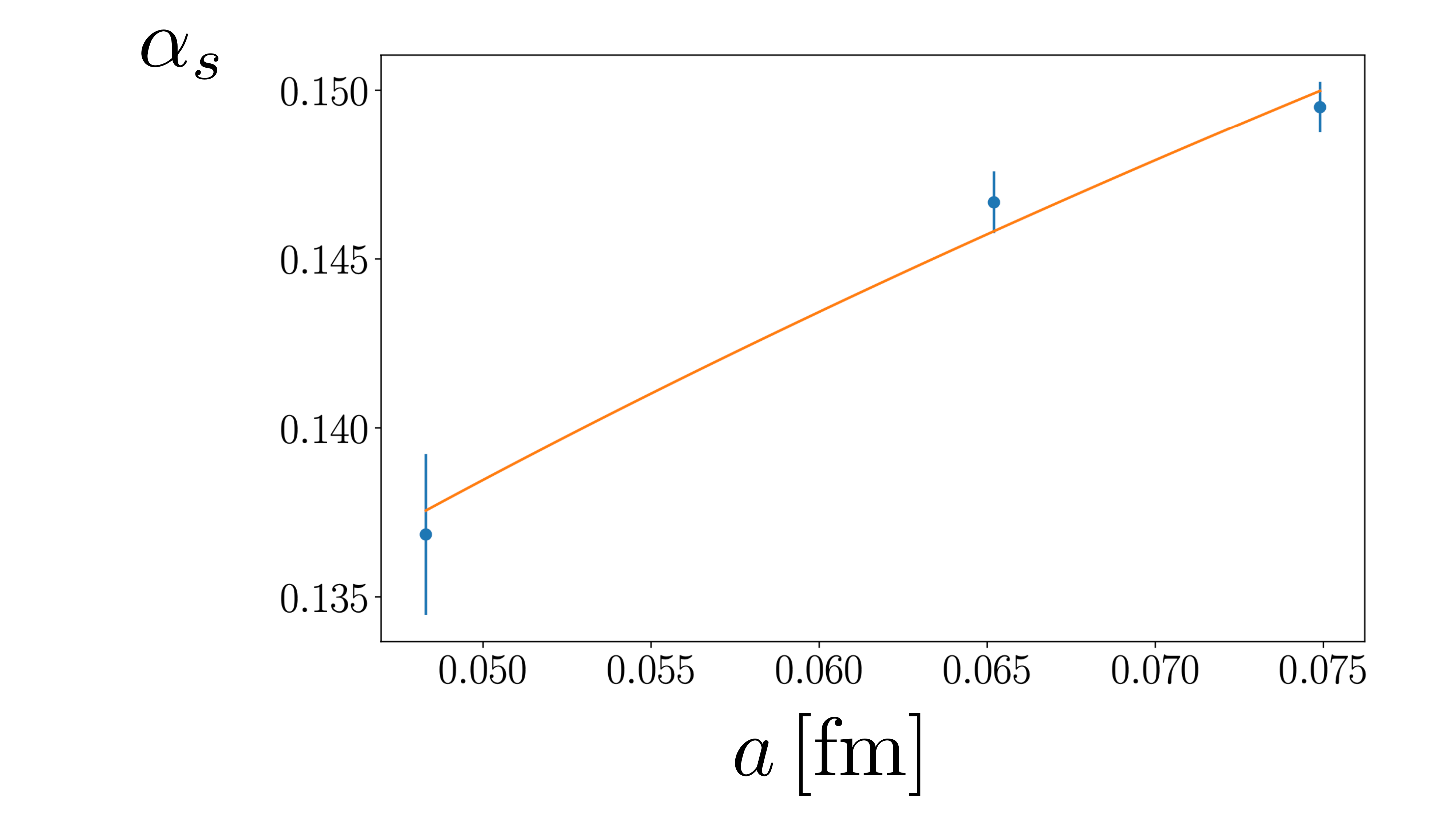}
    \includegraphics[width=0.30\textwidth]{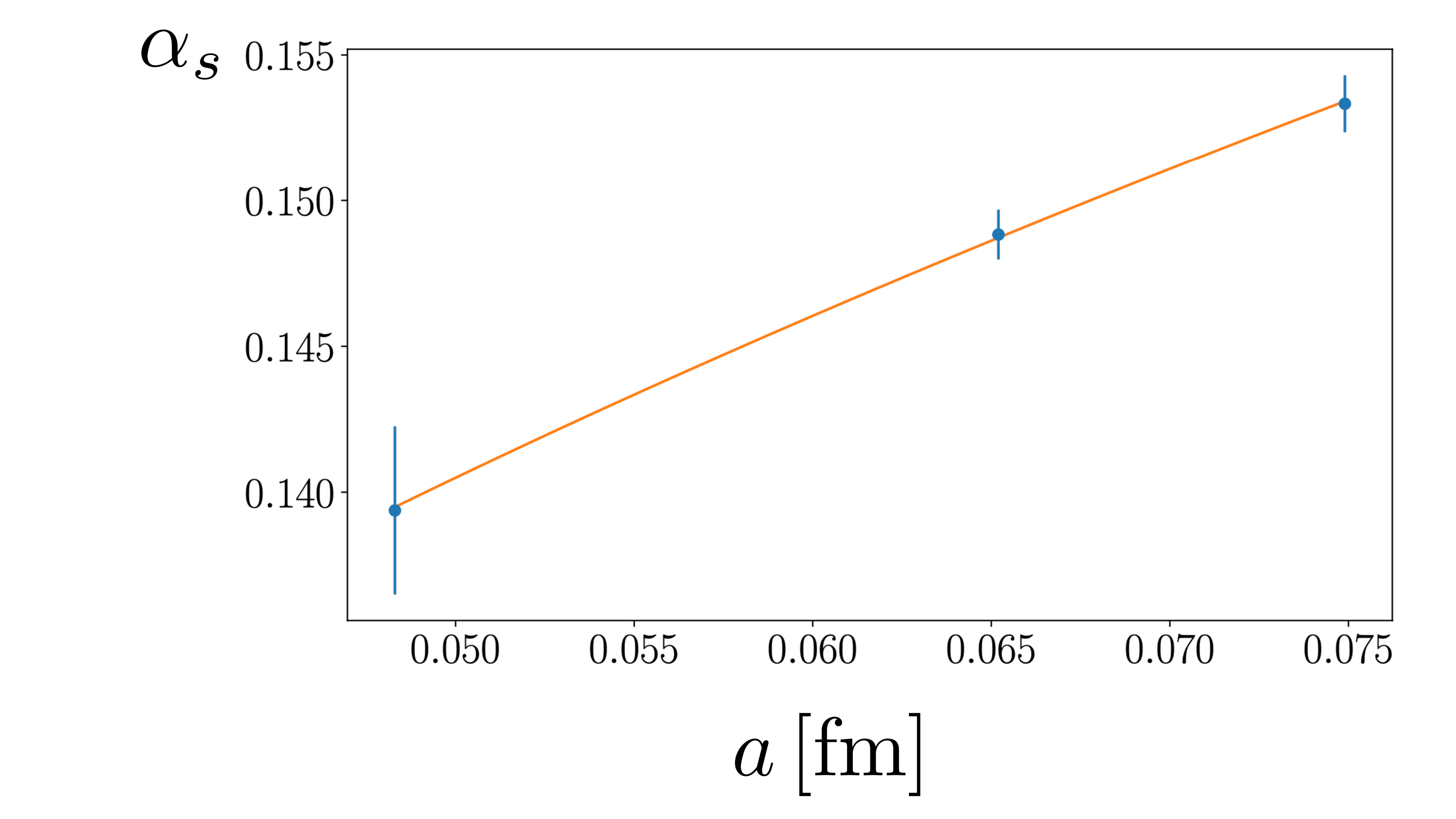}
    \caption{The values of $\alpha_s$  from  fits for $l(z_3,a)$  as a function of lattice spacing   in  a  Polyakov regulated  model containing additional Gaussian term,  with
    $z_{\rm min}=2a$ (left),  and  $z_{\rm min}=3a$ (right). The curve, intended solely to guide the eye, represents a fit to the LO perturbative formula (\ref{eq:running_alpha})  for $\alpha_s$, 
 where $\Lambda_{\rm QCD}$ is the single fit parameter.  }
    \label{fig:alpha_ht_fit}
\end{figure}

\begin{figure}[ht!]
    \centering
    \includegraphics[width=0.30\textwidth]{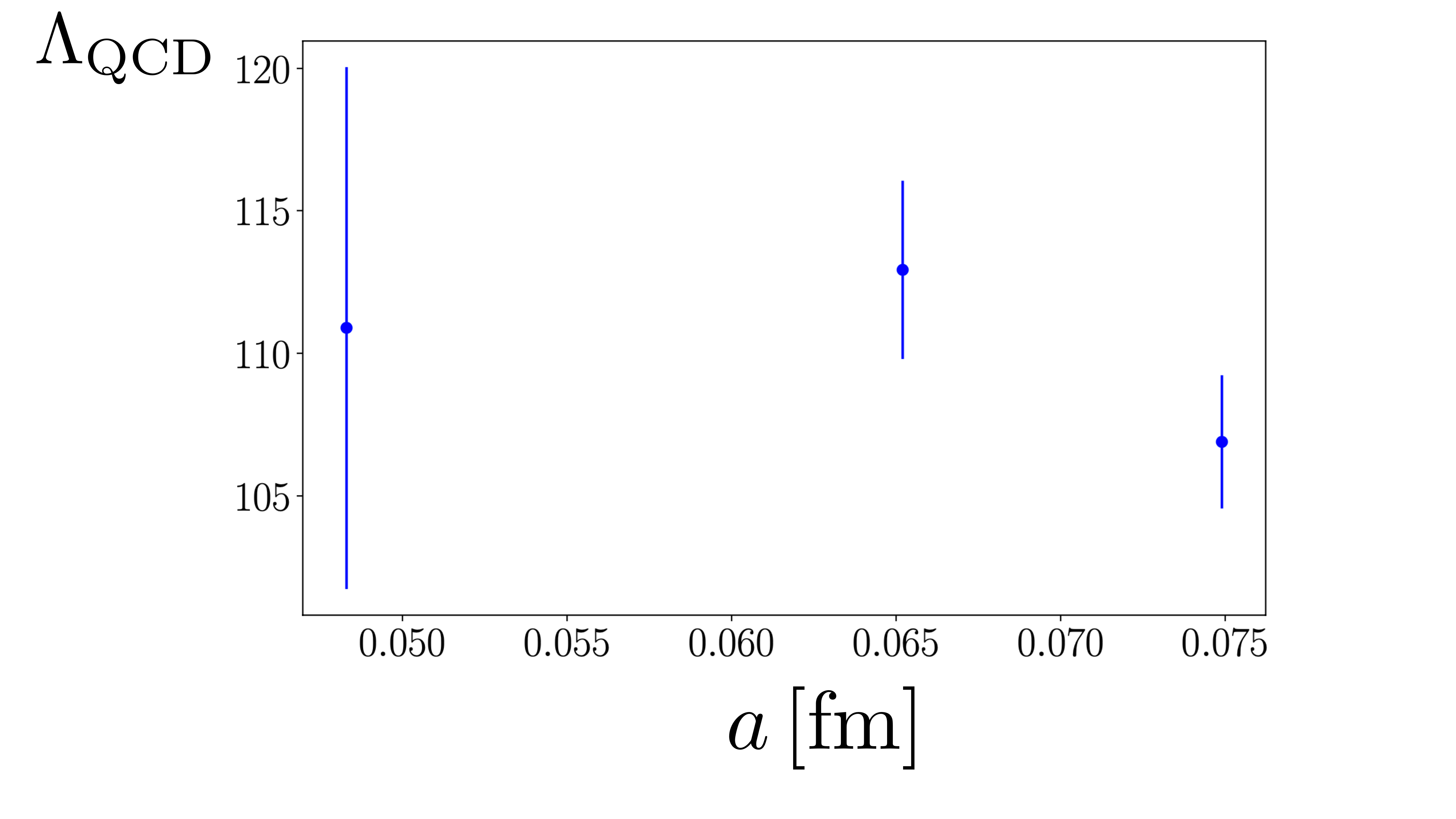}
   \includegraphics[width=0.30\textwidth]{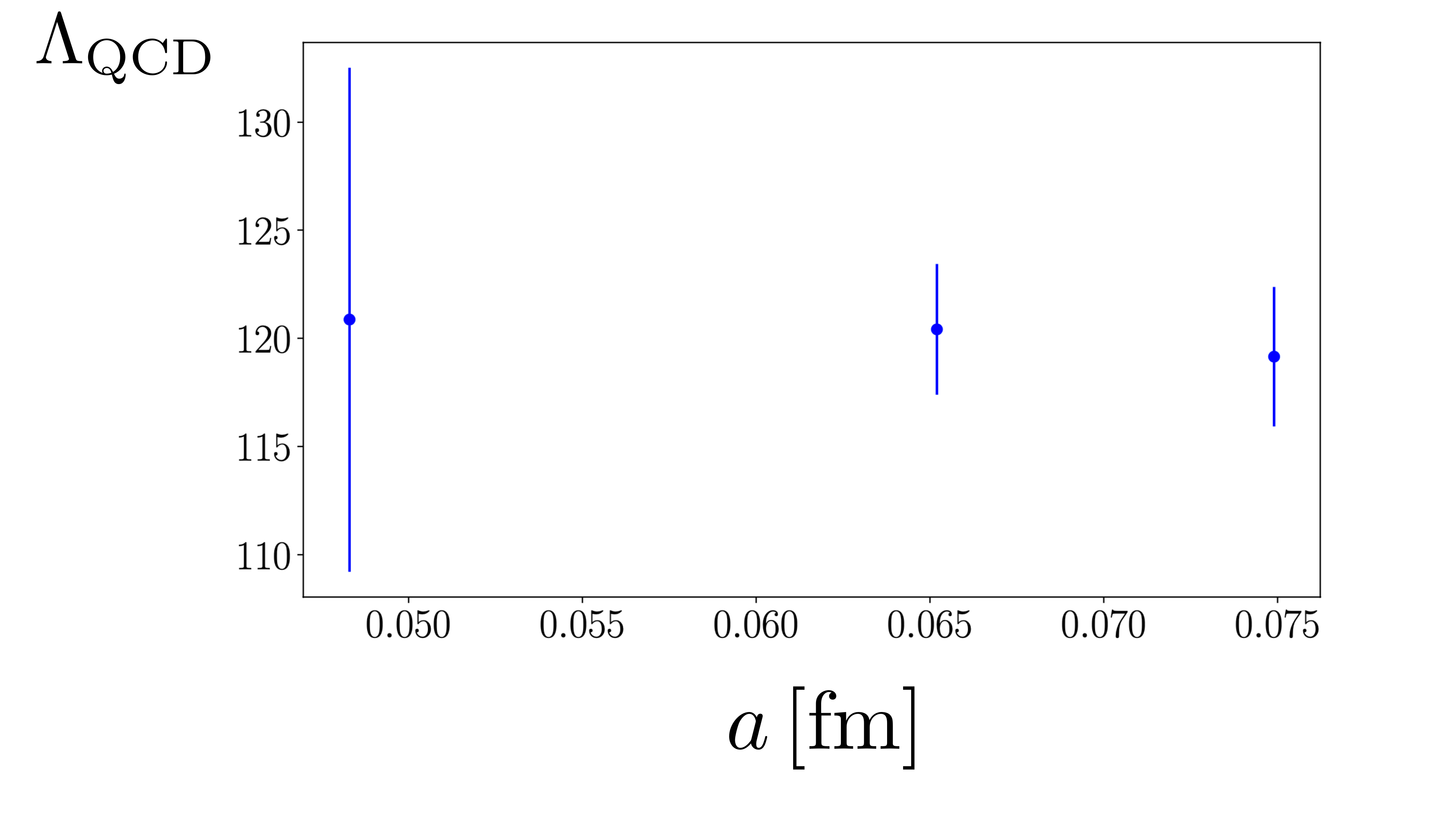}
    \caption{The value of $\Lambda_{\rm QCD}$ extracted from $\alpha_s$ using LO perturbation theory in Fig.~\ref{fig:alpha_ht_fit} with $z_{\rm min}=2a$ (left),  and  $z_{\rm min}=3a$ (right))}
    \label{fig:lam_qcd_ht}
\end{figure}

  In Fig.~\ref{fig:alpha_ht_fit} we show the results for $\alpha_s$. 
We see that the addition of Gaussian term does not change the value of $\alpha_s$ dramatically, shifting it  approximately 
by  ${\cal O}(10\%)$. As before, these values can be translated into a $\Lambda_{\rm QCD}$ scale, shown in Fig.~\ref{fig:lam_qcd_ht}. It should be noted that the Gaussian term  acts in the same direction as the perturbation theory. Thus, substituting the  combination  in Eq.~\eqref{lzaL} by the
 purely perturbative  expression of Eq.~\eqref{lza} results in  larger values of $\alpha_s$. The value of $\Lambda_{\rm QCD}$ from this augmented model appears to be a smoother function of the lattice spacing and  approaches a smaller value of $\sim 120$ MeV.

\begin{figure}[ht!]
    \centering
    \includegraphics[width=0.30\textwidth]{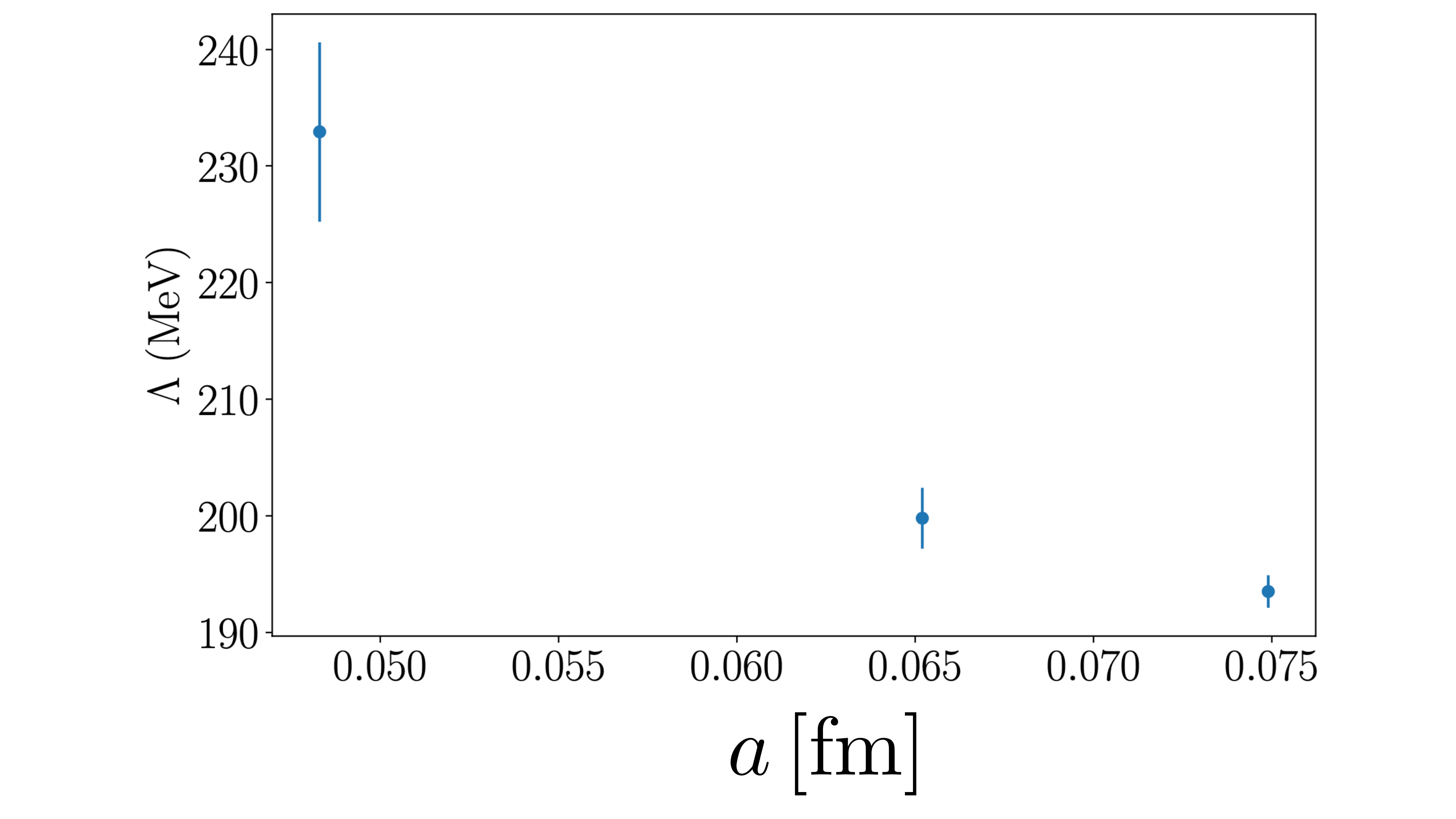}
    \includegraphics[width=0.30\textwidth]{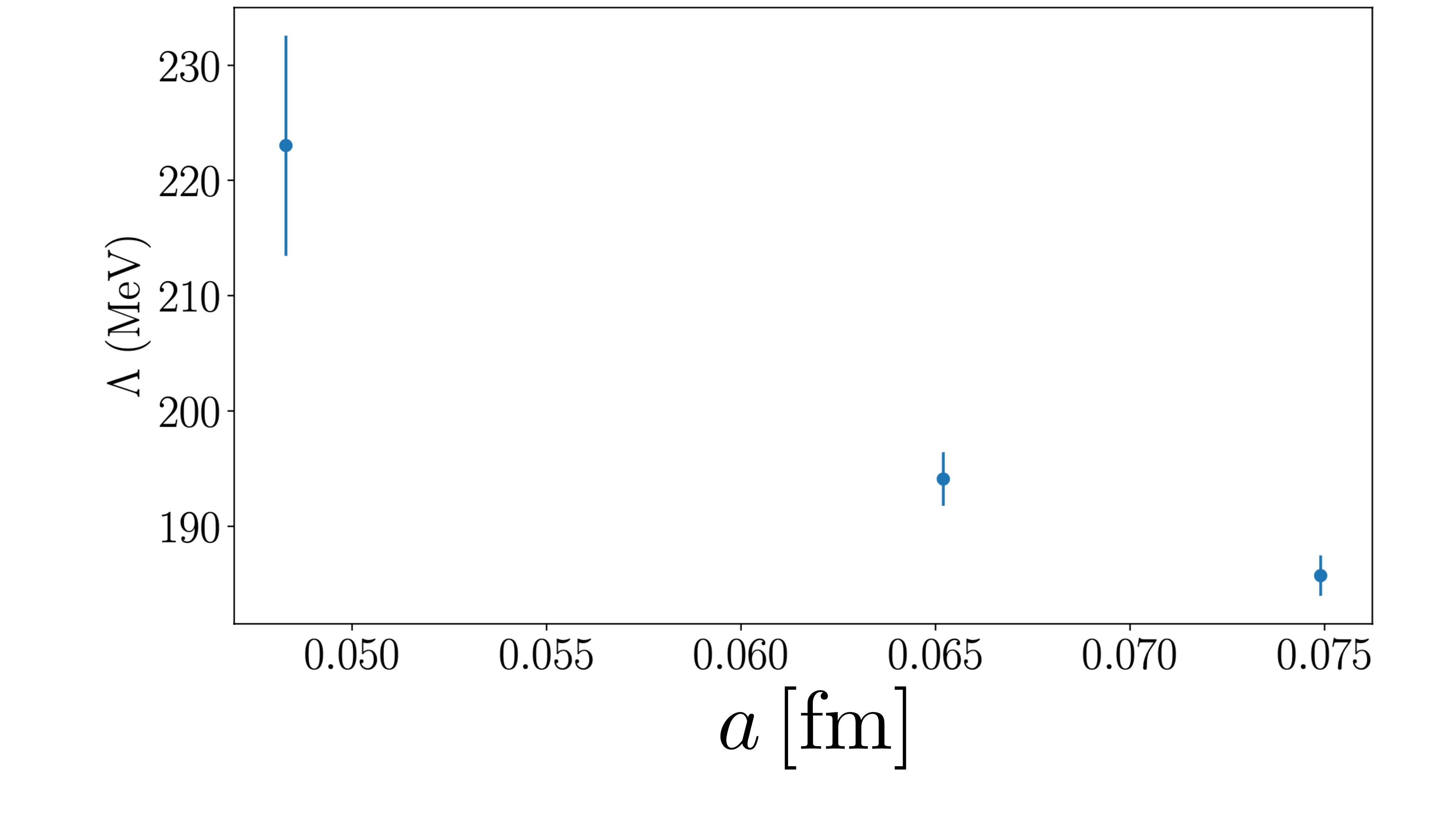}
   \vspace{-3mm} 
    \caption{The fit results for the Gaussian model parameter $\Lambda$ from fits with $z_{\rm min}=2a$ (left), and  $z_{\rm min}=3a$ (right).}
    \label{fig:fit_lam}
\end{figure}

In Fig.~\ref{fig:fit_lam}, the results for the parameter  $\Lambda$ that reflects  the finite size effects is shown. It is rather 
constant in $a$.   
{  For our smallest $a$, its  value approaches  $\sim 230$ MeV.}

 It should also be noted that the  Gaussian term  $ -{z_3^2\Lambda^2}/{4}$ in Eq.~\eqref{lzaL}  includes a  factor of four that arises if we take the $\exp[-k_T^2/\Lambda^2]$   dependence on 
$k_T$ in the  model for  the straight-link TMD.  Thus,  $\Lambda$ may be  interpreted  as an average transverse momentum scale.

\begin{table}[]
    \centering
    \def\arraystretch{2.0}
    \begin{tabular}{p{35pt}ccc ccc ccc}
    \hline\hline
     & ($z_{\rm min}=a$) & ($z_{\rm min}=2a$)& ($z_{\rm min}=3a$) & ($z_{\rm min}=a$) & ($z_{\rm min}=2a$) & ($z_{\rm min}=3a$)& ($z_{\rm min}=a$) &($z_{\rm min}=2a$)& ($z_{\rm min}=3a$) \\
         Ens  & $\alpha_s$  &  $\alpha_s$ &  $\alpha_s$ & $\Lambda$ (MeV)   & $\Lambda$ (MeV)  &  $\Lambda$ (MeV)  & $\chi^2$/dof& $\chi^2$/dof &  $\chi^2$/dof \\\hline
    N5  & 0.1336(15) & 0.1368(23) & 0.1394(29) & 245(6) & 232(8) & 223(10) & 1.5(9)& 0.8(6)& 0.5(4) \\
    E5   & 0.1401(4) & 0.1467(9) & 0.1488(9) & 216(2) & 200(3) & 194(2) & 18(3)& 7(1)& 3.8(8)\\
    A4p5 & 0.1462(5) & 0.1495(8) & 0.1533(10) & 200(1) & 193(1) & 185(2) & 12(2) & 7(1)& 3.0(5)\\\hline\hline
    \end{tabular}
    \caption{Results of fits on all three ensembles to the renormalization constant and a finite size model.}
    \label{tab:fit_res_ht}
\end{table}

The inclusion of a finite size correction dramatically improved the quality of fits for the long distance results. 
  The lattice data is well described in a 2 parameter fit by perturbation theory and a finite size correction up to $z\sim$0.6 fm. Both of these effects could be beneficial in the ratio scheme to cancel the same effects in the moving frame data.


\section{Lattice Perturbation Theory $Z$-Factor \label{sec:lpt}}

To estimate the effects of the lattice regulator at one loop in perturbation theory, we calculate the contributions to the quark diagrams using lattice perturbation theory (LPT). We discretize the fermion and gluon QCD actions using the naive fermion and gluon discretizations. Although the data we analyze in this work are generated with nonperturbatively ${\cal O}(a)$-improved actions, the Wilson-line operator is not improved, leading to the possibility of ${\cal O}(a/z_3)$ contributions in principle. Our choice of discretization is the simplest that enables us to extract such effects and significantly simplifies the perturbative calculation. Feynman rules for the naive discretization can be read from, for example, Ref.~\cite{Capitani:2002mp}.

We calculate the relevant diagrams in a two-step process: the Euclidean gamma matrix structure is relatively straightforward and handled using \verb+Mathematica+. The resulting scalar integrals are evaluated numerically using \verb+vegas+~\cite{Lepage:1977sw}, an adaptive Monte Carlo algorithm, implemented in dedicated \verb+FORTRAN+ routines. The values for the diagram, and their total, is given in Tab.~\ref{tab:latt_pt}.

The dominant diagram is given by the Wilson Line self-energy, which generates the logarithmic divergences and was first determined to leading order in Ref.~\cite{Chen:2016fxx}. Still, the other diagrams provide a significant contribution to the total, particularly at smaller $z$. As with the Polyakov regulator, the logarithmic contributions create curvature beyond the linear divergence relevant for an accurate fit to data with percent or smaller precision particularly. Even at the largest $z/a$ in our study, the diagrams without the linear divergence contribute nearly $10\%$ of the total, and more at lower $z/a$. 

\begin{table}[b]
\centering
\def\arraystretch{2.0}
\begin{tabular}{l |c  c  c  c  c  c  c  c  c }
\hline\hline
\diagbox[height=3em,linecolor=white]{Diagram}{$z/a$~} & 0 &1 & 2 & 3 & 4 & 5 & 6 & 7 & 8 \\\hline
     Sunset & 0 &
0.97346(2)& 2.32308(7)& 3.7762(1)& 5.2709(2)& 6.7871(3)& 8.3163(4)& 9.8550(5)& 11.3998(6)\\
Sail & 0 &
0& 0.54974(3)& 0.98654(6)& 1.2110(1)& 1.3580(1)& 1.4518(2)& 1.5226(2)& 1.5776(2)\\
Vertex & 1.4339(6) &  0.5959(6) & 0.1216(6) & -0.0847(6) &  -0.2047(6) & -0.2717(6) & -0.3307(6) & -0.3725(6) & -0.4115(7) \\\hline
Total $\Gamma_{\rm LPT}$ & 1.4339(6) &  1.5694(6) &  2.9944(6) &  4.6780(6) &  6.2772(6) &  7.8734(6) &
 9.4374(6) & 11.005(6) & 12.5659(7) \\
 \hline\hline
\end{tabular}
\caption{The integrals of 1-loop diagrams in lattice perturbation theory for various $z/a$.}\label{tab:latt_pt}
\end{table}

\subsection{Fits by $Z$-factor only}

This section contains analysis of the three ensembles to the functional form 
\begin{eqnarray}
    l(z_3,a) = \ln \left ( \frac{ Z_{\rm LPT}(z_3/a,\alpha_s)}{ Z_{\rm LPT}(0,\alpha_s)}\right )
\end{eqnarray}
where $Z_{\rm LPT} = \exp{-C_f \alpha_s \Gamma_{\rm LPT}}$ and $\Gamma_{\rm LPT}$ is the total of the diagrams in Tab.~\ref{tab:latt_pt}. Again, $\alpha_s$ is allowed to vary as a fit parameter. Moreover, the stochastic \verb+vegas+ algorithm only gives the renormalization constant to a finite statistical precision, which must be taken into account. To do so, model parameters, $\tilde{Z}$, are introduced for the value of the renormalization constant at each $z/a$. These parameters have a Gaussian prior distribution with central value and width given by the result and error of the \verb+vegas+ integration. With the exception of $z=a$ and $z=2a$, which are poorly described in the fit generally, the fitted renormalization constants are well within the errors of the integral.

\begin{figure}[ht!]
    \centering
    \includegraphics[width=0.30\textwidth]{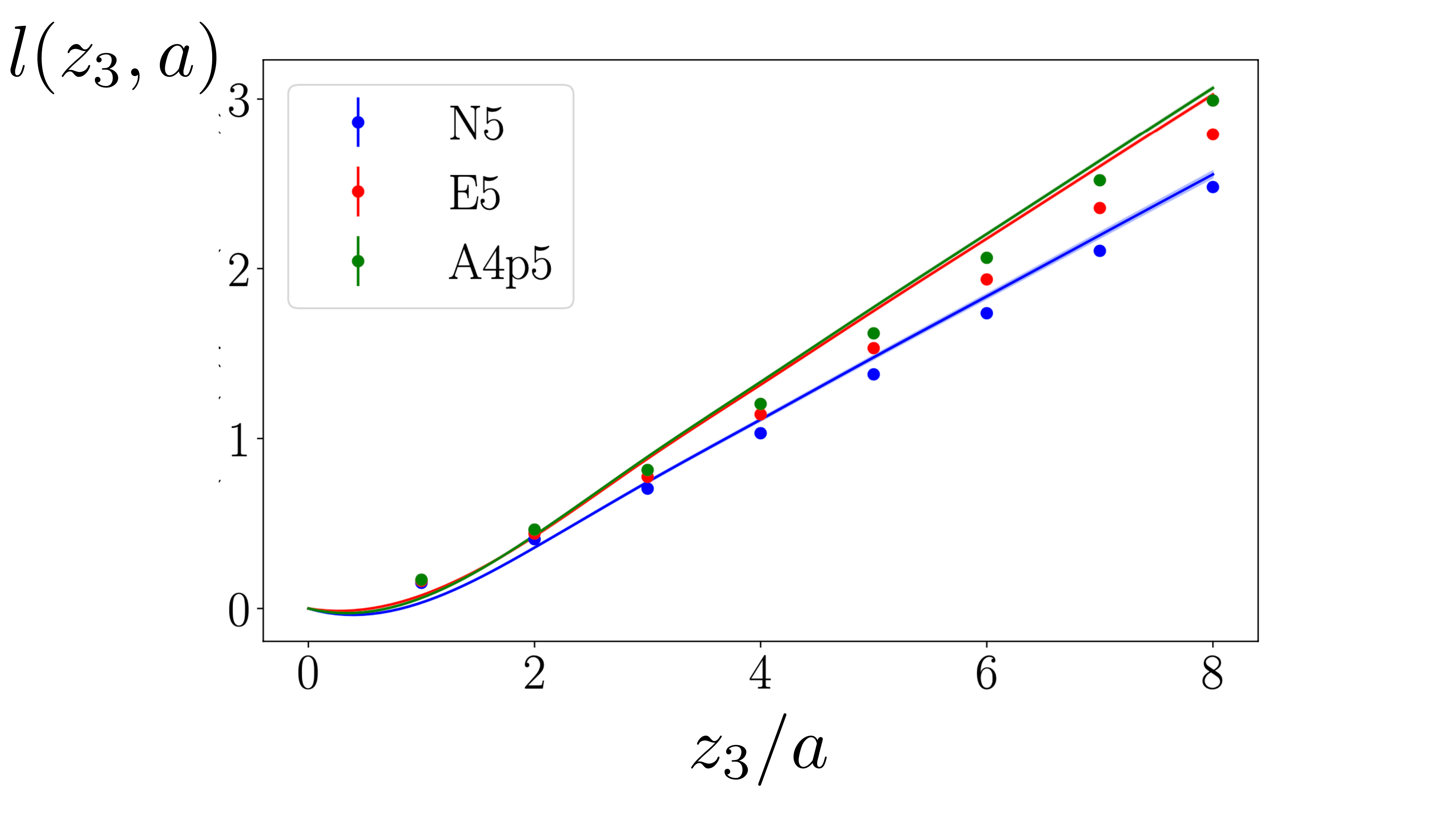}
    \includegraphics[width=0.30\textwidth]{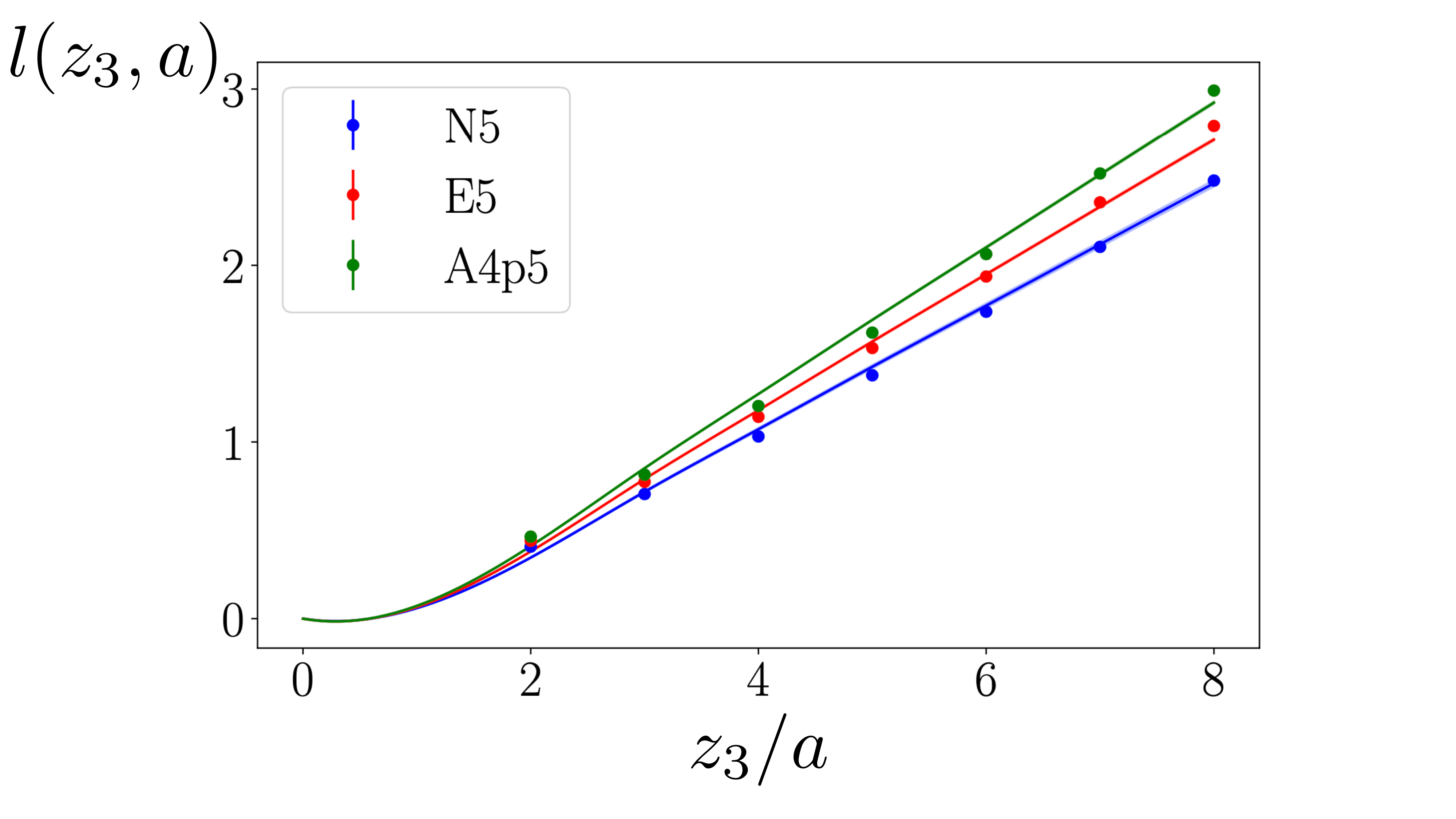}
    \includegraphics[width=0.30\textwidth]{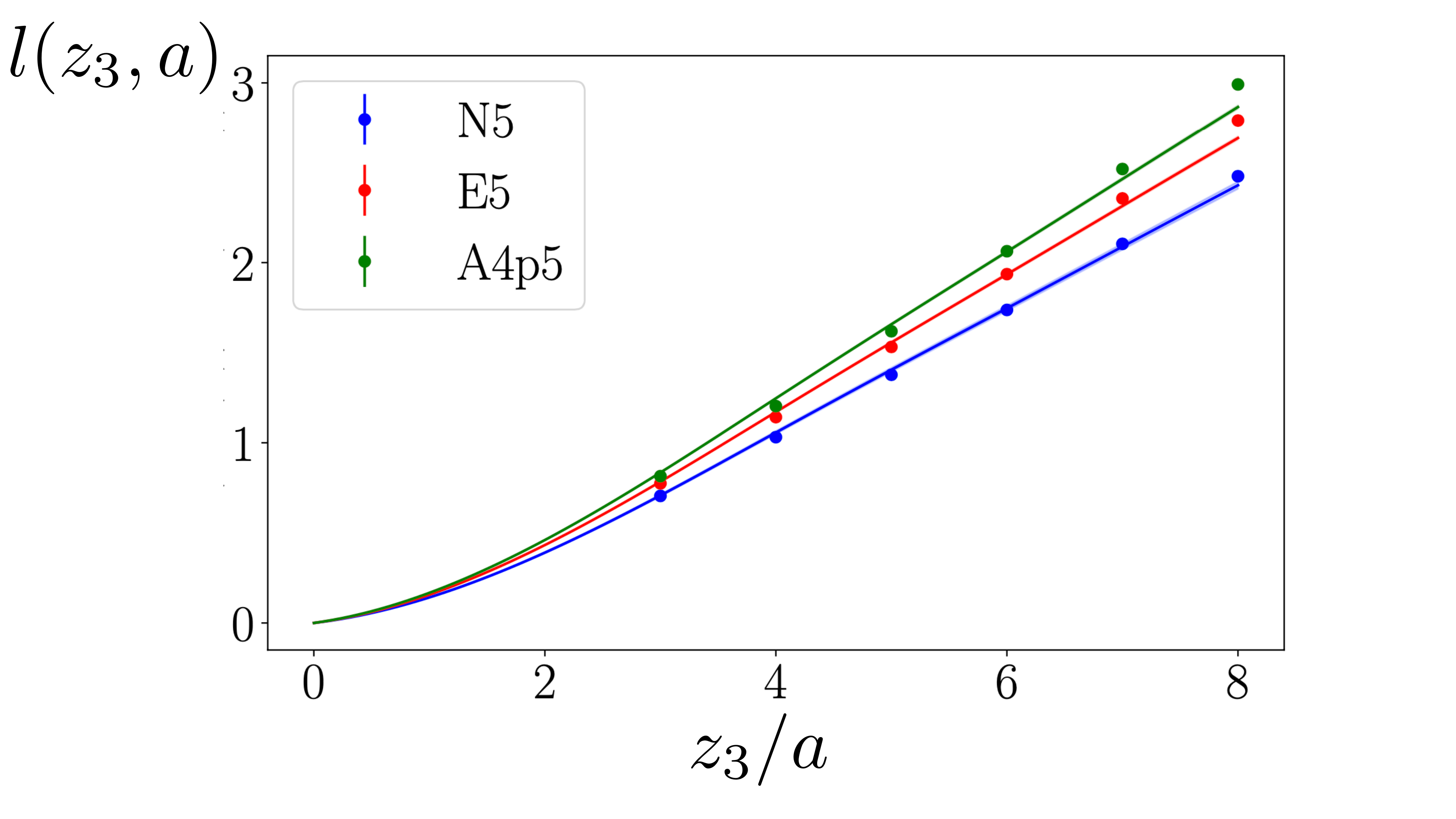}
    \caption{Results of fits on all three ensembles to the lattice regulated renormalization constant with $z_{\rm min}=a$ (left), $z_{\rm min}=2a$ (center), and  $z_{\rm min}=3a$ (right)}
    \label{fig:pt_fit_noht_lattpt}
\end{figure}

\begin{figure}
    \centering
    \includegraphics[width=0.30\textwidth]{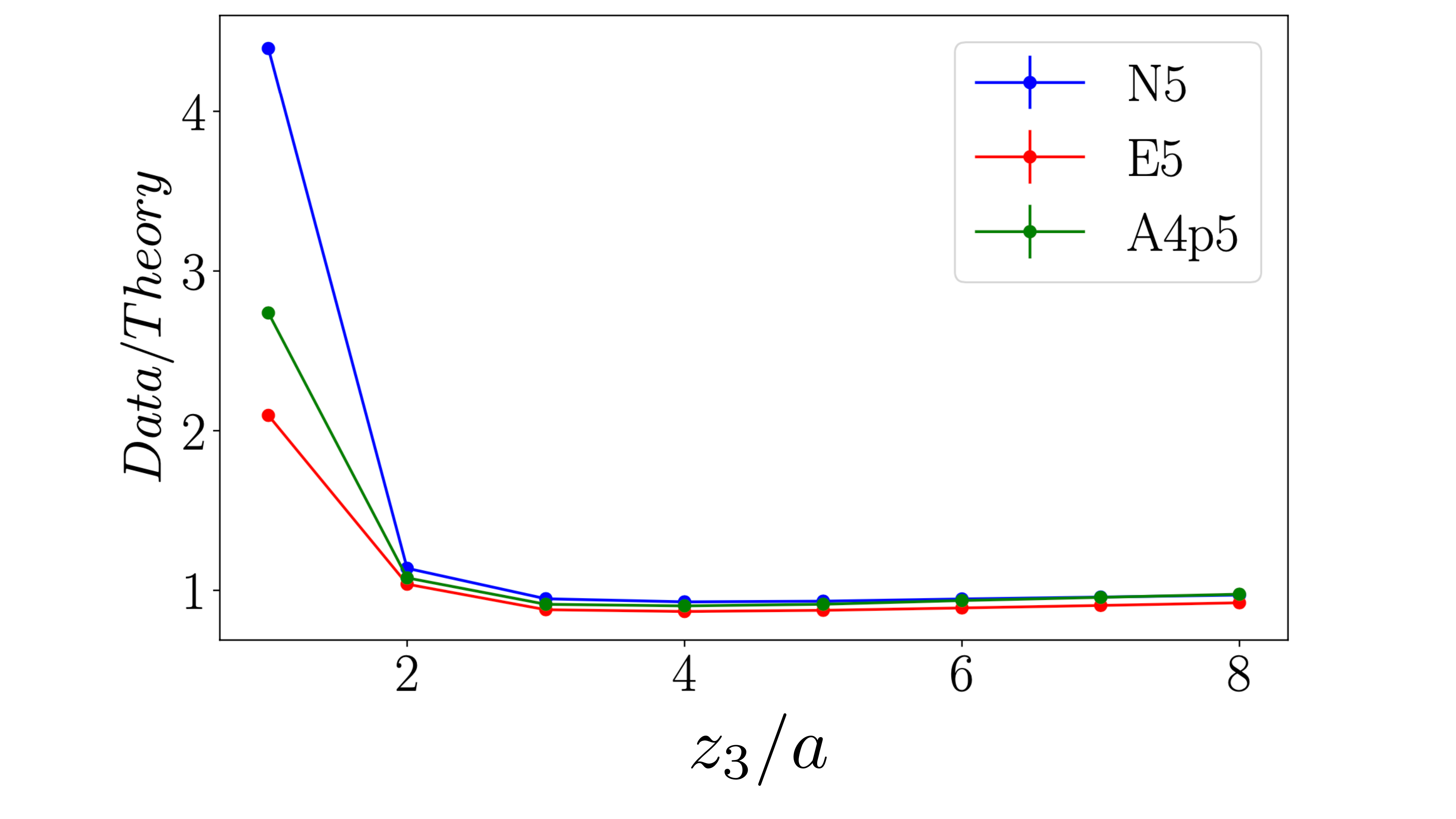}
    \includegraphics[width=0.30\textwidth]{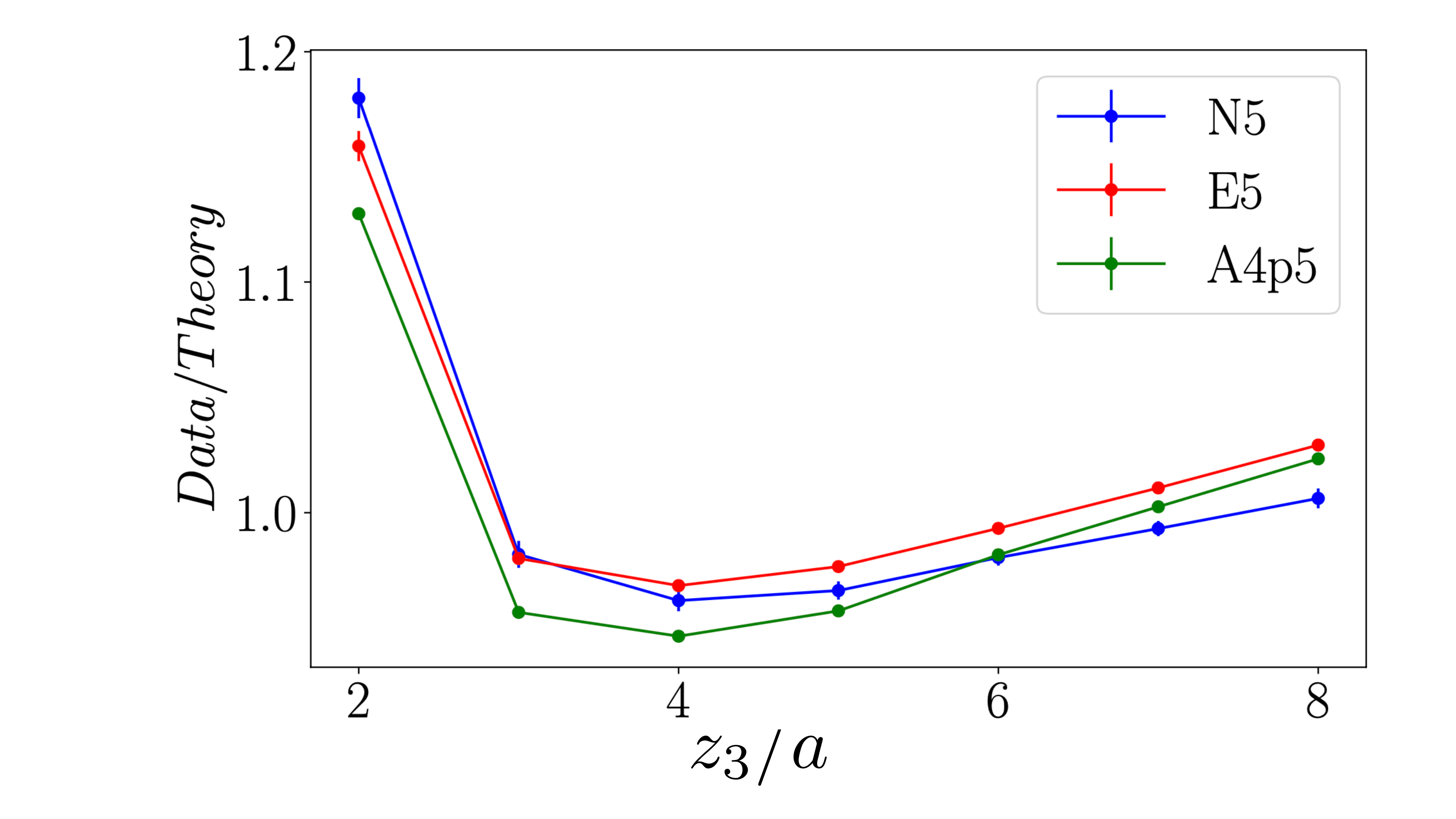}
    \includegraphics[width=0.30\textwidth]{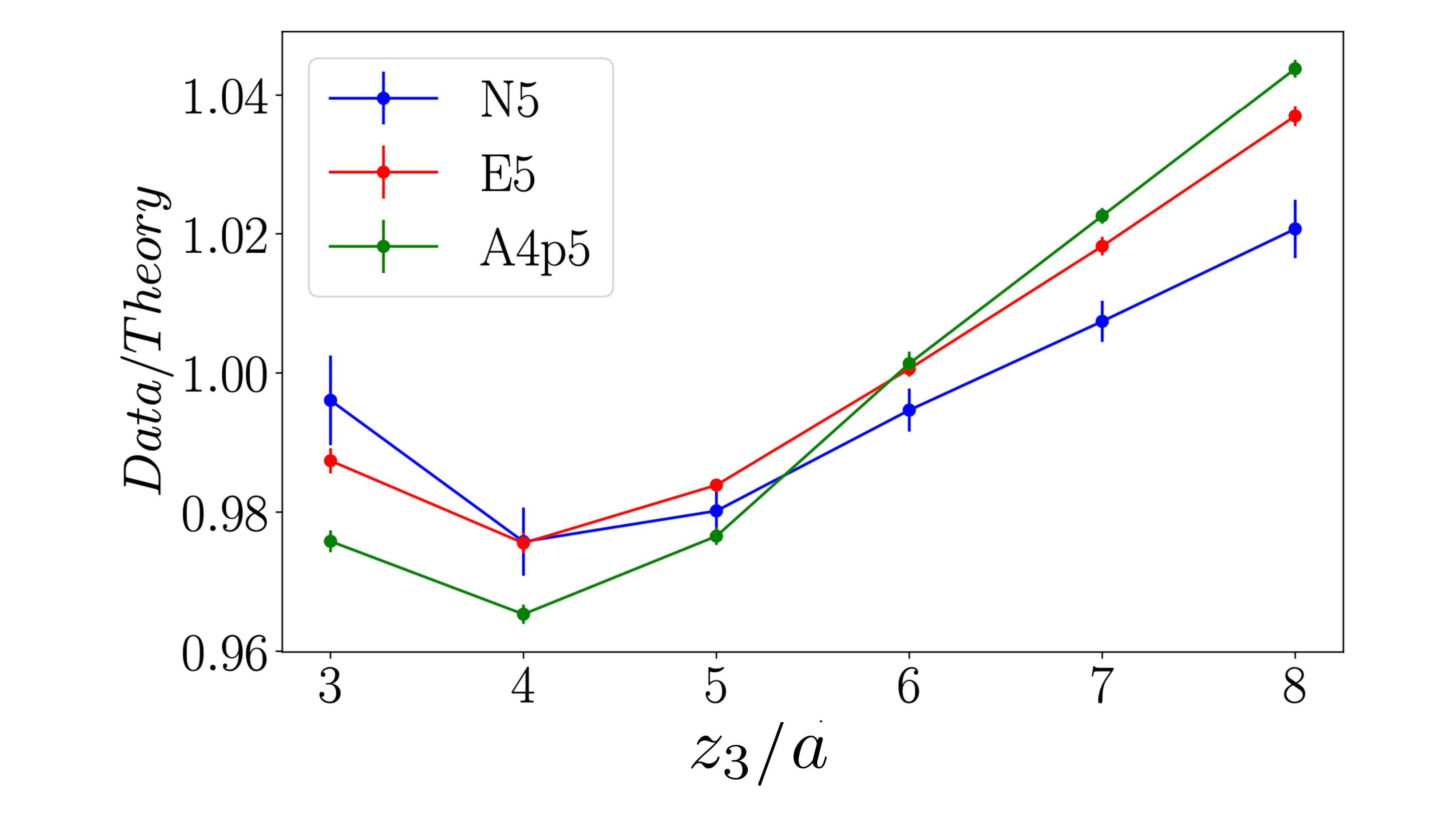}
    \caption{The ratio of the data to the lattice regulated perturbative model results with $z_{\rm min}=a$ (left), $z_{\rm min}=2a$ (center), and  $z_{\rm min}=3a$ (right). }
    \label{fig:dot_noht_lattpt}
\end{figure}

\begin{figure}[ht!]
    \centering
    \includegraphics[width=0.30\textwidth]{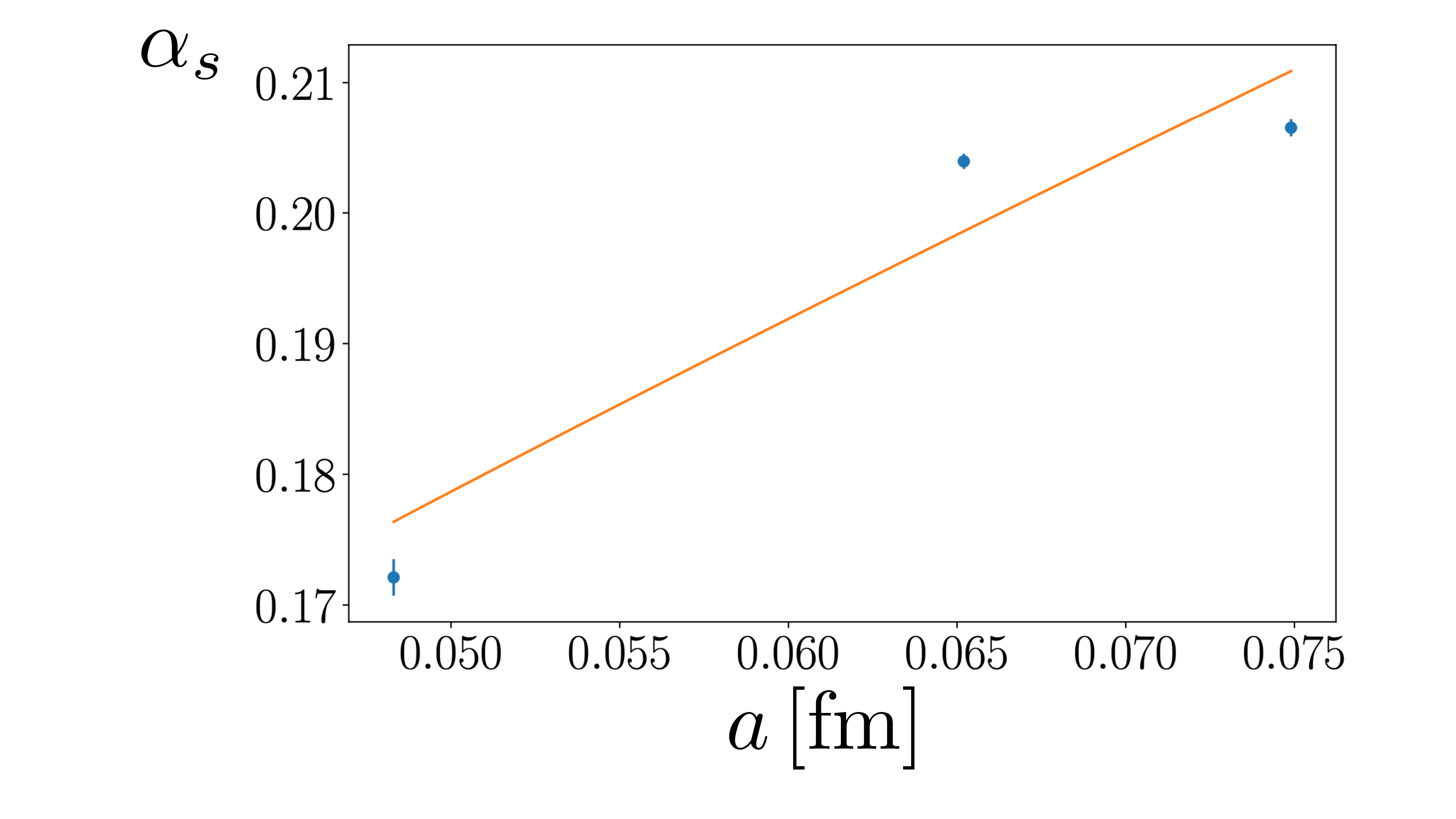}
    \includegraphics[width=0.30\textwidth]{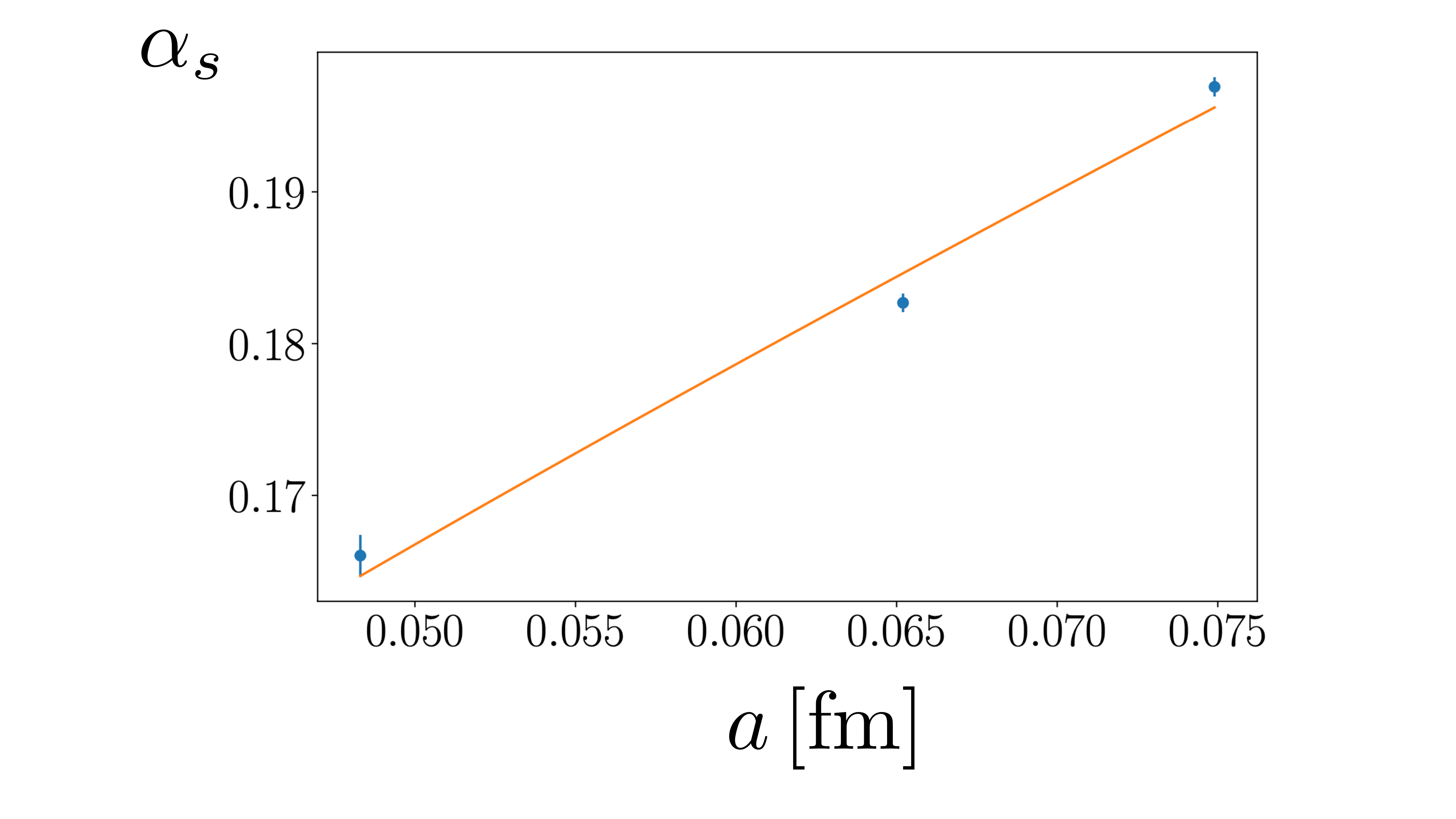}
    \includegraphics[width=0.30\textwidth]{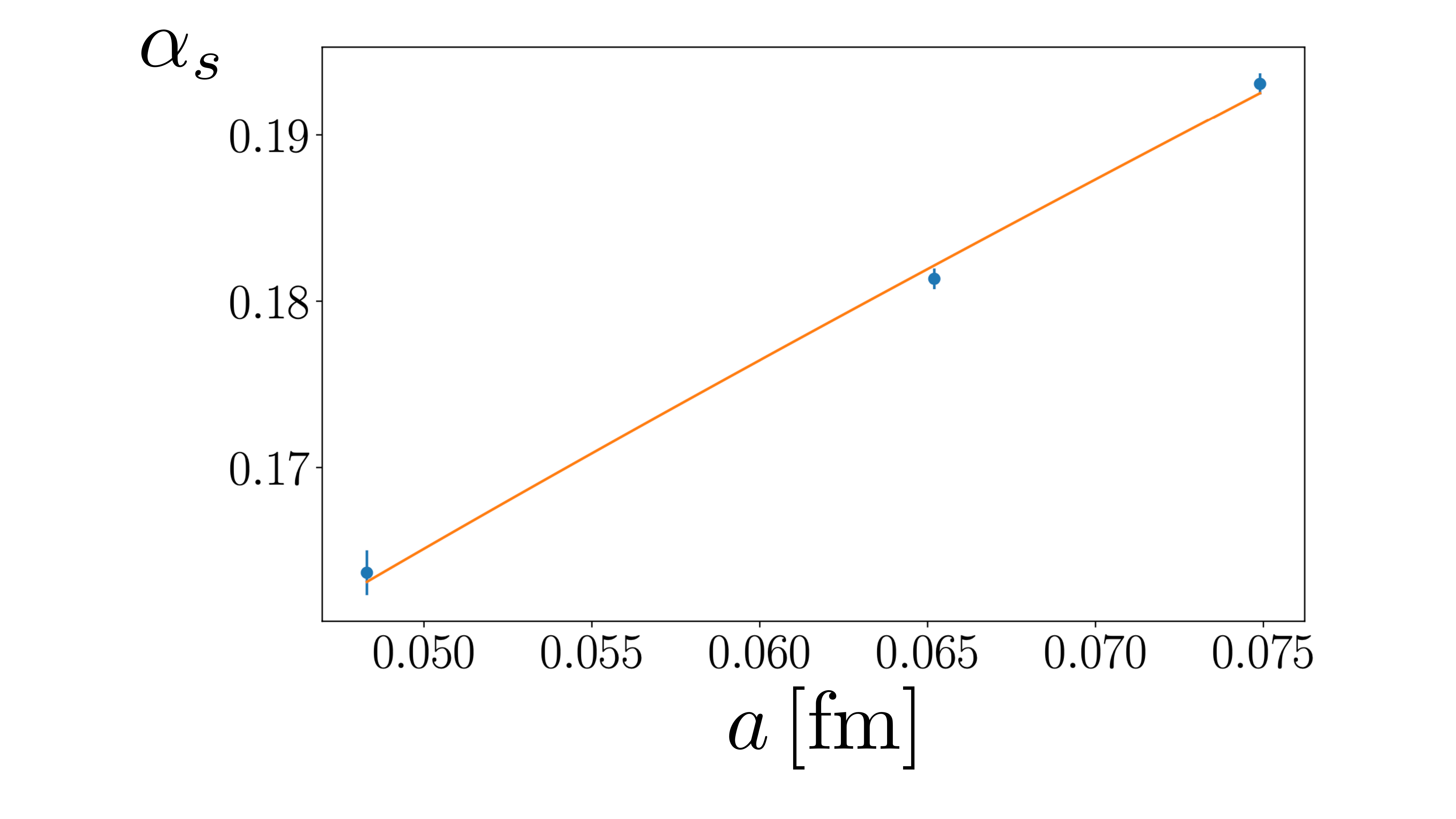}
    \caption{The values of $\alpha_s$ from lattice regulated fits for $l(z_3,a)$  as a function of lattice spacing
    with $z_{\rm min}=a$ (left), $z_{\rm min}=2a$ (center), and  $z_{\rm min}=3a$ (right). The curve, intended solely to guide the eye, represents a fit to the LO perturbative formula (\ref{eq:running_alpha}) 
     where $\Lambda_{\rm QCD}$ was the single fit parameter.  }
    \label{fig:alpha_fit_noht_lattpt}
\end{figure}

\begin{figure}[ht!]
    \centering
    \includegraphics[width=0.30\textwidth]{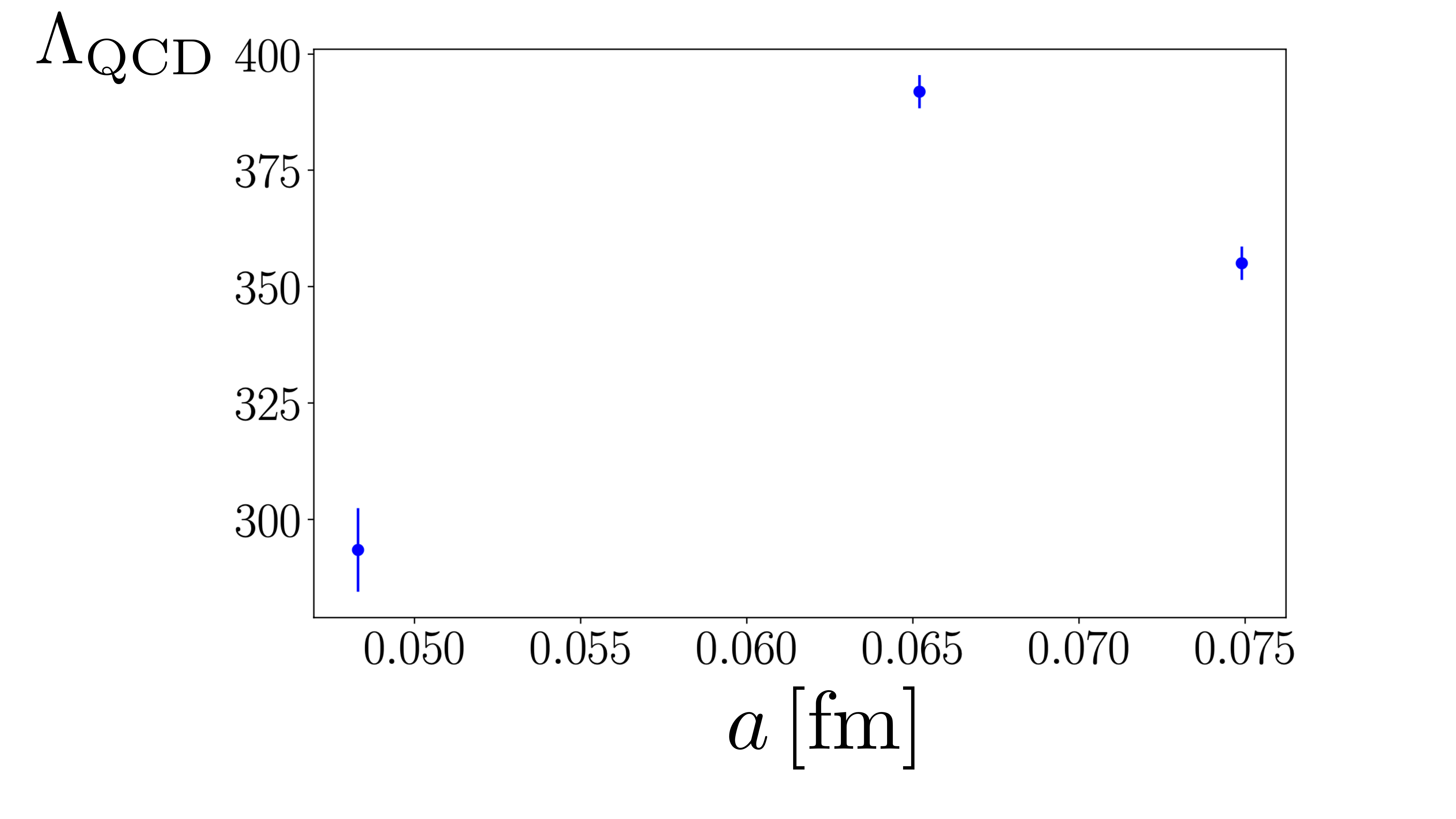}
    \includegraphics[width=0.30\textwidth]{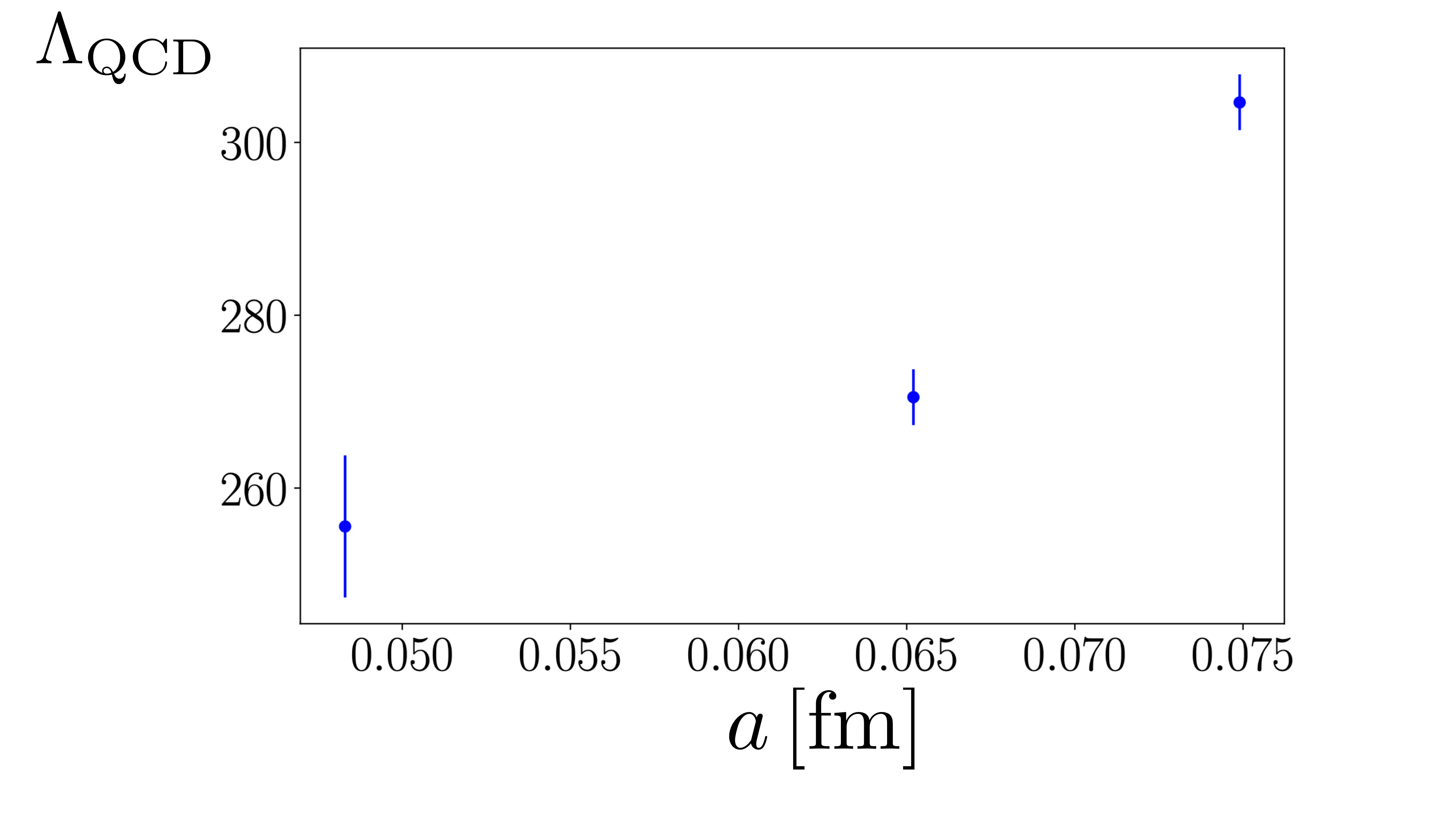}
    \includegraphics[width=0.30\textwidth]{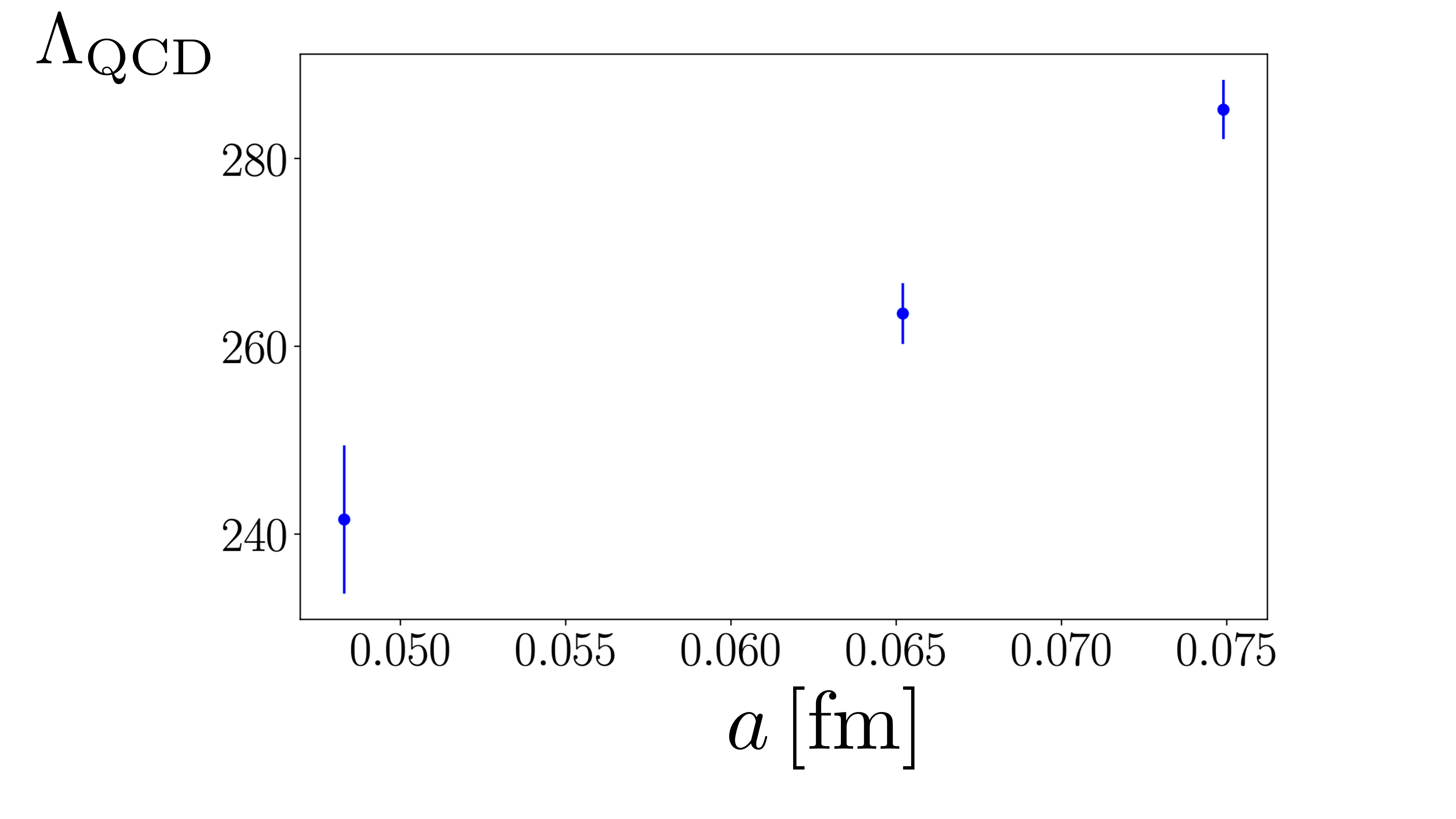}
    \caption{The value of $\Lambda_{\rm QCD}$ extracted from $\alpha_S$ using LO perturbation theory in Fig.~\ref{fig:alpha_fit_noht_lattpt} with $z_{\rm min}=a$ (left), $z_{\rm min}=2a$ (center), and  $z_{\rm min}=3a$ (right))}
    \label{fig:lam_qcd_noht_lattpt}
\end{figure}

The results of the fit are shown in Fig.~\ref{fig:pt_fit_noht_lattpt}. When fitting $z_{\rm min}=3a$, the perturbative results produce the lattice data within a few percent, shown in Fig.~\ref{fig:dot_noht_lattpt}, but the shorter distances create significant discrepancies which spoil their fit results. The fit parameters are given in Tab.~\ref{tab:fit_res_noht_latt_pt}. The results for $\alpha_s$ are shown in Fig.~\ref{fig:alpha_fit_noht_lattpt}. Using LO continuum running of the coupling, the corresponding $\Lambda_{\rm QCD}$ are shown in Fig.~\ref{fig:lam_qcd_noht_lattpt}. The scale used again was $\pi/a$ which represents the largest momentum allowed in the discretization of fermions, whose momenta are $k\in [-\frac\pi a, \frac \pi a]$. The $\Lambda_{\rm QCD}$ are of the same size as the fit with the Polyakov regulator, though they show a stronger lattice spacing dependence. 

\begin{table}[ht!]
    \centering
    \def\arraystretch{2.0}
    \begin{tabular}{p{35pt}ccc ccc }
    \hline\hline
    & ($z_{\rm min}=a$) & ($z_{\rm min}=2a$)& ($z_{\rm min}=3a$) & ($z_{\rm min}=a$) &($z_{\rm min}=2a$) & ($z_{\rm min}=3a$) \\
      Ens & $\alpha_s$  &  $\alpha_s$ &  $\alpha_s$ & $\chi^2$/dof & $\chi^2$/dof &  $\chi^2$/dof \\\hline
    N5  & 0.1721(14) & 0.1660(13)& 0.1637(13) & 1529 & 36 & 3.6(1.4) \\
    E5  & 0.2039(6) & 0.1826(6) & 0.1813(6) & 14811 & 124 & 41(4) \\
    A4p5 & 0.2065(7) & 0.1969(6) & 0.1930(6) & 9086 & 305 & 78(6) \\
    \hline\hline
    \end{tabular}
    \caption{Results of fits on all three ensembles by the LPT perturbative expression.}
    \label{tab:fit_res_noht_latt_pt}
\end{table}

\subsection{Fits involving Gaussian model for finite size effects}

\begin{figure}[ht!]
    \centering
    \includegraphics[width=0.30\textwidth]{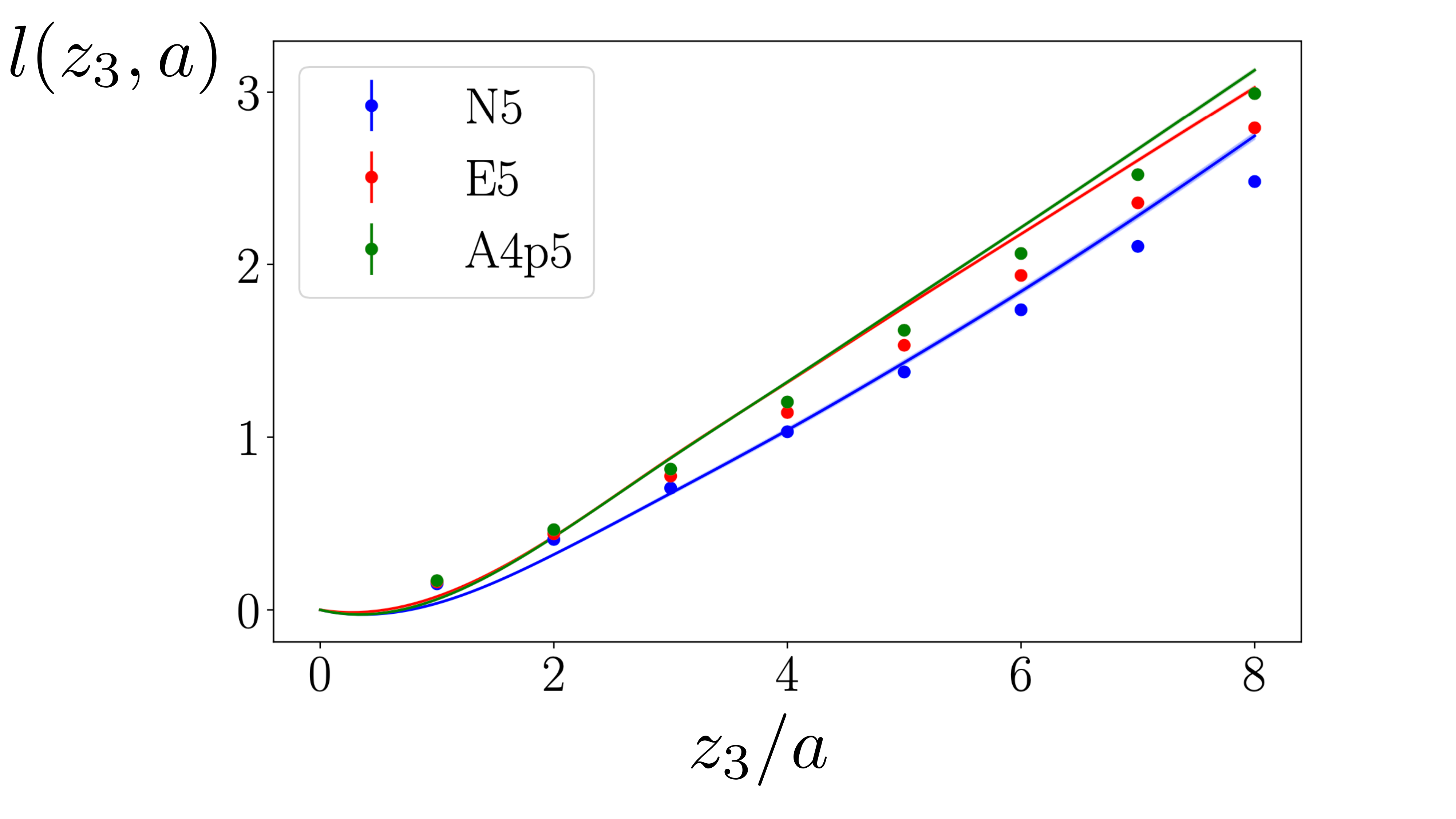}
    \includegraphics[width=0.30\textwidth]{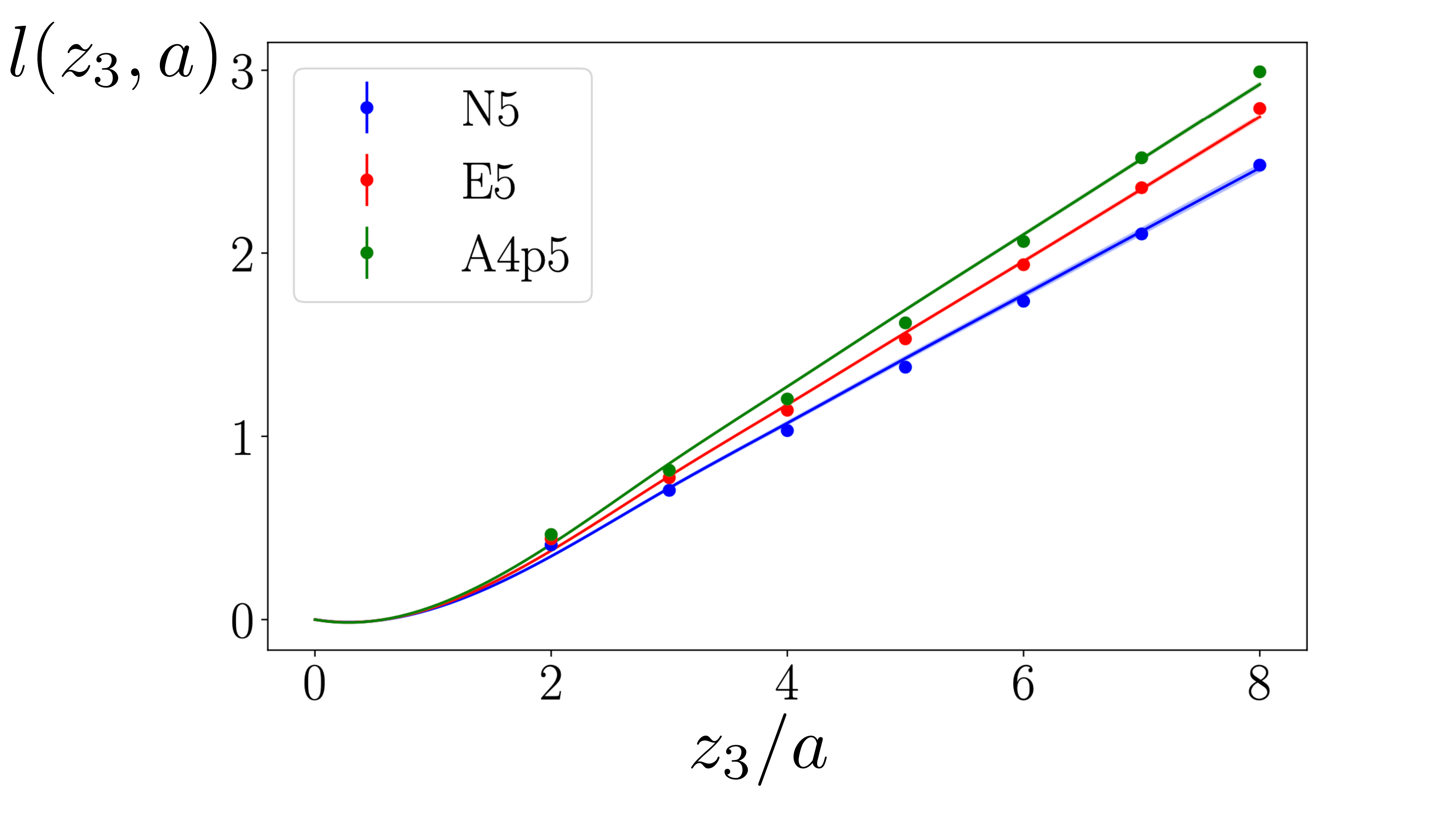}
    \includegraphics[width=0.30\textwidth]{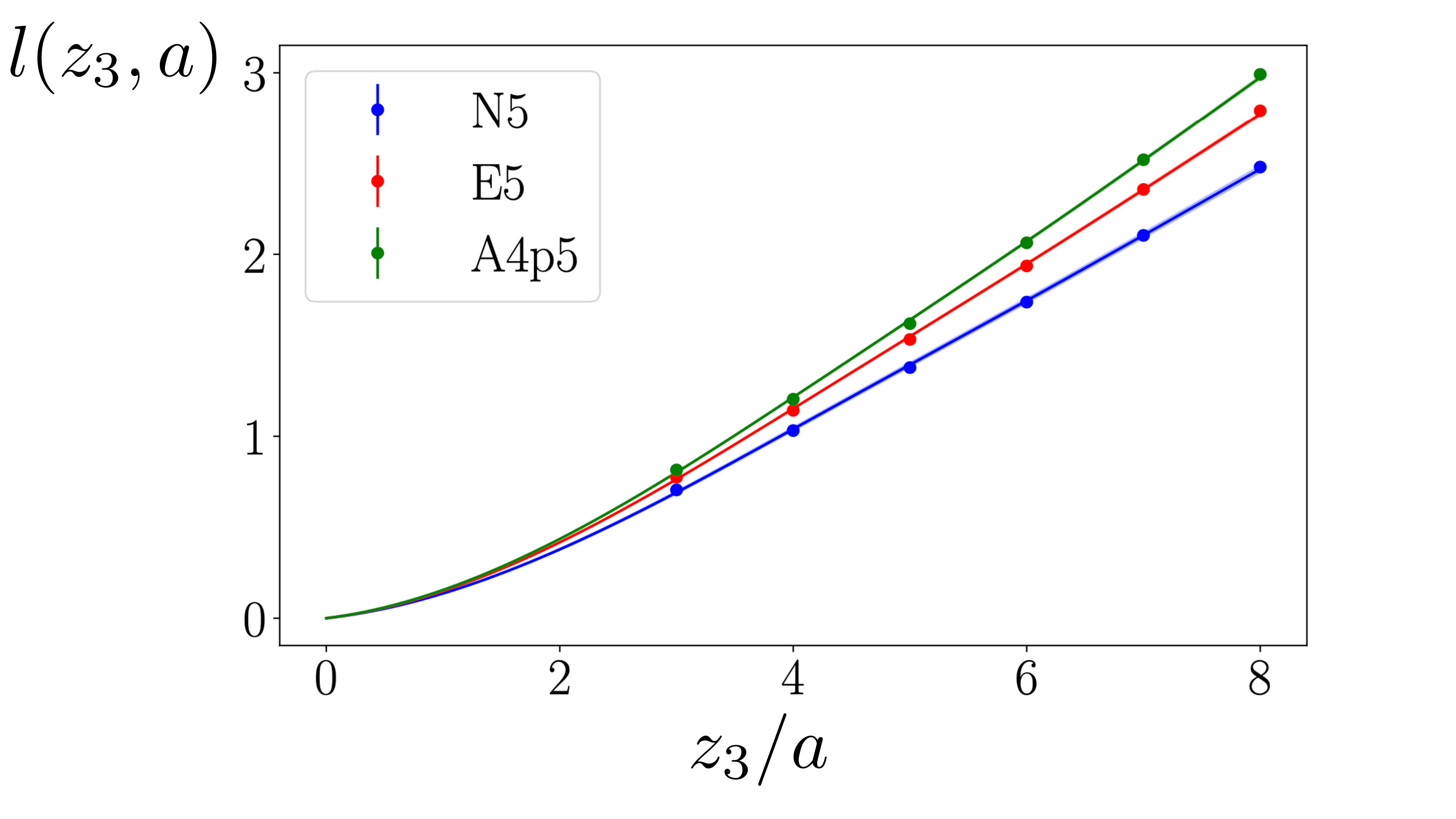}
    \caption{Results of fits for $l(z_3,a)$  on all three ensembles in  a lattice regulated  model containing additional Gaussian factor with $z_{\rm min}=a$ (left), $z_{\rm min}=2a$ (center), and  $z_{\rm min}=3a$ (right).}
       \label{fig:pt_ht_fit_lattpt}
\end{figure}

\begin{figure}
    \centering
    \includegraphics[width=0.30\textwidth]{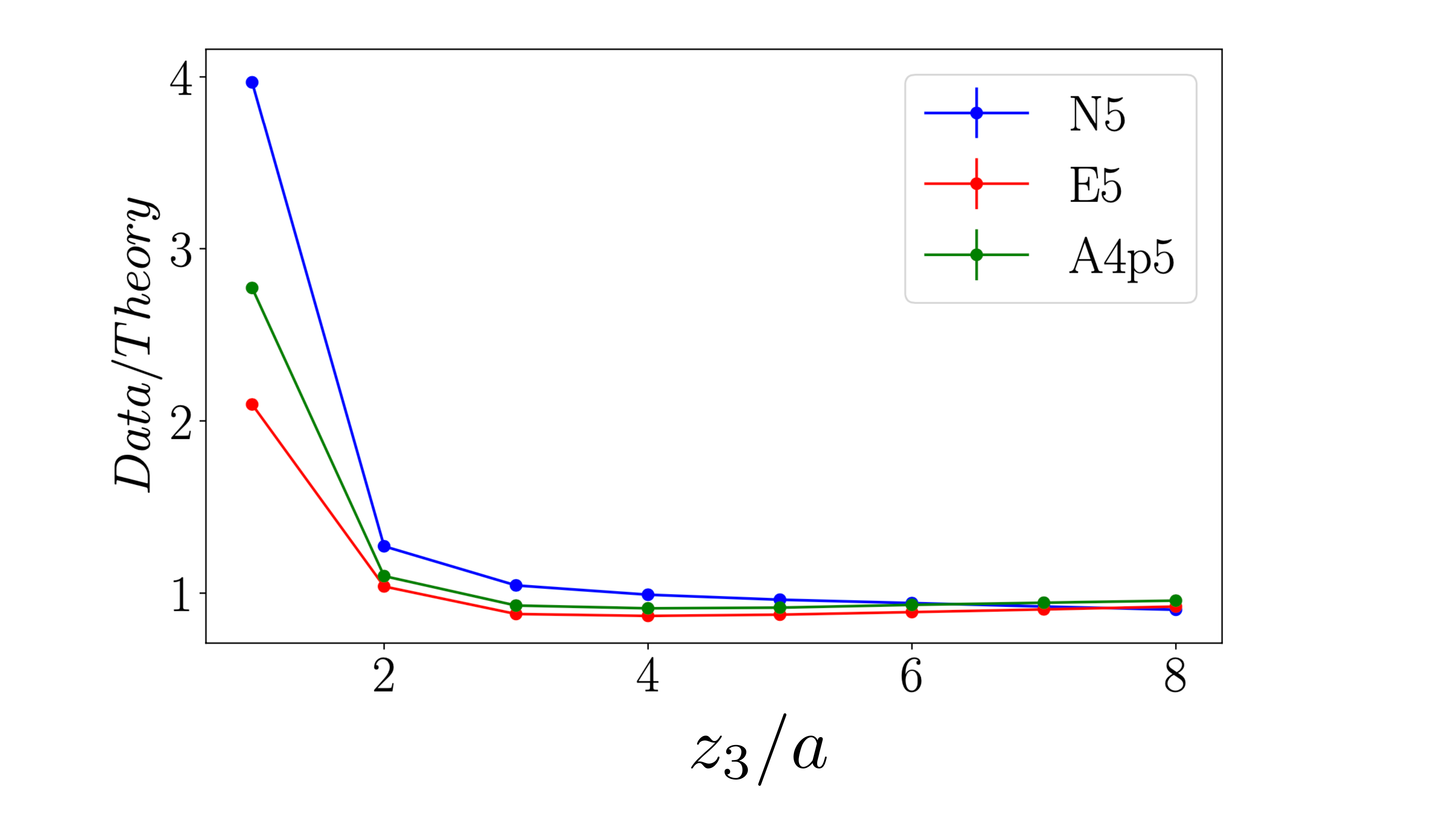}
    \includegraphics[width=0.30\textwidth]{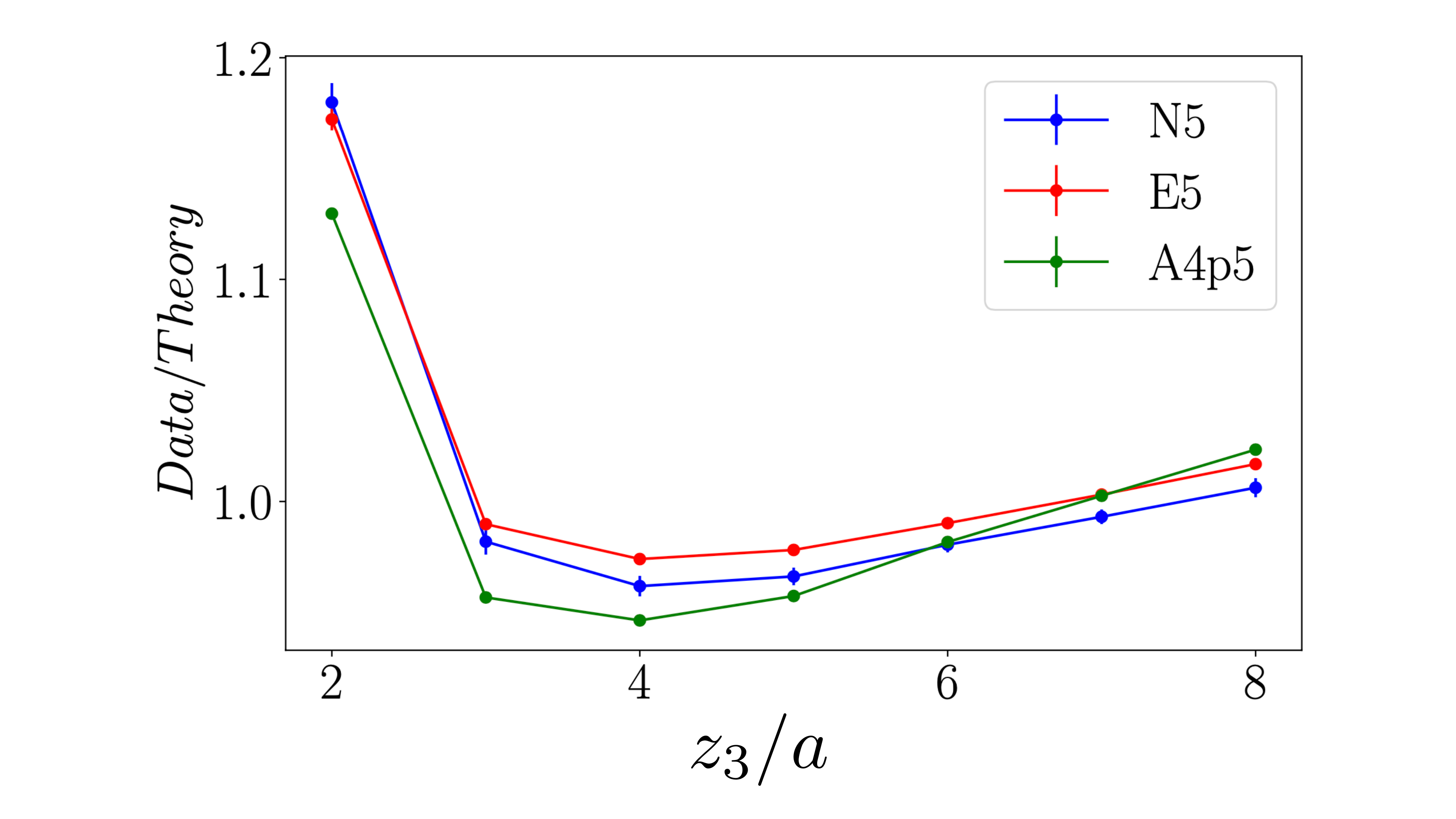}
    \includegraphics[width=0.30\textwidth]{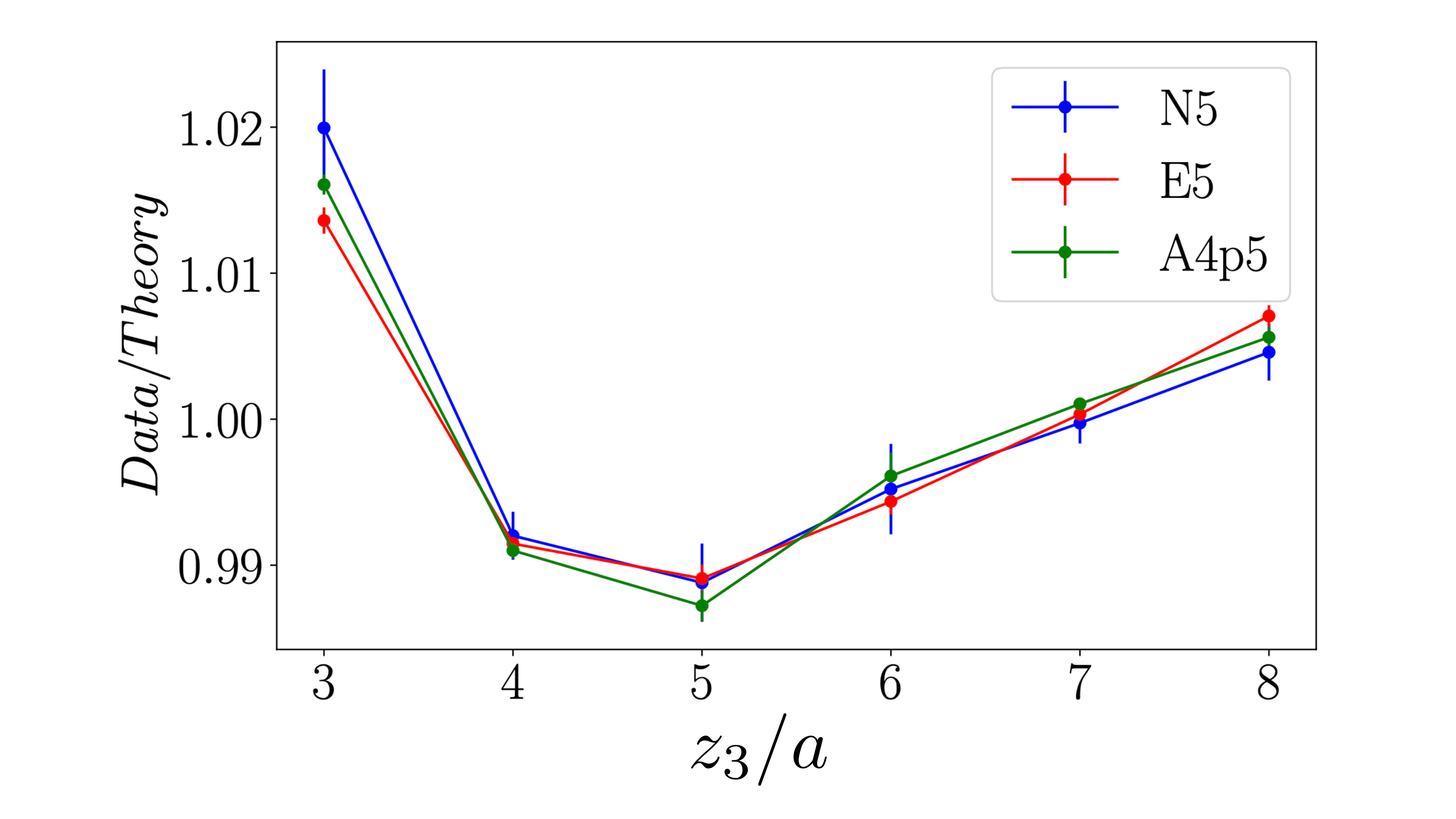}
    \caption{The ratio of the data to the results of  the lattice regulated  model containing additional Gaussian factor, with $z_{\rm min}=a$ (left), $z_{\rm min}=2a$ (center), and  $z_{\rm min}=3a$ (right).  with $z_{\rm min}=a$ (left), $z_{\rm min}=2a$ (center), and  $z_{\rm min}=3a$ (right). }
    \label{fig:dot_ht_lattpt}
\end{figure}

\begin{figure}[ht!]
    \centering
    \includegraphics[width=0.30\textwidth]{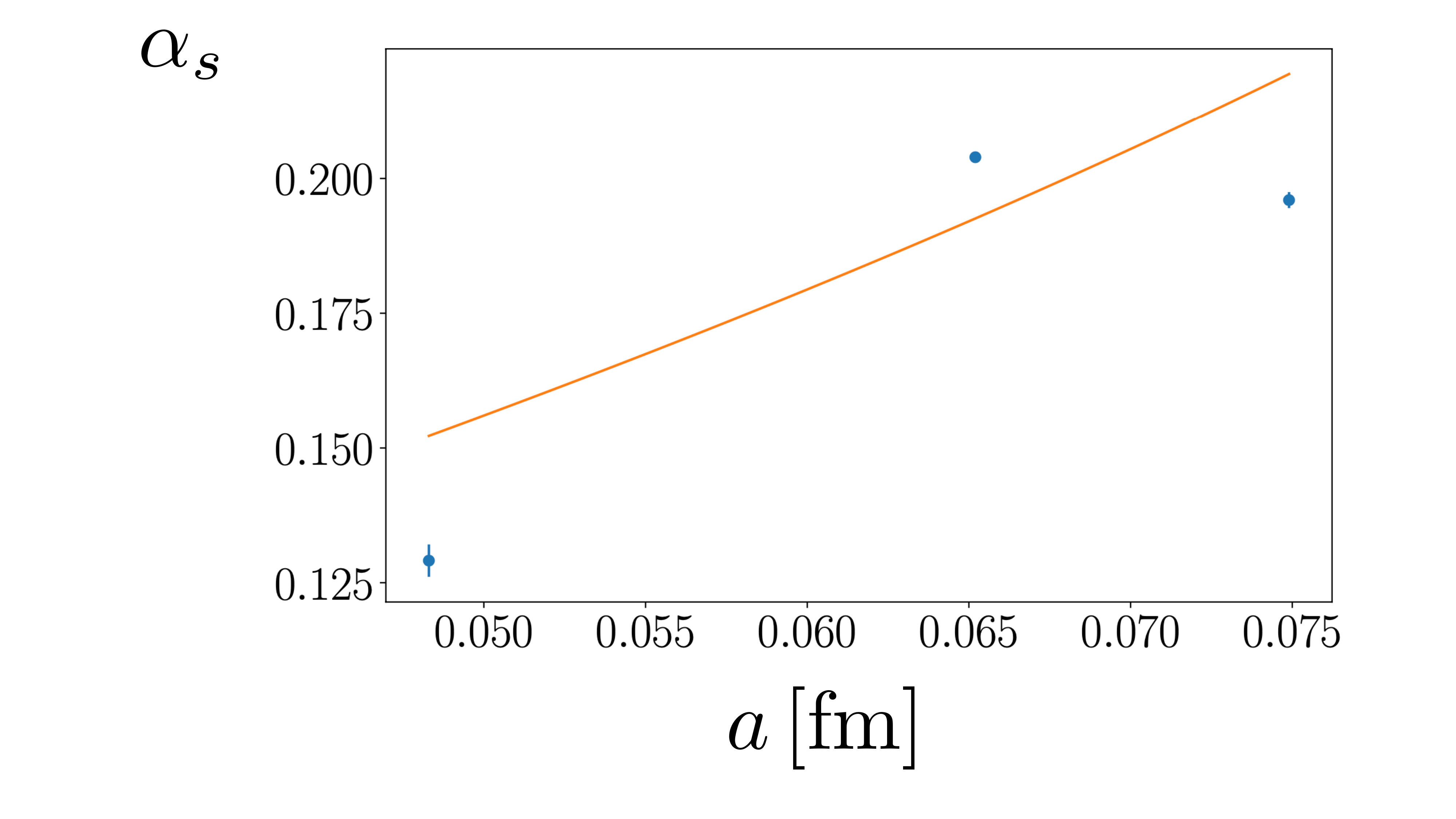}
    \includegraphics[width=0.30\textwidth]{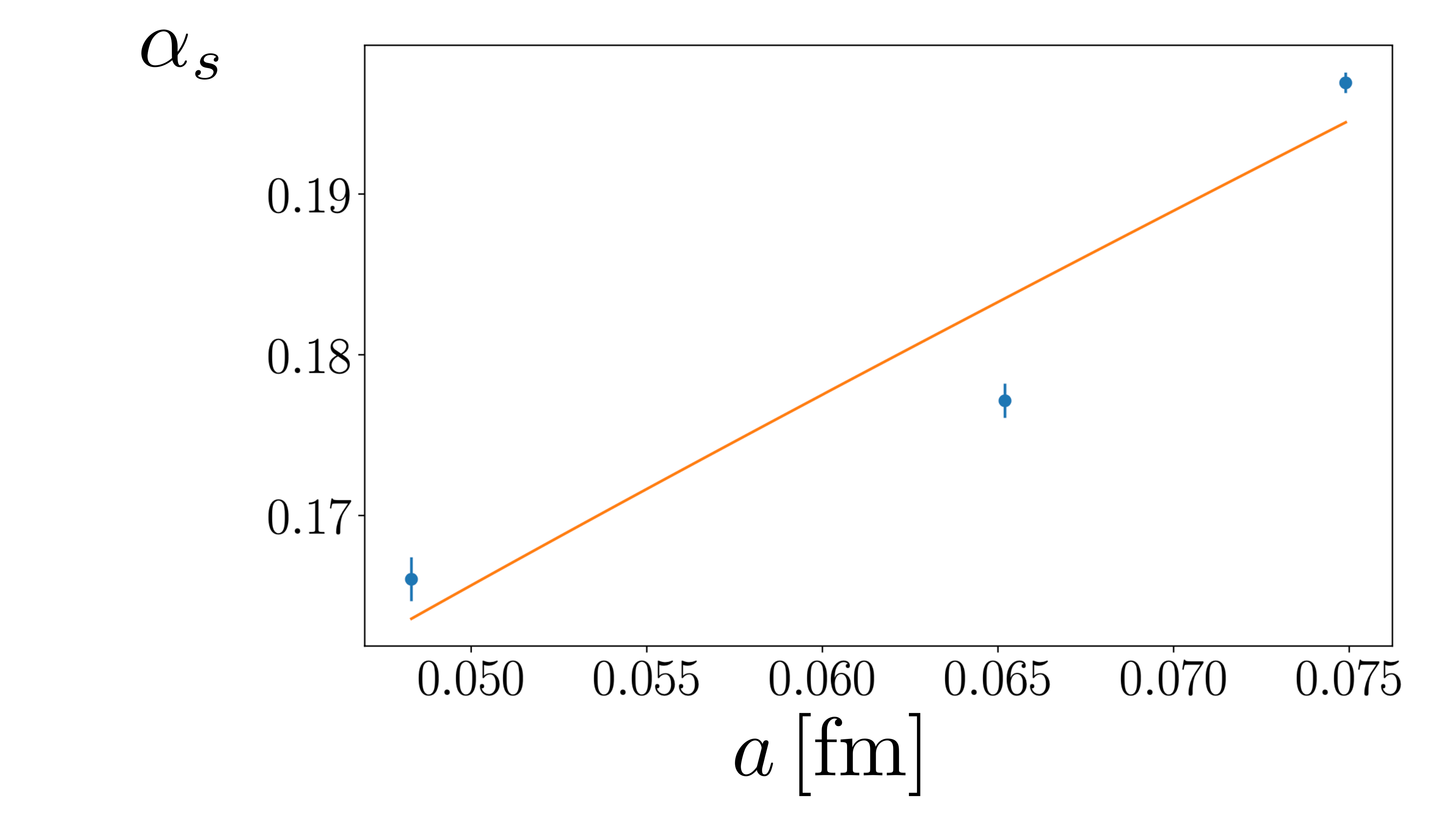}
    \includegraphics[width=0.30\textwidth]{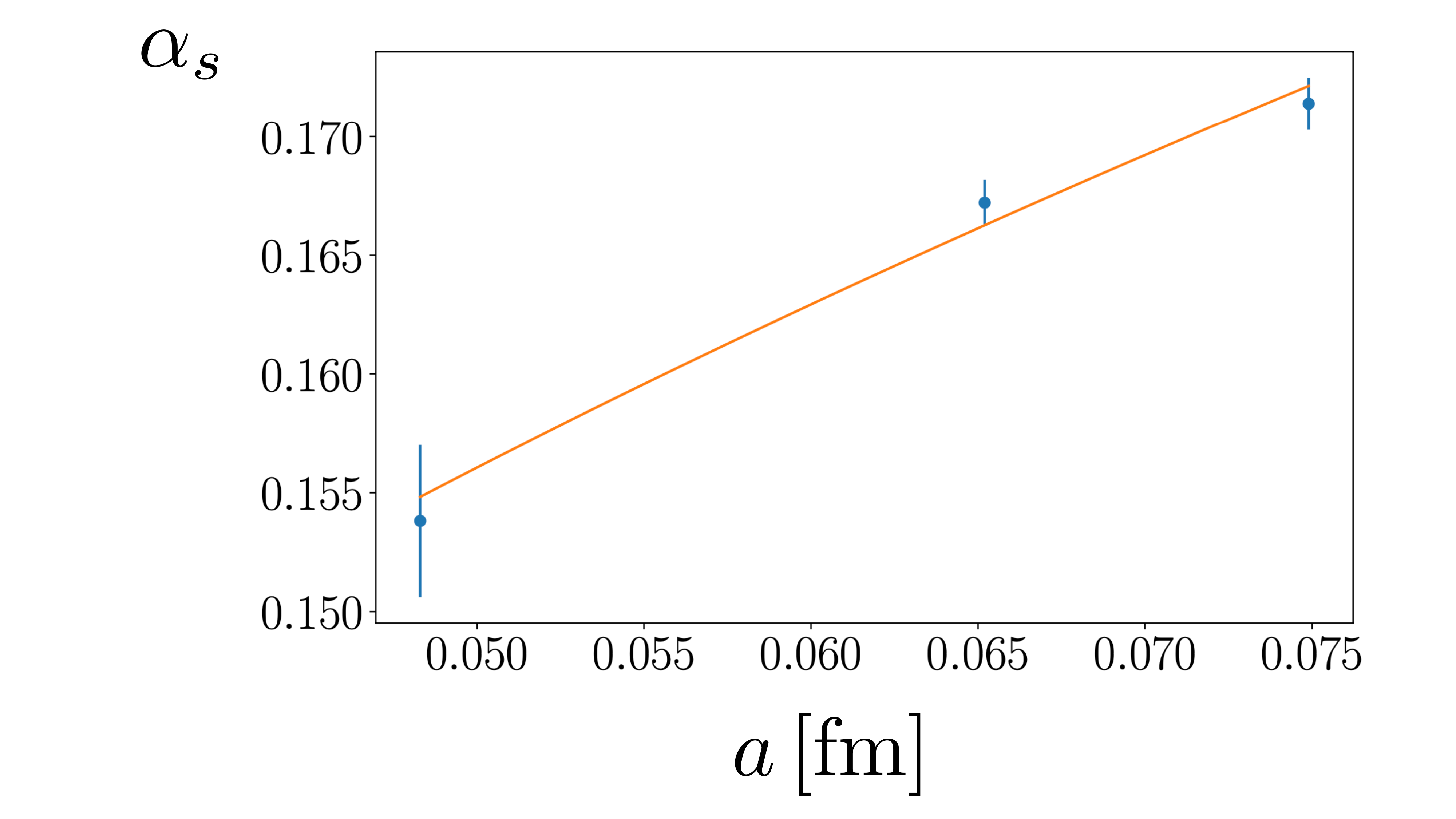}
    \caption{The values of $\alpha_s$  from lattice regulated  fits for $l(z_3,a)$  as a function of lattice spacing   in  a  model containing additional Gaussian factor,  with
    $z_{\rm min}=a$ (left), $z_{\rm min}=2a$ (center), and  $z_{\rm min}=3a$ (right). The curve, intended solely to guide the eye, represents a fit to the LO perturbative formula (\ref{eq:running_alpha}) 
 where $\Lambda_{\rm QCD}$ was the single fit parameter.  }
    \label{fig:alpha_ht_fit_lattpt}
\end{figure}

\begin{figure}[ht!]
    \centering
    \includegraphics[width=0.30\textwidth]{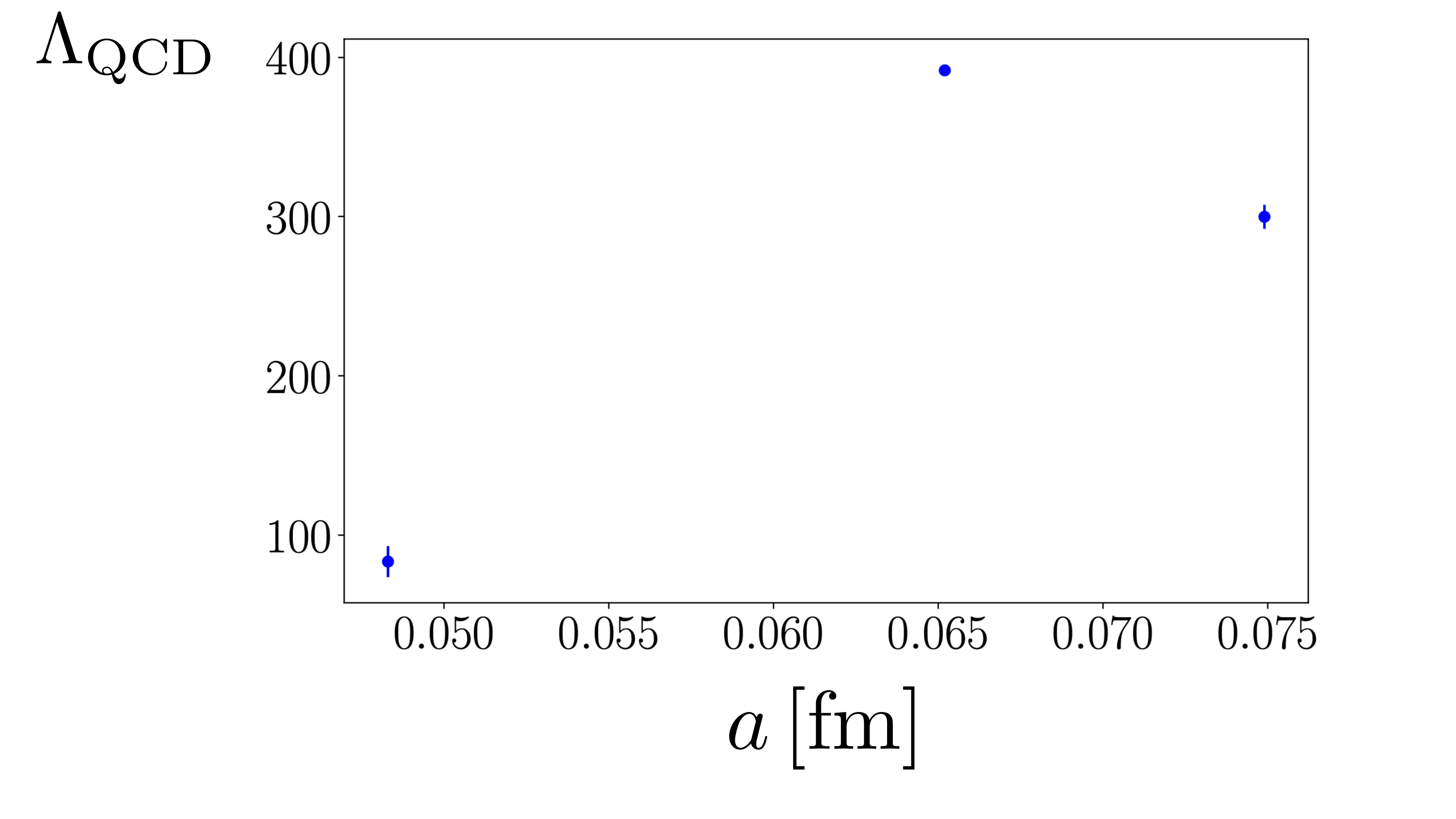}
    \includegraphics[width=0.30\textwidth]{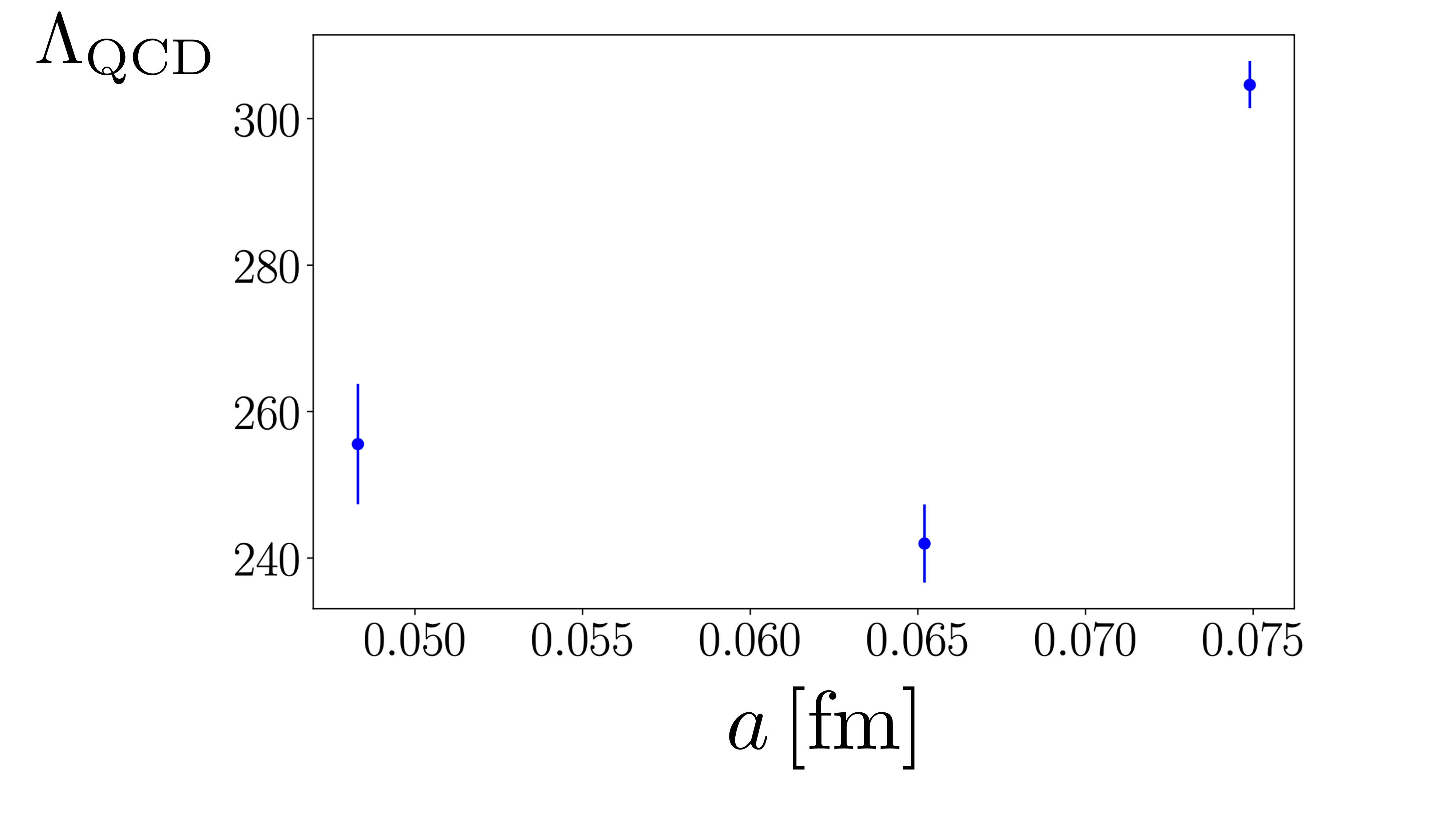}
    \includegraphics[width=0.30\textwidth]{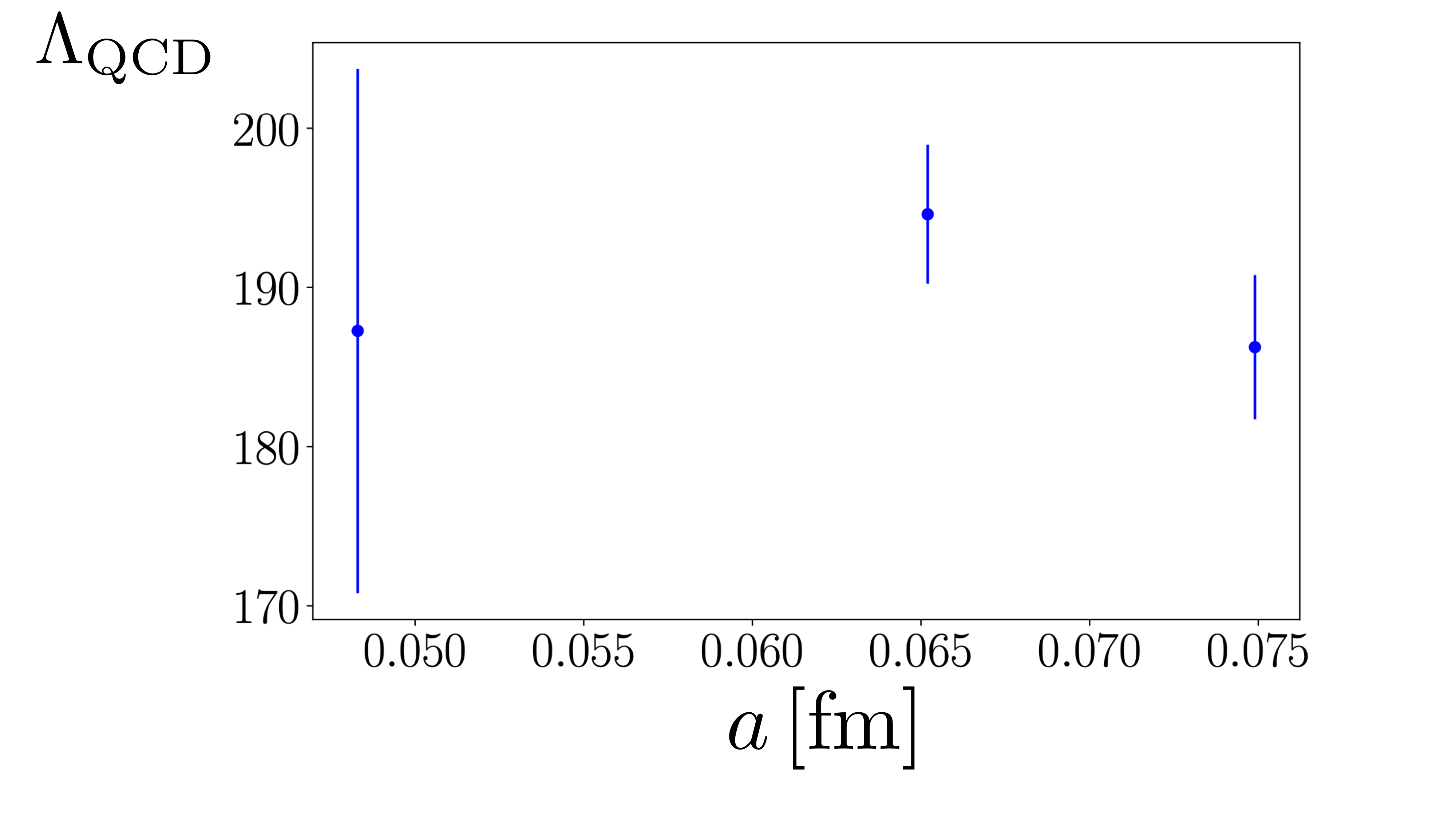}
    \caption{The value of $\Lambda_{\rm QCD}$ extracted from $\alpha_S$ using LO perturbation theory in Fig.~\ref{fig:alpha_ht_fit_lattpt} with $z_{\rm min}=a$ (left), $z_{\rm min}=2a$ (center), and  $z_{\rm min}=3a$ (right))}
    \label{fig:lam_qcd_ht_lattpt}
\end{figure}

\begin{figure}[ht!]
    \centering
    \includegraphics[width=0.30\textwidth]{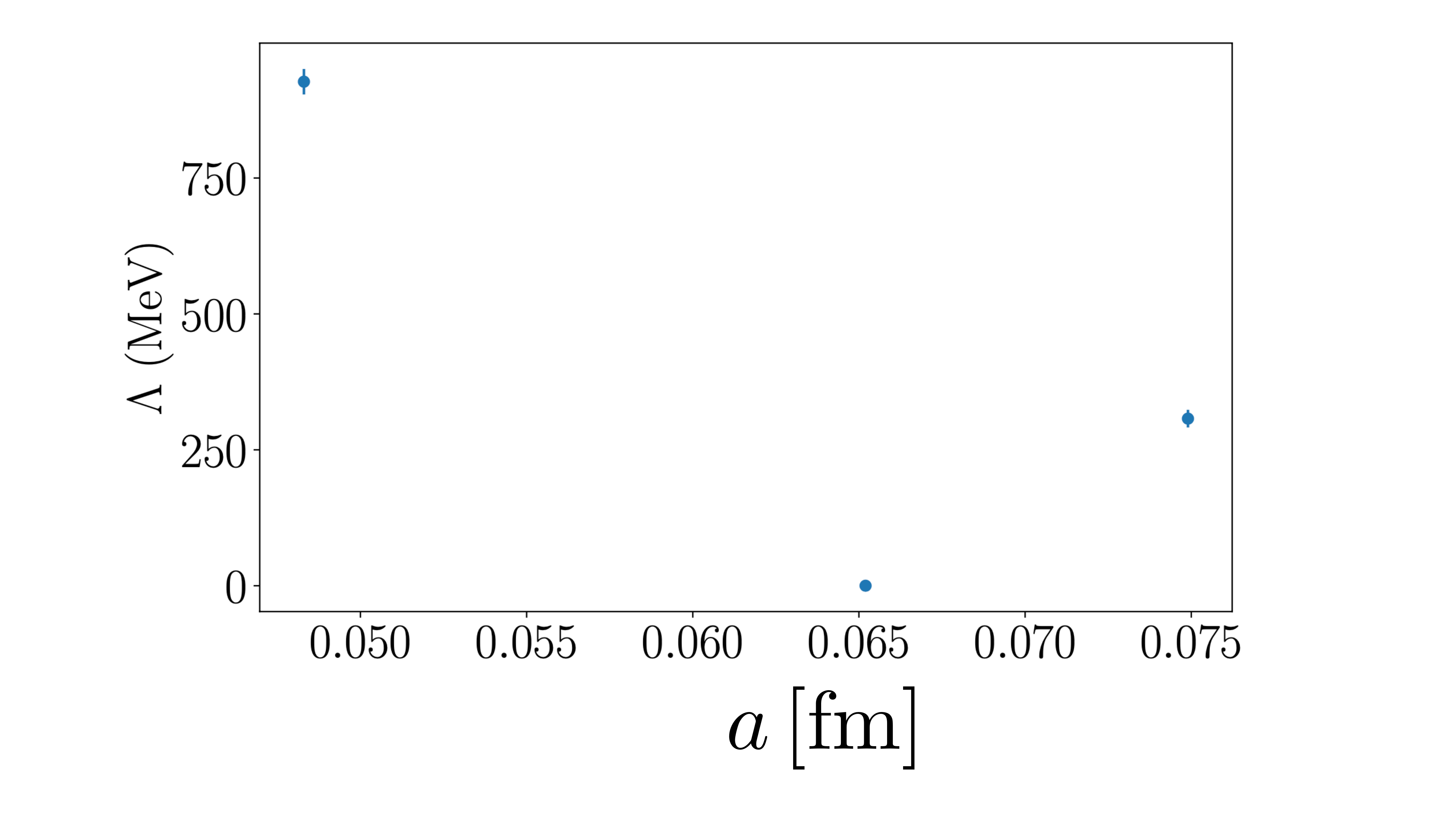}
    \includegraphics[width=0.30\textwidth]{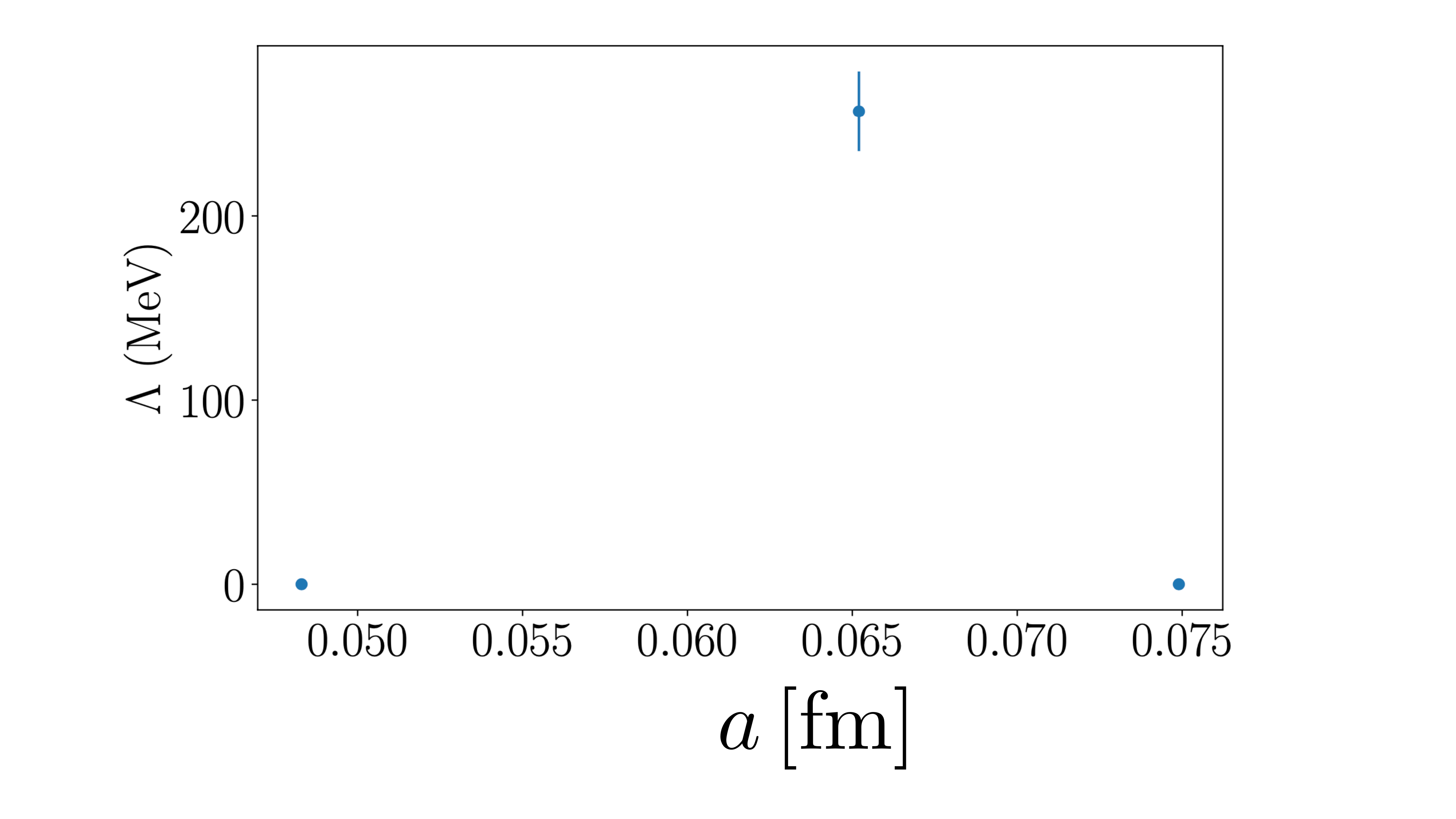}
    \includegraphics[width=0.30\textwidth]{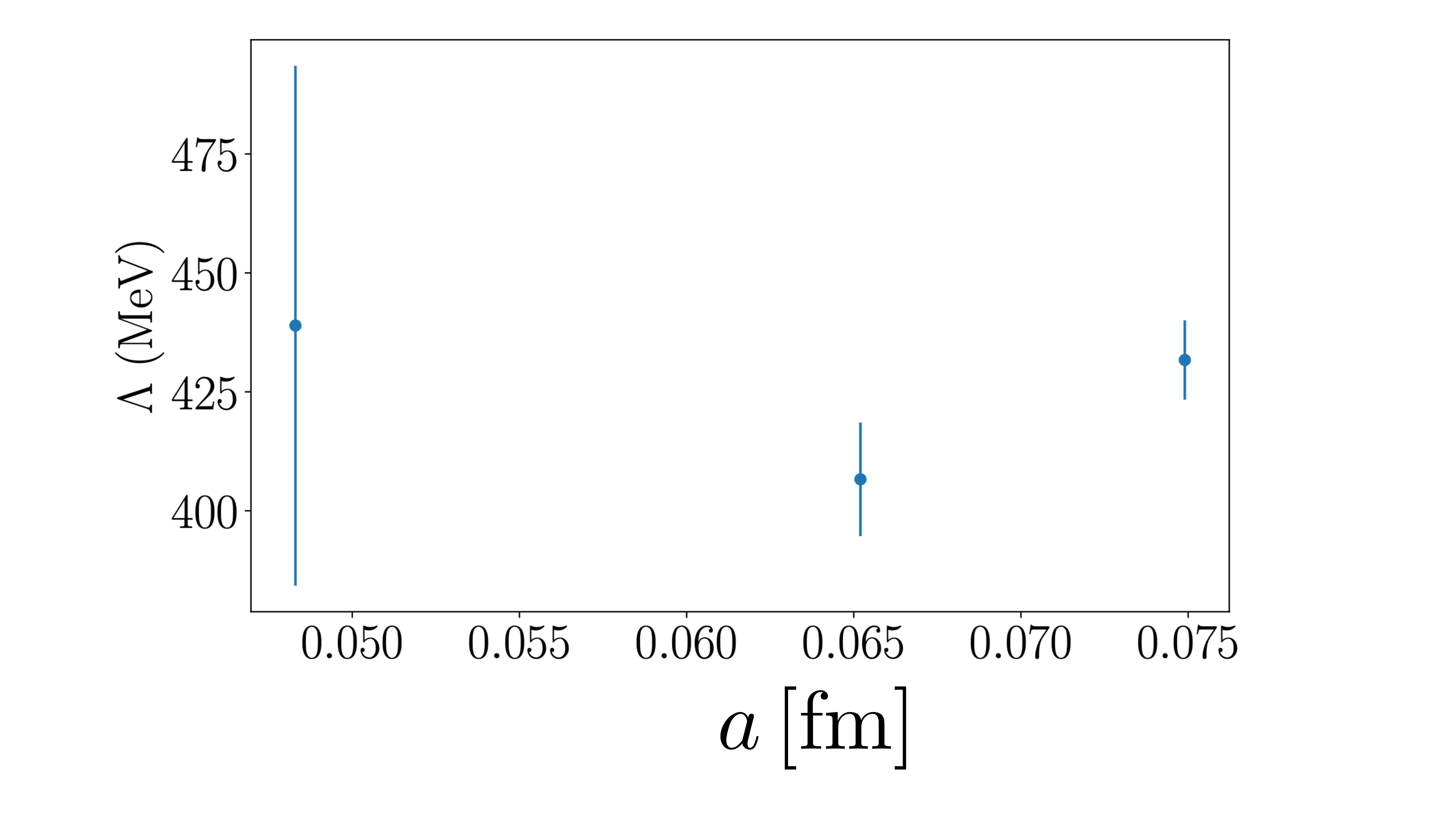}
    \caption{The fit result for the Gaussian model parameter $\Lambda$ from fits with $z_{\rm min}=a$ (left), $z_{\rm min}=2a$ (center), and  $z_{\rm min}=3a$ (right).}
    \label{fig:fit_lam_lattpt}
\end{figure}

This section contains analysis of the three ensembles to the functional form

\begin{eqnarray}
    l(z_3,a) = \ln \left ( \frac{ Z_{\rm LPT}(z_3/a,\alpha_s)}{ Z_{\rm LPT}(0,\alpha_s)}\right )-\frac{z_3^2\Lambda^2}{4}
\end{eqnarray}
where the Gaussian term  with $\Lambda^2>0$ is  assumed to reflect  the finite size of the nucleon 
and produces damping of  the matrix element  for large $z_3$. The results for  its logarithm $l(z_3,a)$ (see Eq.~\eqref{lza}) are shown in Fig.~\ref{fig:pt_ht_fit_lattpt}. The comparison of data to the model is at most a few percent shown in Fig.~\ref{fig:dot_ht_lattpt}. The fit parameters are given in Tab.~\ref{tab:fit_res_ht_latt_pt} Again only the fit with $z_{\rm min}=3a$ gives a reasonable reproduction of the data. In Fig.~\ref{fig:alpha_ht_fit_lattpt} we show the results for $\alpha_s$, which are again slightly smaller than those fit without the finite size corrections. As before, these values can be translated into a $\Lambda_{\rm QCD}$ scale, shown in Fig.~\ref{fig:lam_qcd_ht_lattpt}. In the fit with $z_{\rm min}=3a$, these values show less dependence of lattice spacing than the lattice perturbation only fits. Fig.~\ref{fig:fit_lam_lattpt} shows the results for the parameter  $\Lambda$ that governs the finite size effects.

\begin{table}[]
    \centering
    \def\arraystretch{2.0}
    \begin{tabular}{p{35pt}ccc ccc ccc}
    \hline\hline
     & ($z_{\rm min}=a$) & ($z_{\rm min}=2a$)& ($z_{\rm min}=3a$) & ($z_{\rm min}=a$) & ($z_{\rm min}=2a$) & ($z_{\rm min}=3a$)& ($z_{\rm min}=a$) &($z_{\rm min}=2a$)& ($z_{\rm min}=3a$) \\
         Ens  & $\alpha_S$  &  $\alpha_S$ &  $\alpha_S$ & $\Lambda$ (MeV)   & $\Lambda$ (MeV)  &  $\Lambda$ (MeV)  & $\chi^2$/dof& $\chi^2$/dof &  $\chi^2$/dof \\\hline
    N5  & 0.1291(2) & 0.1660(13) & 0.1538(32) & 927(24) & 0(0) & 438(54) & 1730 & 43 & 1.3(6) \\
    E5   & 0.2039(6) & 0.1771(11) & 0.1672(10) & 0(0) & 256(21) & 406(12) & 17279 & 143 & 8.6(1.1)\\
    A4p5 & 0.1960(14) & 0.1969(6) & 0.1714(11) & 307(16) & 0(0) & 431(8) & 10584 & 367 & 9.9(9) \\
    \hline\hline
    \end{tabular}
    \caption{Results of fits on all three ensembles to the LPT renormalization constant and a finite size model. The values of $\Lambda$ which are 0 occur from failures of the fit procedure to reproduce the data even with 2 parameters.}
    \label{tab:fit_res_ht_latt_pt}
\end{table}

The use of lattice perturbation theory here was crude for the shortest distances. The action used numerically included a non-perturbatively tuned clover term meant to cancel leading lattice spacing errors. The perturbative action was the na\"ive Wilson action, which neglects such improvements. The clover term is a quark-gluon coupling that extends across two lattice spacings, so its absence could easily explain large discrepancies for $z=a$ and $2a$. Since this term is added to remove discretization effects, its absence may explain why the continuum Polyakov scheme had more consistent results when fit to the data than the lattice perturbation theory result. Future perturbative calculations with improved lattice actions should be performed to test the effect of terms neglected in the current analysis.

\section{Conclusion\label{sec:conclusion}}

In this study, we analyze the forward bare matrix elements of the space-like Wilson line operator ${\cal O}^\mu(z)$ in a nucleon's rest frame calculated with numerical non-perturbative lattice QCD. The lattice data were calculated on three lattice spacings at a fixed pion mass allowing for study of the lattice divergences, which appear as functions of $z/a$. The dominant contribution to these matrix elements are from the renormalization constant of the operator.  We fit the logarithm of the rest frame matrix element to one loop results with a Polyakov and a lattice regulator. It is clear that when the lowest two separations are neglected, both of the perturbative results agree with the data 
for  $l(z_3,a)$, the logarithm of the matrix element,  to within 5\%. 
Such deviations can be expected from the accuracy of the models used.

    Beyond the UV dependence from the operator, the renormalized matrix element also contributes to scale dependence of the data. This behavior is governed by the finite size of the hadron. We add a Gaussian $z_3^2$ dependence to model this effect, which is equivalent to a Gaussian $k_T^2$ dependence of the primordial TMD. In each case, this improves the quality of the fit, but adding a second parameter also simply increases the flexibility of the model. As suggested in~\cite{Orginos:2017kos}, if present, this Gaussian dependence may be beneficial in cancelling the dominate power corrections in the moving frame matrix elements required for parton structure calculations.

    Future study is required to more definitively distinguish between perturbative and non-perturbative effects. Naturally more lattice spacings with a wider range of $\zeta$ will create greater sensitivity in the data analysis to the two scales $a$ and $z$. This may be necessary for more complex models of finite size effects. An analysis with higher order of perturbation theory is certainly required given the precision of data. Furthermore, the lattice perturbation theory should be performed with the improved actions used in the generation of the non-perturbative data, because the effects of Symanzik improvement are particularly likely to affect the small $z_3/a$ results.

\section{Acknowledgments}
This project was supported by the U.S.~Department of Energy, Office of Science, Contract No.~DE-AC05-06OR23177, under which Jefferson Science Associates, LLC operates Jefferson Lab. CJM is supported in part by U.S.~DOE ECA \mbox{\#DE-SC0023047}.
AR acknowledges support by U.S.~DOE Grant \mbox{\#DE-FG02-97ER41028}. This work has benefited from the collaboration enabled by the Quark-Gluon Tomography (QGT) Topical Collaboration, U.S.~DOE Award \mbox{\#DE-SC0023646}.
In addition, this work used resources at NERSC, a DOE Office of Science User Facility supported by the Office of Science of the U.S. Department of Energy under Contract \#DE-AC02-05CH11231. The software codes {\tt Chroma} \cite{Edwards:2004sx}, and {\tt QPhiX} \cite{QPhiX2} were used in our work. The authors also acknowledge the Texas Advanced Computing Center (TACC) at The University of Texas at Austin for providing HPC resources, like Frontera computing system~\cite{frontera} that has contributed to the research results reported within this paper. The authors acknowledge William \& Mary Research Computing for providing computational resources and/or technical support that have contributed to the results reported within this paper.



\end{document}